%% file: main.tex
\documentclass[a4paper,11pt]{article}
\pdfoutput=1 
\usepackage{jheppub} 

\usepackage[utf8]{inputenc}
\usepackage{graphicx} 
\usepackage{grffile}
\usepackage{dcolumn}  
\usepackage{makecell}
\usepackage{colordvi}
\usepackage{color}
\usepackage{amssymb}
\usepackage{amsmath}
\usepackage{bm}
\usepackage{slashed}
\usepackage{enumerate}
\usepackage{url}
\usepackage{morefloats}
\usepackage{lineno}
\usepackage{subfig}
\usepackage{graphicx}
\usepackage[percent]{overpic}
\usepackage{natbib}
\usepackage[colorlinks=true,
urlcolor=blue,
linkcolor=blue,
citecolor=blue,
linktocpage=true,
pdfproducer=medialab,
pdfa=true,
anchorcolor=blue]{hyperref}
 \usepackage{footnotebackref}
\usepackage{tabularx,array}

\newcolumntype{C}{>{\centering\arraybackslash}X}
\usepackage{multirow}
\usepackage{lineno}

\usepackage{float}
\usepackage{xcolor}
\usepackage[title]{appendix}

\renewcommand{\thesection}{\arabic{section}}

\usepackage{changepage}
\usepackage{booktabs}

\input belle2sym.tex

\preprint{}
\makeatletter
\gdef\@fpheader{}
\makeatother
\begin{document}
\raggedbottom
\title{Analysis note: one-point charge correlator with DELPHI Open Data}
\author[a]{Jingyu Zhang,}
\author[a]{Yi Chen,}
\author[b]{Kyle Lee,}
\author[c]{Ian Moult,}
\author[d]{Cristian Baldenegro,}
\author[d]{Hannah Bossi,}
\author[d]{Yen-Jie Lee}

\affiliation[a]{Department of Physics and Astronomy, Vanderbilt University, Nashville, TN, USA}
\affiliation[b]{High Energy Physics Division, Argonne National Laboratory, Lemont, IL, USA}
\affiliation[c]{Department of Physics, Yale University, New Haven, CT, 06511, USA}
\affiliation[d]{Laboratory for Nuclear Science, Massachusetts Institute of Technology, Cambridge, MA, USA}

\abstract{
We present the first measurement of the one-point charge correlator, the angular flux of electric charge in hadronic final states, using DELPHI Open Data collected at LEP-1 at $\sqrt{s} = 91.2$~GeV during 1994 and 1995. The data, corrected for detector effects, exhibit a clear $\sin(2\theta)$ modulation, consistent with the parity-violating hadronic charge flow that the chiral structure of the $Z$ couplings imprints on the final state. The measurement demonstrates the experimental feasibility of the observable and establishes strategies for controlling associated detector effects, thereby motivating a new program to measure charge-flux observables. This note documents the experimental details supporting the companion experimental paper~\cite{Zhang:2026vdo} and the joint theory--experiment Letter~\cite{Zhang:2026emt}.
}

\maketitle


\section{Introduction}
\label{sec:intro}

Measurements of patterns of asymptotic energy flux produced in colliders have provided one of the key means of uncovering the microscopic structure of the Standard Model. With the advent of the LHC, this has evolved into the sophisticated jet substructure program, with applications ranging from searches for beyond the Standard Model physics, to nuclear physics measurements \cite{Larkoski:2017jix}. In the last several years, there has been a program to reformulate the study of asymptotic energy flux in terms of correlation functions, instead of jets. This is particularly natural in the case of electron-positron colliders. Instead of working with jets, one can consider correlation functions of the energy flow operator
\cite{Sterman:1975xv,Korchemsky:1997sy}
\begin{align}
\mathcal{E}(\vec n) = \lim_{r\to\infty} r^2 \int_0^\infty dt\, n^i\, T_{0i}(t, r\vec n)\,.
\end{align}
Correlation functions of the energy flow operator exhibit many elegant theoretical properties. Most importantly, they can be directly measured. They have therefore enabled a direct connection between theory and experiment \cite{Moult:2025nhu}.

It has long been realized that properties beyond energy flux carry crucial information about the underlying microscopic interactions. In particular, the study of jet charge at colliders was first proposed by Feynman and Field \cite{Feynman:1972xm,Field:1977fa}. The original definition by Feynman and Field \cite{Field:1977fa}
\begin{align}
Q_\kappa = \sum_{h\in \text{jet}} z^\kappa Q\,,
\end{align}
used an energy weight $z^\kappa$, where $z$ denotes the energy fraction of the hadron to suppress low energy radiation, which is difficult to control both experimentally, and theoretically. The jet charge was first used in early DIS measurements \cite{Erickson:1979wa,Fermilab-Serpukhov-Moscow-Michigan:1979zgc,EuropeanMuon:1984xji} and first applied to the hadronic forward--backward charge asymmetry in $e^+e^-$ annihilation at TRISTAN~\cite{AMY:1989fku}. Relevant for this article, it then played a major role at LEP and SLC, including measurements of the inclusive hadronic charge asymmetry and of $\sin^2\theta_{\rm W}^{\rm eff}$ \cite{ALEPH:1991fba,L3:1991gfs,DELPHI:1991mqi,OPAL:1992jsm,ALEPH:1996qlh,L3:1998jet,SLD:1996gjt}, measurements of heavy-flavour forward--backward asymmetries with jet-charge and inclusive charge tags \cite{OPAL:1993wua,ALEPH:2001mdb,L3:1992fsb,DELPHI:2004wvq,OPAL:2002ddm,L3:2000vgx,SLD:2005gev}, measurements of $B-\bar B$ mixing~\cite{ALEPH:1992net,DELPHI:1993vqf,DELPHI:1996hzb,OPAL:1994xvz,L3:1994fxw}, 
and measurements of the triple-gauge boson coupling \cite{ALEPH:1997agc}. More recently, the jet charge has been studied \cite{Krohn:2012fg,Waalewijn:2012sv}, and measured \cite{ATLAS:2015rlw} at the LHC.

Due to the success in formulating questions about energy flux in terms of energy correlators, it is interesting to understand if this can also be achieved for charge flux. This would open up applications of correlator based techniques to a wide variety of physics applications, in particular, at electron-positron colliders, the study of forward-backward asymmetries. Correlators of charge were defined in \cite{Hofman:2008ar}
\begin{align}
\label{eq:chargedet}
\mathcal{Q}(\vec n) = \lim_{r\to\infty} r^2 \int_{0}^\infty dt\, n_i\, J^i_Q(t, r\vec n)\,,
\end{align}
where $J$ denotes the current for the associated charge. Theoretically, they exhibit a number of interesting properties \cite{Hofman:2008ar}. Similar to the Feynman-Field definition, charge correlators with an energy weight were introduced in \cite{Lee:2023npz,Lee:2023tkr}, and measured in \cite{ALICE:2026zyx}. However, the study of the charge correlator in Eq.~\eqref{eq:chargedet} is special due to the fact that it is associated with a conserved charge, while the energy-weighted charge is not conserved. This implies a number of beautiful theoretical properties of the charge detector. While these have long been theoretically appreciated \cite{Hofman:2008ar}, no measurement of the charge correlator has ever been performed.

In this analysis note, we discuss in detail the experimental challenges of measuring charge correlators at electron--positron colliders, and present the first measurement of a one-point charge correlator in QCD, performed at the $Z$ pole with archival DELPHI data. The measured correlator exhibits a pronounced asymmetry, imprinted by the chiral structure of the $Z$-boson couplings. Most importantly, it demonstrates the experimental feasibility of charge correlator measurements, forging a new connection between theory and experiment. 

The measurement uses DELPHI Open Data~\cite{DELPHI:2024opendata} recorded at LEP-1 at $\sqrt{s} = 91.2$~GeV in 1994 and 1995. A detailed study of the DELPHI Open Data, including measurements of event-shape observables, is presented in Ref.~\cite{Zhang:2025delphiEEC}. This note focuses on the charge-sensitive observables and the corrections they require. Section~\ref{sec:observable} provides a more detailed definition of the observable. Section~\ref{sec:data_samples} describes the data and simulated samples. The raw data are corrected with a combined MC-based and data-driven strategy: the MC-based correction for track reconstruction efficiency is described in Section~\ref{sec:mc_correction}, and the residual data--simulation differences in charge reconstruction are quantified in a control region of $e^+e^- \to \tau^+\tau^-$ events, as detailed in Section~\ref{sec:data_driven}. Charge calibrations at LEP were traditionally derived from the hadronic sample itself and supplemented by extensive calibrations of the detector response, such as the track sagitta and the detector material. Self-correction methods are explored in Appendix~\ref{sec:hadronic_correction}. In the present measurement, the effect of detector-response mismodeling is instead bounded by the tag-and-probe measurement, with dedicated detector-response calibrations left to future precision measurements. Sections~\ref{sec:systematics} and~\ref{sec:results} describe the systematic uncertainties and the final results, respectively, and further cross-checks are presented in the remaining appendices.

\section{Definition of the observable}
\label{sec:observable}

At the $Z$ pole, the Born-level differential cross section for $e^+e^- \to q\bar{q}$ takes the form
\begin{equation}
\frac{\mathrm{d}\sigma}{\mathrm{d}\cos\theta} \;=\;
\frac{3\,\sigma_{q\bar{q}}}{8}\left(1 + \cos^2\theta\right)
\;+\; \sigma_{q\bar{q}}\,A_{\rm FB}^{q}\,\cos\theta ,
\label{eq:diffxsec}
\end{equation}
where $\theta$ is the polar angle of the outgoing quark with respect to the $e^-$ beam direction and $\sigma_{q\bar{q}}$ is the total $Z\to q\bar{q}$ cross section. The two angular terms have distinct symmetry under $\theta \leftrightarrow \pi - \theta$. The $(1+\cos^2\theta)$ component is parity-even and dominates the rate, while the parity-odd $\cos\theta$ component carries the entirety of the parity-violating information, with coefficient the forward--backward asymmetry 
\begin{equation}
A_{\rm FB}^{q} = \frac{\sigma_{\rm F} - \sigma_{\rm B}}{\sigma_{\rm F} + \sigma_{\rm B}},
\end{equation}
where $\sigma_{\rm F}$ and $\sigma_{\rm B}$ are the cross sections for the quark to be produced in the forward and backward hemispheres.

The quark direction appearing in Eq.~\eqref{eq:diffxsec} is not itself an observable: confinement means that only the final state hadrons reach the detector. The one-point charge correlator is the expectation value of the charge flux operator, defined in Eq.~\eqref{eq:chargedet}, $\langle \mathcal{Q}(\vec{n})\rangle$, averaged over the $Z$ decay final state hadrons.
Experimentally, measured against the polar angle $\theta$, instead of the solid angle pointed by $\vec{n}$, this comes out as a normalized charge-weighted cross section $\mathcal{Q}(\theta)$, and is a direct sum over reconstructed charges,
\begin{equation}
    \mathcal Q(\theta)
    =
    \left\langle \sum_{i\in X} q_i\,\delta(\theta-\theta_i) \right\rangle
    =
    \frac{1}{N_{\rm evt}}
    \sum_{e=1}^{N_{\rm evt}}
    \sum_{i\in e} q_i\,\delta(\theta-\theta_i), 
    \label{eq:qdef}
\end{equation}
where $N_{\rm evt}$ denotes the total numbers of events. 
The measurement evaluates this continuous distribution in finite angular bins. For a bin $a$ of width $\Delta\theta_a$, let
\begin{equation}
    N_a^{\pm}
    \equiv
    \frac{Y_a^{\pm}}{N_{\rm evt}\,\Delta\theta_a}
    \label{eq:charged_densities}
\end{equation}
denote the positive- and negative-track densities, per event and per unit polar angle, where $Y_a^{\pm}$ are the corresponding total track yields. The measured value in the bin is the bin average of Eq.~\eqref{eq:qdef},
\begin{equation}
    \mathcal Q_a
    \equiv
    \frac{1}{\Delta\theta_a}
    \int_{\mathrm{bin}\,a} \mathcal Q(\theta)\,\mathrm d\theta
    = N_a^+ - N_a^-,
    \label{eq:qNN}
\end{equation}
and the corresponding total charged-particle density is
\begin{equation}
    N_{{\rm tot},a} \equiv N_a^+ + N_a^-.
    \label{eq:ntot}
\end{equation}
Thus, $\mathcal Q_a$ is the experimentally measured, binned representation of the charge correlator $\mathcal Q(\theta)$. In the following, $\mathcal Q(\theta)$, $N^{\pm}(\theta)$, and $N_{\rm tot}(\theta)$ always denote these per-event densities per unit polar angle, plotted at the angular-bin centers. Bin widths are expressed in radians.

As formulated in the companion paper, the asymptotic charge flux inherits the symmetries of the Born-level cross section. The parity-even $(1+\cos^{2}\theta)$ term that dominates Eq.~\eqref{eq:diffxsec} contributes to the total charged-particle density but cancels in the charge flux, which retains only the parity-odd $\cos\theta$ structure. Per unit solid angle the flux is therefore modulated as $\cos\theta$; expressed as a density in the polar angle, as in Eq.~\eqref{eq:qdef}, the $\sin\theta$ Jacobian of the angular measure converts this into a $\sin(2\theta)$ modulation. Hadrons in the forward hemisphere, $\theta < 90^\circ$, carry on average a net charge correlated with that of the quark, and those in the backward hemisphere with that of the antiquark, since the quark is produced preferentially in the forward direction. The charge-symmetric part of fragmentation instead produces charge-balanced pairs that are nearby in angle and compensate locally. The size of the modulation is proportional to the flavor-weighted combination $A_{\rm FB}=\sum_q R_{q}\,Q_q\,A_{\rm FB}^{q}$, where $R_q = \Gamma_{q\bar{q}}/\Gamma_{\rm had}$ is the hadronic branching fraction of flavor $q$, $Q_q$ its electric charge, and $A_{\rm FB}^{q}$ its forward--backward asymmetry.

\clearpage
\section{Detector, datasets, simulations, and selections}
\label{sec:data_samples}
DELPHI is one of the four large general-purpose detectors at LEP and collected data from 1989 to 2000. A comprehensive description of its design and performance can be found in Refs.~\cite{DELPHI:1990cdc, DELPHI:1995dsm}. It comprises over 20 sub-detectors arranged in a cylindrical geometry around the interaction point, with charged-particle tracking performed in a 1.2~T solenoidal magnetic field parallel to the beam axis. The tracking system, the component most critical for this measurement, consists of the silicon Microvertex Detector (upgraded in 1994 to double-sided sensors enabling three-dimensional impact-parameter reconstruction), the Inner Detector drift chamber, the Time Projection Chamber (TPC), and the Outer Detector, complemented in the forward regions by the FCA and FCB planar drift chambers, which extend the tracking coverage down to polar angles of about $11^\circ$. The TPC, the main tracking device, reconstructs tracks with up to 16 space points between polar angles of $39^\circ$ and $141^\circ$, and with at least four space points down to $21^\circ$ and up to $159^\circ$, motivating the polar-angle acceptance of the track selection used in this analysis. Importantly for this measurement, the apparatus is symmetric in the azimuthal angle $\phi$ and, to good approximation, forward--backward symmetric under $\theta \to 180^\circ - \theta$: the nominal detector geometry therefore does not by itself induce a parity-odd modulation of the type targeted here, and instrumental contributions can enter only through residual charge- and $\theta$-dependent response differences, which are the subject of the corrections described in Sections~\ref{sec:mc_correction} and~\ref{sec:data_driven}.

This analysis uses data collected by DELPHI at a center-of-mass energy of $\sqrt{s} = 91.2$~GeV during 1994 and 1995~\cite{DELPHI:OpenData:short94_c2, DELPHI:OpenData:short95_d2}. The 1994 data correspond to an integrated luminosity of 46~pb$^{-1}$. In 1995, LEP concluded its $Z$ operations with an energy scan. Only the on-peak data from this scan, corresponding to 15~pb$^{-1}$, are used in this analysis. The total integrated luminosity is 61~pb$^{-1}$. The DELPHI Open Data release~\cite{DELPHI:2024opendata} and data re-use policy~\cite{DELPHI:2024policy} make these datasets publicly available.

The primary MC samples used for correction are generated with \textsc{PYTHIA}~8.3~\cite{Bierlich:2022pfr} with the Monash Tune~\cite{Skands:2014pea}, processed through the full DELPHI detector simulation using DELSIM~\cite{DELSIM}. The DELPHI detector and offline software differ slightly between the two years, denoted by the tags $94\_\text{c}$ and $95\_\text{d}$ in the DELSIM program. Separate samples of 5 million hadronic $Z \to q\bar{q}$ events per year are therefore produced for the two configurations. The simulated events include the full chain of parton showering, hadronization, and detector response. To ensure consistent treatment of detector effects, the analysis is performed independently on the two datasets; the fully corrected distributions are then combined to obtain the final results. Leptonic samples of $Z \to \tau^+\tau^-$, $Z \to \mu^+\mu^-$, and $Z \to e^+e^-$, of one million events each per year, are generated with the same \textsc{PYTHIA}~8.3 configuration and processed through the identical DELSIM detector simulation. These samples provide the signal and background estimates of the $\tau^+\tau^-$ control region (Section~\ref{sec:data_driven}).
Two additional MC samples from the publicly released DELPHI production 
are used to evaluate systematic uncertainties. From the most 
recent legacy DELPHI campaign~\cite{DELPHI:OpenData:kk2f_pythia_94, 
DELPHI:OpenData:kk2f_pythia_95}, these samples use 
\textsc{KK}2f~\cite{Jadach:1999vf} for the hard 
$e^+e^- \to f\bar{f}$ process and 
\textsc{PYTHIA}~6.1/\textsc{JETSET}~7.4~\cite{Sjostrand:1995vx} for 
parton showering and hadronization.

\begin{figure}[ht!]
    \centering
    \includegraphics[width=0.8\textwidth]{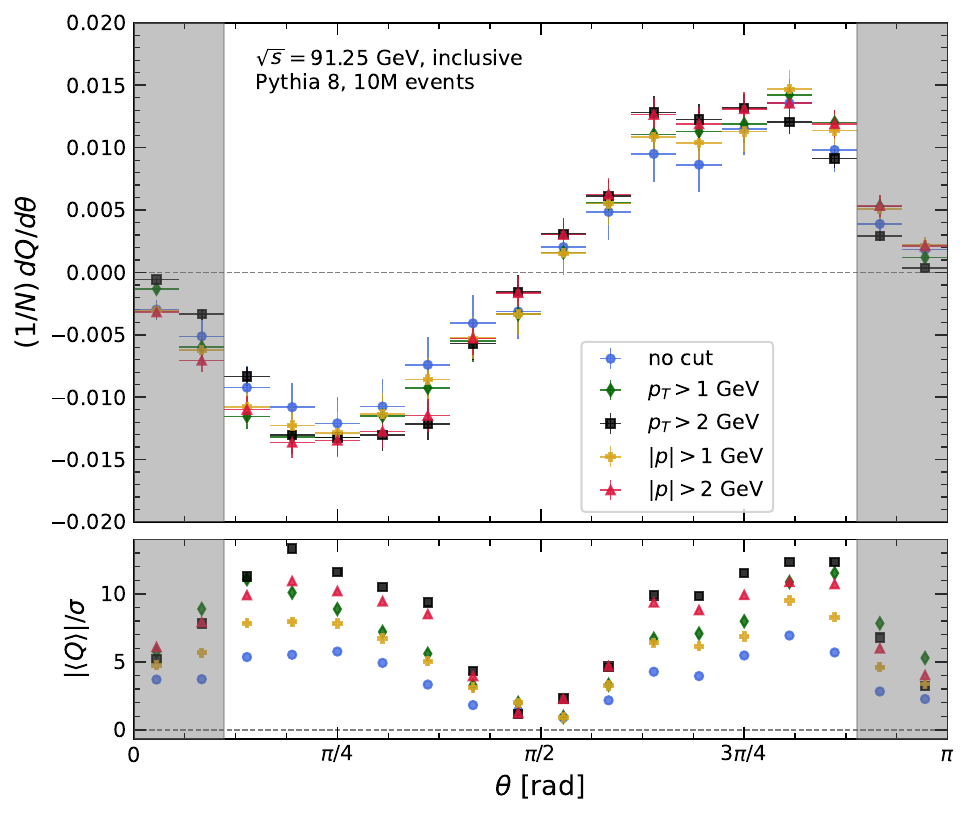}
    \caption{Generator-level comparison of the one-point charge correlator for the fully inclusive selection and for the $p_{\rm T} > 1$, $p_{\rm T} > 2$, $|p| > 1$, and $|p| > 2$~GeV selections at $\sqrt{s} = 91.2$~GeV. Top panel shows the charge correlator as a function of $\theta$, with the dashed curves indicating $A\sin(2\theta)$ fits. Bottom panel shows the per-bin significance $|{\rm data}|/\sigma_{\rm stat}$. The gray bands indicate the regions outside the detector acceptance. The $p_{\rm T}$-based selections yield higher significance than the $|p|$-based ones across the full angular range, while preserving the $\sin(2\theta)$ modulation shape.}
    \label{fig:pt_vs_p_significance}
\end{figure}

The baseline track and event selections follow those established in Refs.~\cite{DELPHI:2003yqh, DELPHI:2004wvq} and utilized in Ref.~\cite{Zhang:2025delphiEEC}.
Charged particles are required to lie within the polar-angle acceptance $20^\circ \le \theta \le 160^\circ$, to have a measured track length greater than 30~cm, and to satisfy a relative momentum uncertainty $\Delta p/p \le 1.0$.
Two requirements are tightened relative to the baseline to protect the charge measurement.
First, the track impact parameters are restricted to $|d_0| \le 0.6$~cm and $|z_0| \le 1.0$~cm, which removes displaced and secondary tracks whose charge is most prone to misreconstruction; the resulting improvement in charge balance is shown in Figure~\ref{fig:app_d0z0_compare} of Appendix~\ref{app:additional_datamc}.
Second, the track transverse momentum is raised to $p_{\rm T} > 2$~GeV.
This cut suppresses soft tracks that are sensitive to charge mismodeling from multiple Coulomb scattering, further improving the charge balance (Figure~\ref{fig:app_pcut_compare} of Appendix~\ref{app:additional_datamc}), and it also enhances the statistical sensitivity to the charge-asymmetry signal, as discussed below (Figure~\ref{fig:pt_vs_p_significance}).
Together, these tighter requirements also bring the signal-region track kinematics closer to those of the $\tau^+\tau^-$ control region used for the data-driven charge-misreconstruction correction (Section~\ref{sec:data_driven}), strengthening the transfer of that correction to the hadronic sample.
Hadronic events are selected by requiring at least 7 good charged tracks, total reconstructed energy $E_{\rm vis} \ge 0.5 E_{\rm cm}$, and a thrust-axis polar angle in the range $30^\circ \le \theta_{\rm thrust} \le 150^\circ$.
These selections suppress background from leptonic final states and two-photon collisions, and remove the tail of radiative-return events in which hard initial-state radiation lowers the effective annihilation energy $\sqrt{s'}$. Approximately 1.6 million hadronic events are selected across both years. 
The visible-energy distribution before and after the event selection is shown in Figure~\ref{fig:app_evis} of Appendix~\ref{app:additional_datamc}.

Because the charge correlator is constructed as the small difference between two large track densities, $\mathcal{Q}(\theta) = N^+(\theta) - N^-(\theta)$, the measurement is intrinsically statistics-limited. The two densities are both of order $N \sim N^+(\theta) \sim N^-(\theta)$ in each $\theta$ bin, while the parity-violating signal of interest is a much smaller residual $\mathcal{Q}(\theta) = \mathcal{O}(\varepsilon\,N)$ with $\varepsilon \ll 1$. A naive Poisson estimate of the per-bin precision then gives $\sigma_{\rm stat}(\mathcal{Q}) \sim \sqrt{2N}$, so the achievable per-bin significance scales as $\varepsilon\,N / \sqrt{2N} = \varepsilon\sqrt{N/2}$ and is set by the degree of input track sample self-cancelling, in addition to the available integrated luminosity.

Physically, the bulk of $N^+(\theta)$ and $N^-(\theta)$ comes from the soft fragmentation products of the parton shower. Each $g \to q\bar{q}$ or $q \to q\,g \to q\,q'\bar{q}'$ branching produces a charged pair of equal and opposite charge, so by construction these contributions cancel exactly in $\mathcal{Q}(\theta)$ on average and contribute only through Poisson fluctuations. These fragmentation tracks, being charge-symmetric, act as a background that degrades the statistical significance of the modulation without shifting it, so the charge correlator is evaluated requiring transverse momentum $p_{\rm T} > 2$~GeV to increase per-bin statistical sensitivity. 

As an example, Figure~\ref{fig:pt_vs_p_significance} compares the generator-level one-point charge correlator for the fully inclusive selection and for the $p_{\rm T} > 1$, $p_{\rm T} > 2$, $|p| > 1$, and $|p| > 2$ selections, using 10 million generator-level \textsc{PYTHIA}~8.3 events. The per-bin significance $|{\rm data}|/\sigma_{\rm stat}$ is shown in the lower panel. As the threshold increases, the significance improves systematically across the full $\theta$ range, indicating that higher-$p_{\rm T}$ selections more effectively suppress tracks from the soft-fragmentation background in the forward and backward regions, which are very sensitive to detector mismodeling due to multiple Coulomb scattering, while preserving the charge-carrying signal tracks and the $\sin(2\theta)$ modulation shape of the inclusive selection. However, raising the threshold further would distort the $\sin(2\theta)$ modulation: to first order, the asymmetry information is carried by the primary high-$p_{\rm T}$ hadrons that retain the charge of the primary quark, while the soft hadrons from gluon fragmentation are pair-produced with zero net charge. A moderate threshold therefore removes only the charge-symmetric soft background, but an overly tight threshold begins to cut into the charge-carrying signal hadrons themselves, sculpting the modulation shape. The threshold $p_{\rm T} > 2$~GeV is therefore adopted as a more optimal working point. A less optimal $|p| > 2$~GeV selection serves as a cross-check, with results provided in Appendix~\ref{app:fiducial_check}. A complete summary of all particle and event selection requirements is given in Table~\ref{tab:SelectionSummary}.

\begin{figure}[ht!]
    \centering
    \includegraphics[width=0.55\textwidth]{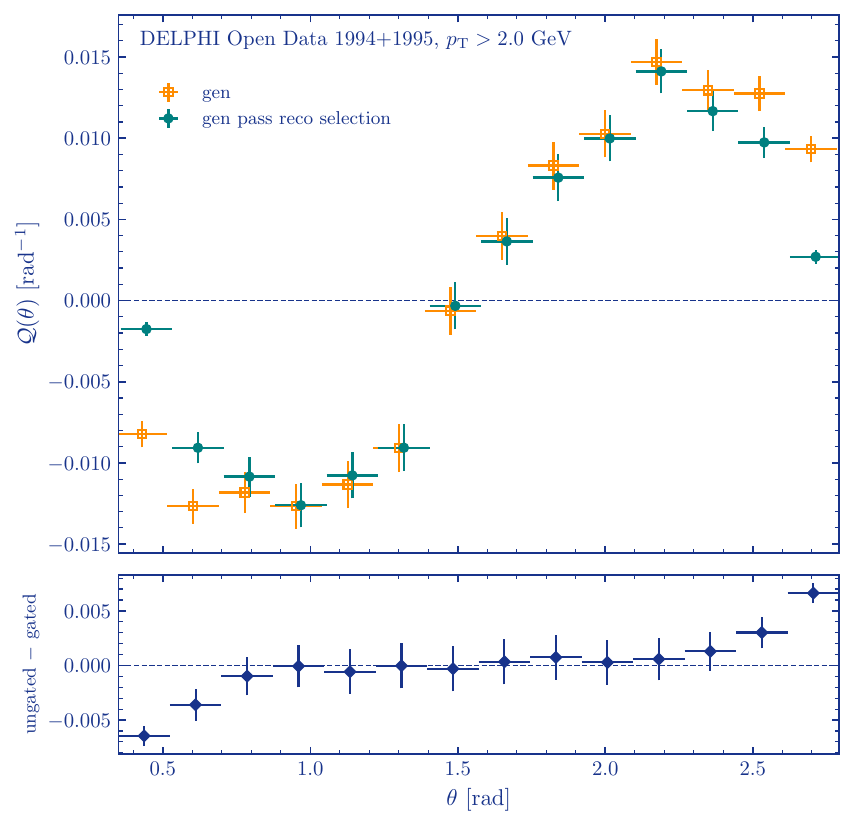}
    \caption{Generator-level one-point charge correlator from \textsc{PYTHIA}~8.3 for tracks with $p_{\rm T} > 2$~GeV, combining the 1994 and 1995 configurations, for all generated events (orange) and for the subset of events passing the reconstruction-level hadronic event selection of Table~\ref{tab:SelectionSummary} (teal). The lower panel shows the difference between the two distributions. The event selection leaves the central region unchanged but suppresses the modulation in the bins nearest the acceptance edges, dominated by the thrust-axis polar-angle requirement. This shape change is included as a correction when the final results are compared to theory predictions (Section~\ref{sec:results}).}
    \label{fig:evtsel_gating}
\end{figure}

While the track-level $p_{\rm T}$ requirement thus preserves the $\sin(2\theta)$ modulation shape of the inclusive selection, the event selection does not. Figure~\ref{fig:evtsel_gating} compares the generator-level charge correlator before and after the reconstruction-level event selection: the two agree in the central region, but the selection suppresses the modulation in the bins nearest the edges of the angular acceptance, primarily through the thrust-axis polar-angle requirement, which removes events whose quark axis points close to the beam direction. Since the theoretical predictions are computed for inclusive kinematics, this selection-induced shape change is included as a correction when the final corrected results are compared to theory predictions in Section~\ref{sec:results}.

\begin{table}[ht]
\centering
\begin{tabularx}{\textwidth}{l|l}
\hline\hline
\multicolumn{2}{l}{Charged particles}  \\
\hline
Acceptance              & $20^\circ\le\theta\le160^\circ$ \\
Transverse momentum     & $p_{\rm T} > 2~\text{GeV}$ \\
High quality tracks     & measured track length $\ge 30~\text{cm}$ \\
                        & $\Delta p/p \le1.0 $ \\
Impact parameter        & $|d_0|\le0.6$~cm, $|z_0|\le1.0$~cm \\
\hline\hline
\multicolumn{2}{l}{Event selection}  \\
\hline
Hadronic events         & $30^\circ \le \theta_{\rm thrust} \le 150^\circ$\\
                        & at least 7 good tracks \\
                        & $E_{\rm vis} \ge 0.5E_{\rm cm}$ \\
\hline\hline
\end{tabularx}
\caption{Summary of particle and event selections used in this analysis. The acceptance and transverse momentum cuts define the fiducial phase space of this analysis.}
\label{tab:SelectionSummary}
\end{table}

\begin{figure}[h]
    \centering
    \includegraphics[width=0.45\textwidth]{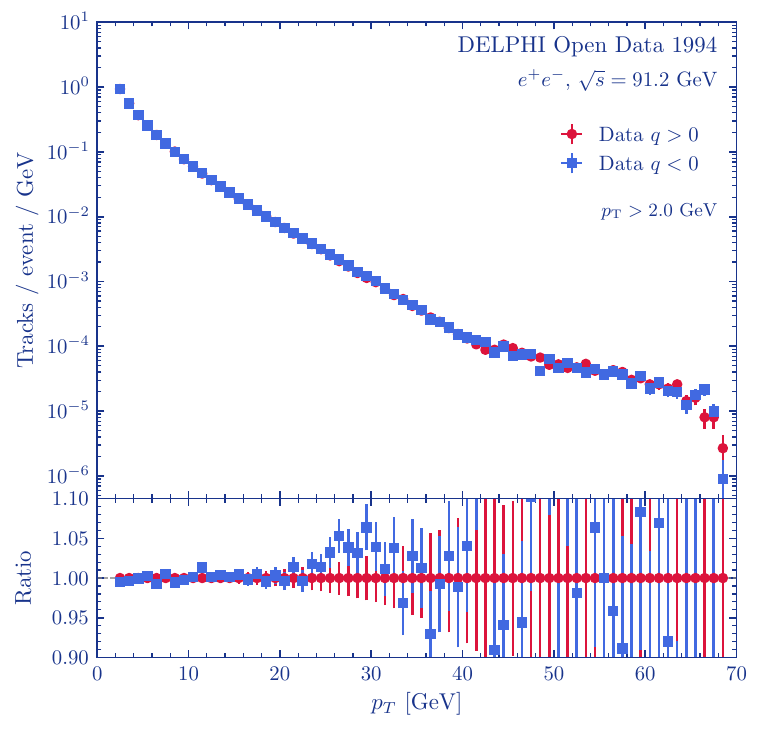}
    \includegraphics[width=0.45\textwidth]{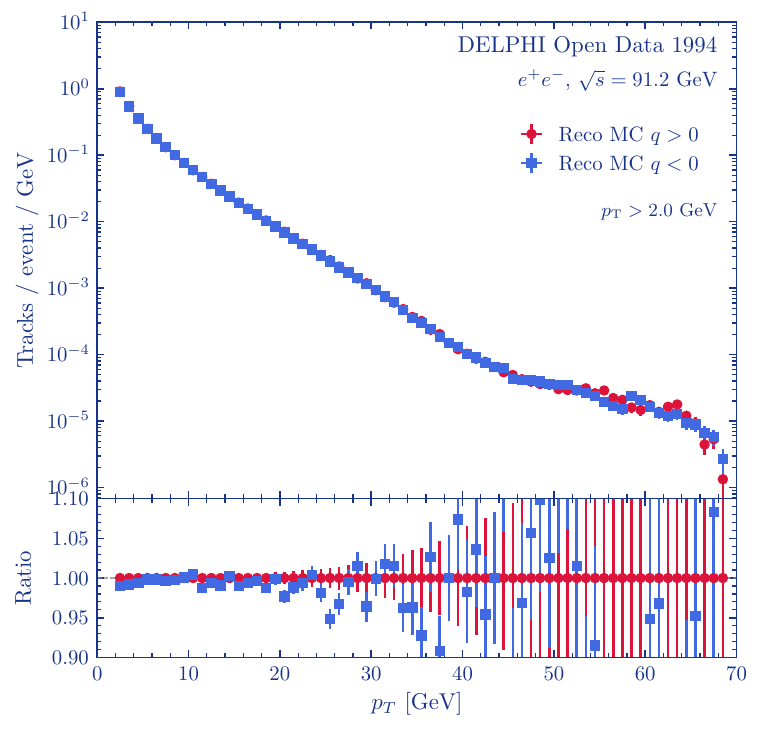}
    \caption{Charged-track momentum spectra in 1994 data (left) and reconstructed \textsc{PYTHIA}~8.3 MC (right), separated by reconstructed charge, as a function of $p_{\rm T}$. }
    \label{fig:data_mc_theta}
\end{figure}

\begin{figure}[ht!]
    \centering
    \includegraphics[width=0.48\textwidth]{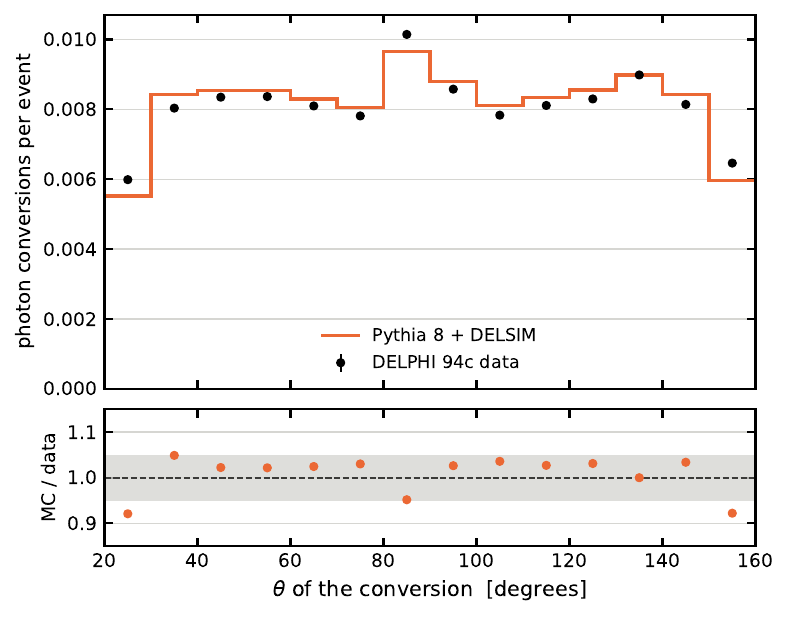}
    \includegraphics[width=0.48\textwidth]{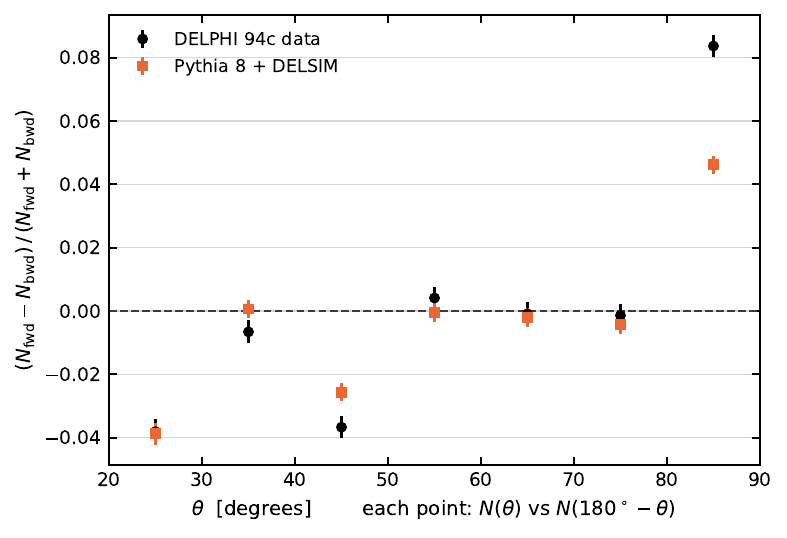}
    \caption{Photon conversions in the 1994 DELPHI Open Data sample (black) compared to the reconstructed \textsc{PYTHIA}~8.3 simulation (orange). Left: rate of reconstructed photon conversions per event as a function of the polar angle of the conversion, with the simulation-to-data ratio in the lower panel; the shaded band indicates $\pm 5\%$. Right: forward--backward asymmetry of the conversion rate, $(N_{\rm fwd} - N_{\rm bwd})/(N_{\rm fwd} + N_{\rm bwd})$, where each point compares the bin at $\theta$ with its mirror bin at $180^\circ\!-\theta$. The conversion rate probes the amount of detector material, and its forward--backward asymmetry the parity-odd component of the material distribution. The bin at $85^\circ$ contains the TPC cathode plane.}
    \label{fig:det_conversions}
\end{figure}

\begin{figure}[ht!]
    \centering
    \includegraphics[width=0.48\textwidth]{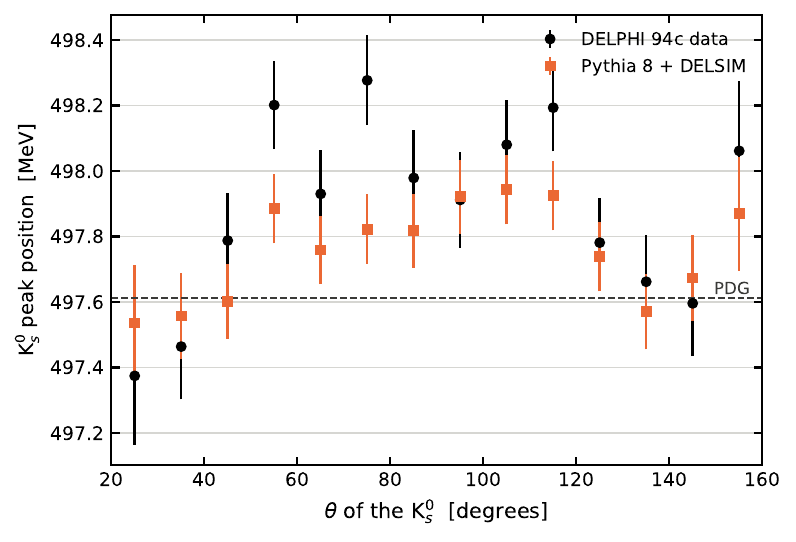}
    \includegraphics[width=0.48\textwidth]{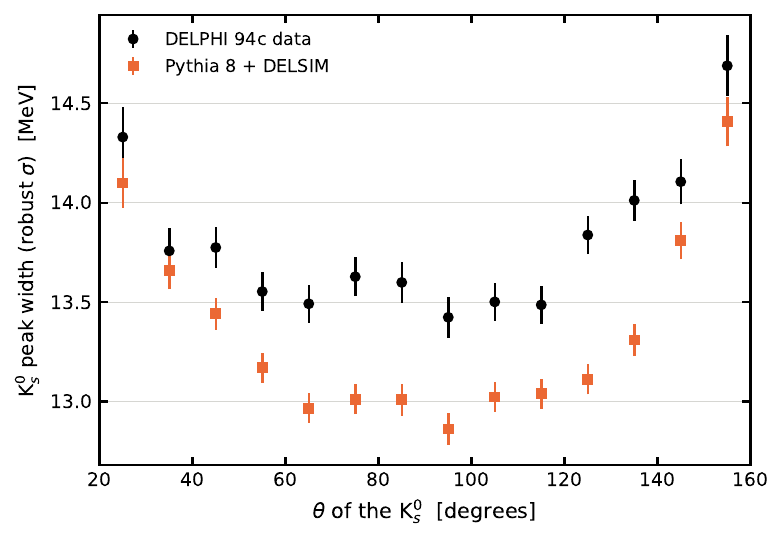}
    \caption{Position (left) and width (right) of the $K^0_S \to \pi^+\pi^-$ mass peak as a function of the polar angle of the $K^0_S$ in the 1994 DELPHI Open Data sample (black) and in the reconstructed \textsc{PYTHIA}~8.3 simulation (orange). The dashed line in the left panel marks the world-average $K^0$ mass~\cite{ParticleDataGroup:2024cfk}. The peak position probes the track momentum scale and the peak width, quoted as a robust standard deviation, the momentum resolution.}
    \label{fig:det_k0s}
\end{figure}

Comparisons between data and the reconstructed \textsc{PYTHIA}~8.3 simulation are performed for the key kinematic distributions used in this analysis. For brevity, the figures in this section display only the 1994 dataset; the corresponding 1995 distributions are provided in Appendix~\ref{app:1995_check}. Figure~\ref{fig:data_mc_fbcc} illustrates the polar-angle distribution of the raw one-point charge correlator, $\mathcal{Q}(\theta) = N^+(\theta) - N^-(\theta)$, at the detector level for the nominal $p_{\rm T} > 2$~GeV selection, with the data-to-simulation ratio shown in the lower panel. Figure~\ref{fig:data_mc_theta} presents the track transverse momentum spectra for both data and reconstructed MC, separated by reconstructed charge. While overall good agreement is observed across the kinematic phase space, a notable discrepancy exists in the charge balance: the reconstructed MC exhibits a small but clear excess of positive tracks over negative tracks, most pronounced at low $p_{\rm T}$ ($p_{\rm T} < 5$~GeV) and in the backward hemisphere, that is absent in the data. 
The origin of this MC charge imbalance, likely tied to imperfect modeling of the detector response, the tracking and material description, the magnetic field, or space-charge distortions, is itself worth a dedicated investigation. As a first step, the charge-sensitive detector response is examined with control quantities of the kind used by the previous LEP and SLC measurements to calibrate it~\cite{ALEPH:1991fba, ALEPH:1996qlh, OPAL:2002ddm, SLD:1996gjt}: the detector material and its forward--backward asymmetry, probed with photon conversions, and the track momentum scale and resolution, probed with the $K^0_S \to \pi^+\pi^-$ mass peak. Figure~\ref{fig:det_conversions} shows the rate of reconstructed photon conversions per event as a function of the polar angle of the conversion, together with the forward--backward asymmetry of this rate, $(N_{\rm fwd} - N_{\rm bwd})/(N_{\rm fwd} + N_{\rm bwd})$, in which each bin at $\theta$ is compared with its mirror bin at $180^\circ\!-\theta$. The simulation reproduces the conversion rate to within about 5\% over $30^\circ \le \theta \le 150^\circ$, with deviations of about 8\% in the outermost bins. The forward--backward asymmetry of the material, the quantity that could feed a parity-odd charge bias, is well reproduced by the simulation across the acceptance, except in the bin containing the TPC cathode plane and in one bin at $45^\circ$, where data and simulation differ by about 1\%. Figure~\ref{fig:det_k0s} shows the position and width of the $K^0_S$ mass peak as a function of the polar angle of the $K^0_S$. The peak position agrees between data and simulation to within 0.5~MeV in every bin, corresponding to a momentum-scale difference at the $10^{-3}$ level, and is symmetric under $\theta \to 180^\circ\!-\theta$ within the statistical uncertainty in both. The peak width, which measures the momentum resolution, follows the same $\theta$ dependence in data and simulation, degrading toward the edges of the acceptance where tracks traverse more material and multiple Coulomb scattering increases; the simulated resolution is better than the measured one by about 4\% in the central region, with the difference shrinking toward the acceptance edges. Importantly, the width is, to good approximation, symmetric under $\theta \to 180^\circ\!-\theta$: forward and backward mirror bins agree to within about 2\% in both data and simulation.

These checks indicate that the forward--backward asymmetry of the charge-averaged detector response is well modeled by the simulation, with residual data--simulation differences confined to a few bins, while its forward--backward-symmetric part, such as the absolute conversion rate near the acceptance edges or the absolute momentum resolution, is reproduced less precisely, which does not mimic the parity-odd signal. Both probes are, however, charge-averaged: they do not constrain a difference between the response to positive and to negative tracks, which is the quantity that enters the charge correlator. A per-charge calibration of the detector response, along the lines of the sagitta and material calibrations of the previous measurements, is deferred to future precision measurements. The present analysis instead bounds the residual data--simulation difference in the per-charge response directly with the tag-and-probe measurement of Section~\ref{sec:data_driven}, whose precision is set by the size of the $\tau^+\tau^-$ control sample.

\begin{figure}[ht!]
    \centering
    \includegraphics[width=0.55\textwidth]{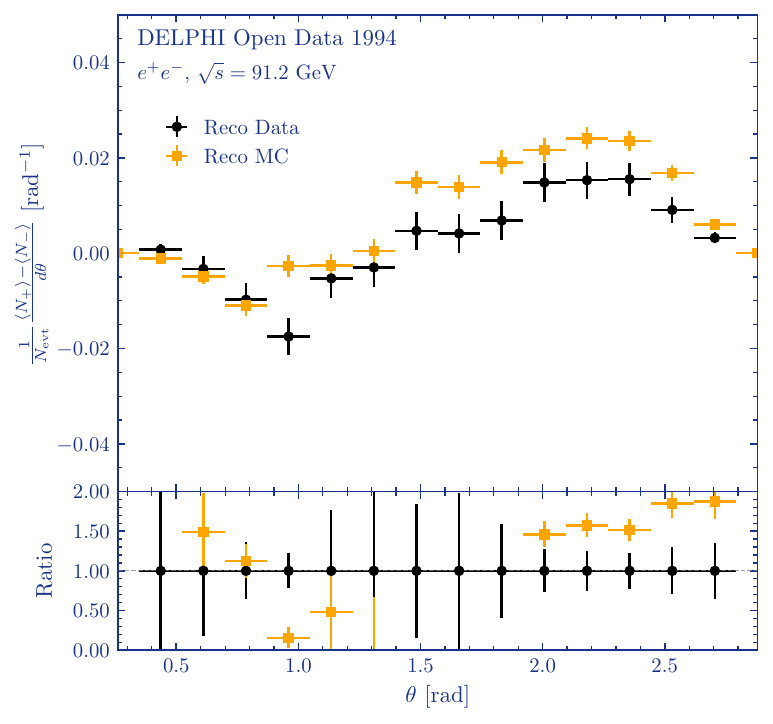}
    \caption{Comparison of the raw one-point charge correlator $\mathcal{Q}(\theta) = N^+(\theta) - N^-(\theta)$ between 1994 data (black) and reconstructed \textsc{PYTHIA}~8.3 MC (yellow), for the nominal $p_{\rm T} > 2$~GeV selection. The lower panel shows the data/MC ratio. }
    \label{fig:data_mc_fbcc}
\end{figure}

\clearpage
\section{Efficiency correction}
\label{sec:mc_correction}

The efficiency correction is derived from the MC samples. A correspondence between reconstructed and generator-level tracks is established using the Hungarian algorithm~\cite{hungarianMatching}, with the track opening angle as the cost metric and a maximum matching distance of $\Delta\theta < 0.05$~rad. Charge agreement is then imposed on top of this geometric association: a reconstructed track that is geometrically matched and carries the same charge as its generator-level partner is classified as truth-matched, whereas one that is geometrically matched but carries the opposite charge is classified as charge misidentification. A reconstructed track with no geometric match is classified as a fake, and a generator-level track with no reconstructed partner indicates track finding inefficiency. These quantities are evaluated separately for the two charge signs, the efficiency relative to the number of generator-level tracks, and the fake and misidentification rates relative to the number of reconstructed tracks.

The correction is applied to the data in two steps. First, the combined contribution of charge-misidentified and fake tracks, estimated from simulation, is subtracted from the reconstructed positive- and negative-track yields in each angular bin. Second, the resulting yields are corrected for same-charge track finding efficiency and converted to per-event track densities using the normalization of Eq.~\eqref{eq:charged_densities}. The simulation-corrected charge correlator, denoted $\mathcal{Q}_{\rm eff}(\theta)$, is obtained from the difference of the corrected positive- and negative-track densities.

\begin{figure}[ht!]
    \centering
    \includegraphics[width=0.45\textwidth]{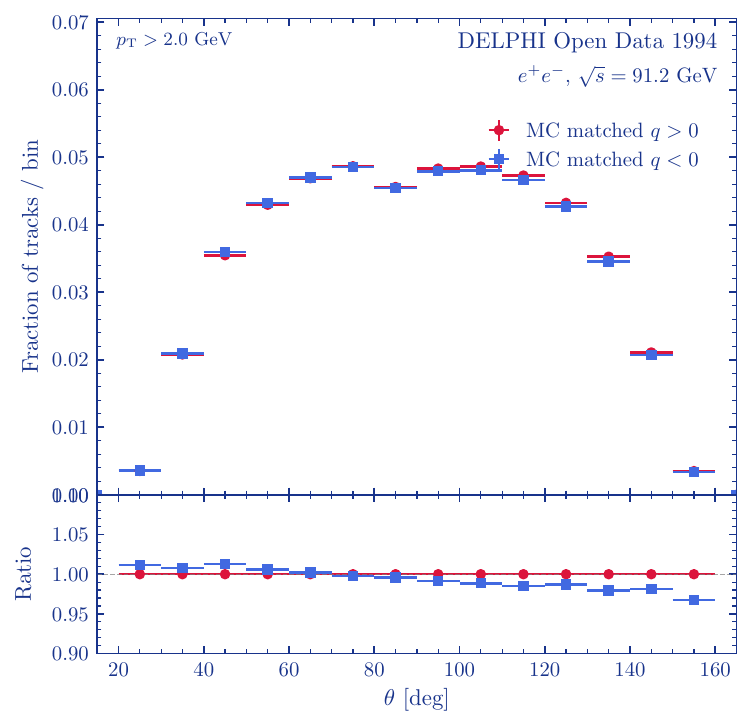}
    \includegraphics[width=0.45\textwidth]{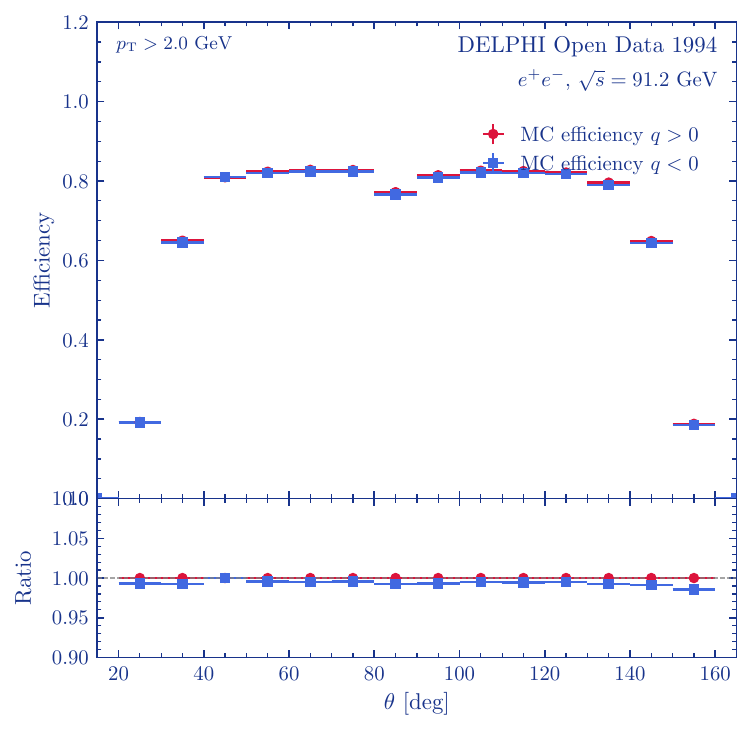}
    \includegraphics[width=0.45\textwidth]{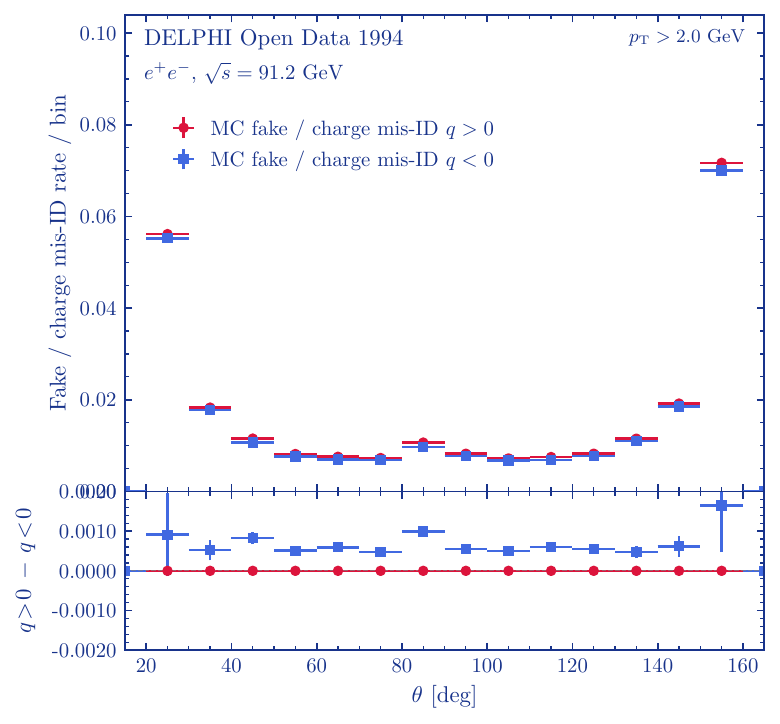}
    \caption{Polar-angle distributions of positively ($q>0$, red) and negatively ($q<0$, blue) charged tracks in the reconstructed \textsc{PYTHIA}~8.3 simulation, for the $p_{\rm T} > 2$~GeV selection: truth-matched reconstructed tracks (top left), the corresponding truth-matched generator-level tracks giving the matching efficiency (top right), and the charge-unmatched component comprising charge-misidentified and fake tracks (bottom).}
    \label{fig:mc_fake_match}
\end{figure}

Figure~\ref{fig:mc_fake_match} shows the polar-angle distributions of positively ($q>0$, red) and negatively ($q<0$, blue) charged tracks, decomposed into three components: the truth-matched sample at detector level (top left), the same matched tracks at generator level (top right), and the charge-unmatched component, comprising charge-misidentified and fake tracks (bottom). The truth-matched detector-level sample exhibits a clear charge asymmetry, as expected. The asymmetry is smaller in the forward region than in the backward region, consistent with Figure~\ref{fig:data_mc_fbcc}. A further breakdown in bins of $p_{\rm T}$, presented in Appendix~\ref{app:additional_datamc}, shows that this asymmetry arises predominantly from the low-$p_{\rm T}$ region, also consistent with Figure~\ref{fig:data_mc_theta}. The matching efficiency is approximately $85\%$ throughout the bulk of the detector acceptance and is slightly higher for positive than for negative tracks, also as expected. The combined charge-misreconstruction and fake rate, dominated by the fake component, is at the $1$--$5\%$ level, depending on $\theta$. 

The charge-asymmetric component is shown in the lower panel. This is the critical input for this measurement. Because the observable $\mathcal{Q}(\theta) = N^{+}(\theta) - N^{-}(\theta)$ is itself the difference between the positive and negative track yields, any charge-dependent reconstruction asymmetry maps directly onto the signal and constitutes a potential bias\footnote{Spurious forward--backward asymmetries have a storied history: an apparent asymmetry in the first Frascati data on the newly discovered $J/\psi$---which did not survive further scrutiny---motivated its tentative interpretation as a neutral weak boson rather than a charmonium state~\cite{Altarelli:1974wm}.}, whereas charge-symmetric effects act equally on $N^{+}$ and $N^{-}$ and do not generate a spurious asymmetry. While the MC-based correction removes the asymmetry predicted by the simulation, the residual difference between data and simulation is the dominant effect and is corrected using the data-driven method described in Section~\ref{sec:data_driven}.

\begin{figure}[ht!]
    \centering
    \includegraphics[width=0.6\textwidth]{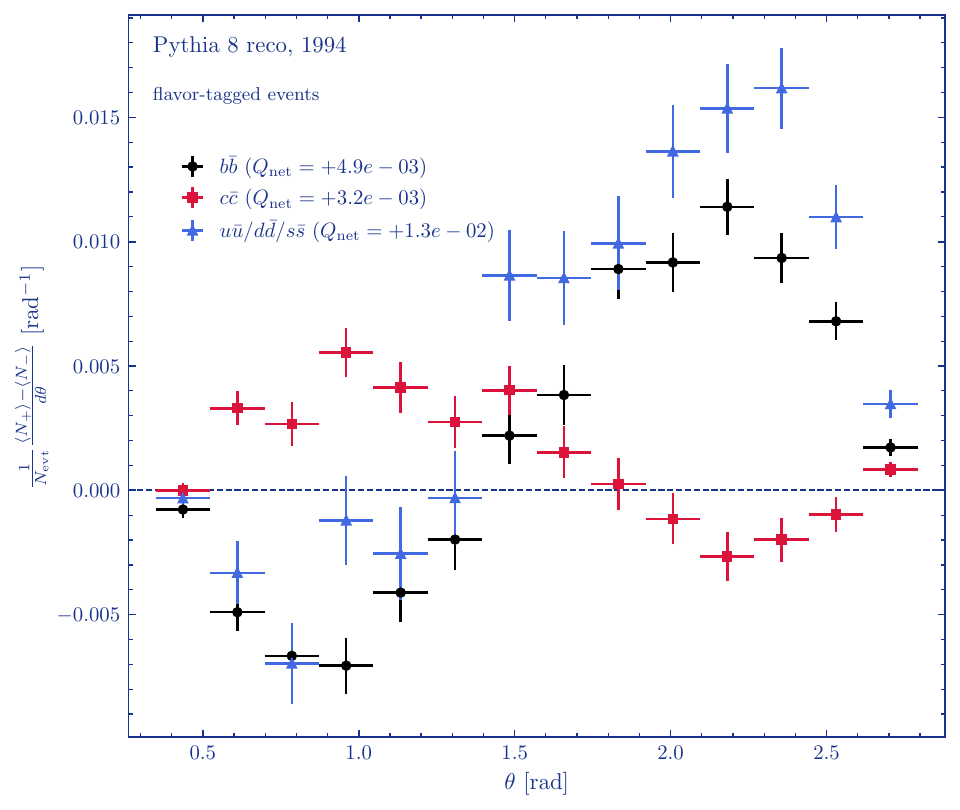}
    \caption{Reconstructed per-event charge difference as a function of the polar angle $\theta$ in the \textsc{PYTHIA}~8.3 simulation (1994 detector configuration), for the $p_{\rm T} > 2$~GeV selection, with events classified by the generator-level primary quark flavor: $b\bar{b}$ (black), $c\bar{c}$ (red), and $u\bar{u}/d\bar{d}/s\bar{s}$ (blue). The $\theta$-integrated net charge imbalance per event, $Q_{\rm net}$, is quoted in the legend. A net positive imbalance of comparable size is observed for all flavor classes, while the differential shapes reflect the flavor-dependent physics asymmetries.}
    \label{fig:flavor_universality}
\end{figure}

The universality of this charge-asymmetric response is examined by decomposing the simulated per-event charge difference by the flavor of the primary quark pair and by the species of the final-state hadron. Figure~\ref{fig:flavor_universality} shows the reconstructed distribution separately for $b\bar{b}$, $c\bar{c}$, and $u\bar{u}/d\bar{d}/s\bar{s}$ events, together with the $\theta$-integrated net charge imbalance per event, $Q_{\rm net}$, quoted in the legend. The differential shapes differ among the flavor classes as expected from the flavor-dependent physics asymmetries; in particular, the parity-violating modulation enters with opposite sign for up-type quarks, visible as the inverted $\theta$ dependence of the $c\bar{c}$ class. Because this physics modulation is parity-odd and largely cancels upon integration over the approximately symmetric $\theta$ acceptance, $Q_{\rm net}$ isolates the detector-induced charge bias. All three flavor classes exhibit a net imbalance of the same (positive) sign, $Q_{\rm net} = +3.2\times10^{-3}$ ($c\bar{c}$), $+4.9\times10^{-3}$ ($b\bar{b}$), and $+1.3\times10^{-2}$ ($u\bar{u}/d\bar{d}/s\bar{s}$). The imbalance is therefore not strictly flavor-universal: it spans a factor of about four across the three classes, and is largest for the light-flavor class. What the decomposition does establish is weaker but sufficient for the present purpose, namely that the detector-induced imbalance is common to all flavors rather than generated by one of them, and that the class carrying the largest weight in the inclusive hadronic sample is also the one that dominates the observed imbalance. The correction derived below is therefore representative of the bulk of the charged-particle yield, and a residual flavor dependence of this size is covered by the generator-model systematic uncertainty of Section~\ref{sec:systematicsSource}, which varies the flavor composition and track kinematics of the sample used to derive the correction.

\begin{figure}[ht!]
    \centering
    \includegraphics[width=0.45\textwidth]{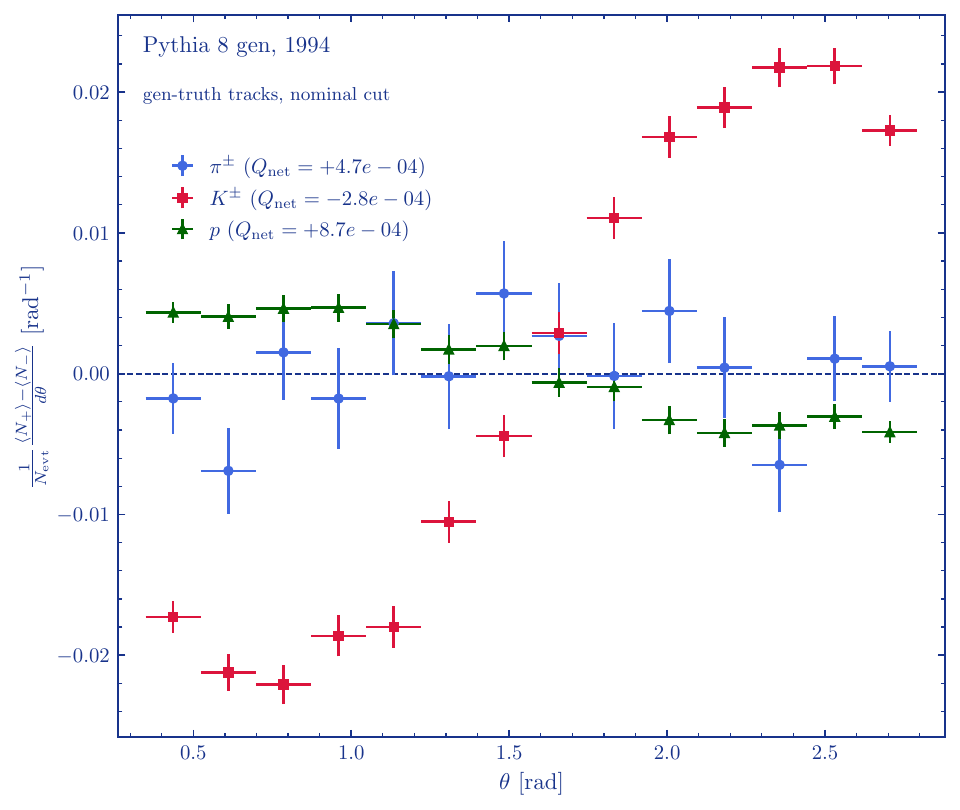}
    \includegraphics[width=0.45\textwidth]{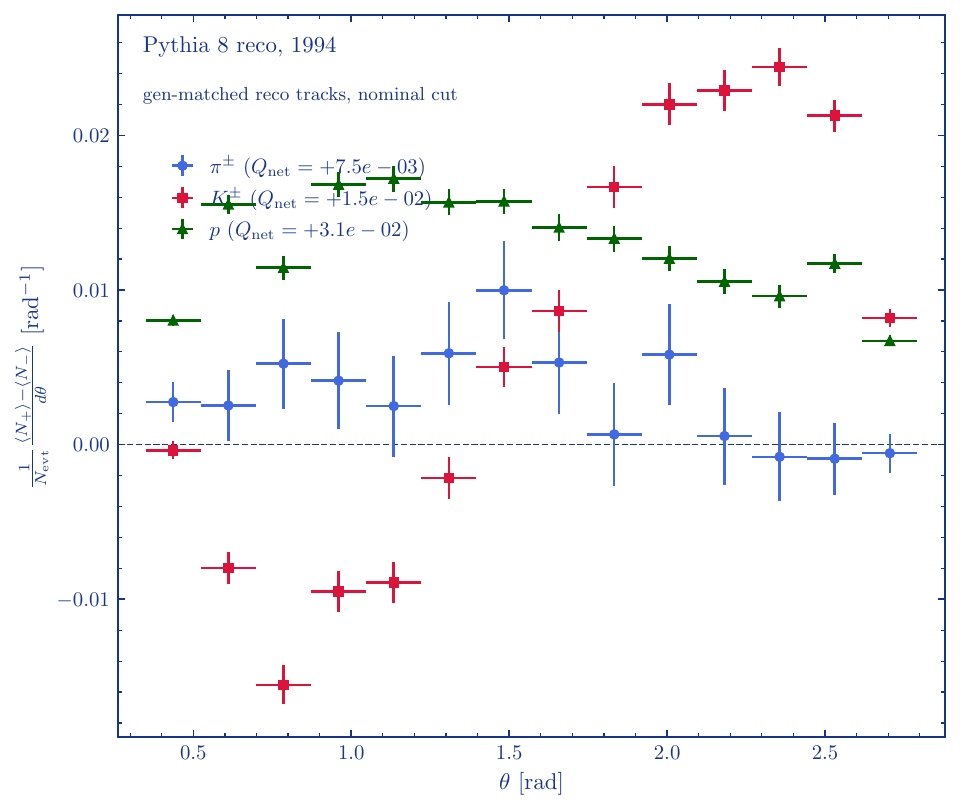}
    \caption{Per-event charge difference as a function of the polar angle $\theta$ in the \textsc{PYTHIA}~8.3 simulation (1994 detector configuration), broken down by hadron species ($\pi^\pm$ blue, $K^\pm$ red, $p/\bar{p}$ green), for the $p_{\rm T} > 2$~GeV selection, at generator level (left) and for truth-matched reconstructed tracks (right). The $\theta$-integrated net charge imbalance $Q_{\rm net}$ of each species is quoted in the legends. The net imbalance is compatible with zero at generator level, while after reconstruction each species acquires a positive imbalance, largest for protons owing to the larger $\bar{p}$ inelastic interaction cross section in the detector material.}
    \label{fig:species_universality}
\end{figure}

Figure~\ref{fig:species_universality} shows the corresponding decomposition by final-state hadron species ($\pi^\pm$, $K^\pm$, $p/\bar{p}$), at generator level (left) and for truth-matched reconstructed tracks (right). The kaon distribution carries a sizable asymmetry already at generator level, which is of physics origin; consistently, the generator-level net imbalance is compatible with zero for all species, $|Q_{\rm net}| \lesssim 9\times10^{-4}$. After reconstruction, each species acquires a net positive imbalance: $Q_{\rm net} = +7.5\times10^{-3}$ for pions, $+1.5\times10^{-2}$ for kaons, and $+3.1\times10^{-2}$ for protons. This ordering follows the expected hierarchy of the particle--antiparticle difference in inelastic interaction cross sections with the detector material: smallest for pions, larger for kaons ($K^-$ versus $K^+$), and largest for protons, where antiproton annihilation depletes the $\bar{p}$ yield relative to the $p$ yield. The pion imbalance is nonetheless sizable, smaller than the proton one by only a factor of four, although pions are the species for which this material asymmetry is smallest. A component of the reconstruction-induced imbalance is therefore common to all species and is already present in pions, consistent with an origin in residual distortions of the reconstructed track geometry rather than in the hadronic interaction cross sections. Taken together, the common sign and comparable magnitude of the imbalance across flavor classes in Figure~\ref{fig:flavor_universality} and the species ordering above, with its sizable pion component, support the strategy of Section~\ref{sec:data_driven}: the species-common part of the charge imbalance is measured directly in the pion-dominated $\tau^+\tau^-$ control sample and transferred to the inclusive hadronic sample, in which pions likewise dominate the charged-particle yield.

\begin{figure}[ht!]
    \centering
    \includegraphics[width=0.45\textwidth]{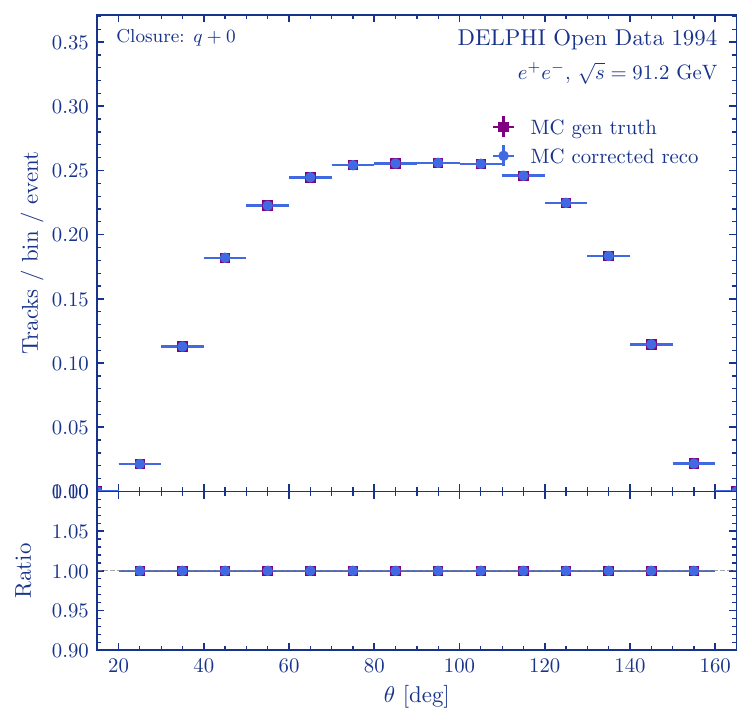}
    \includegraphics[width=0.45\textwidth]{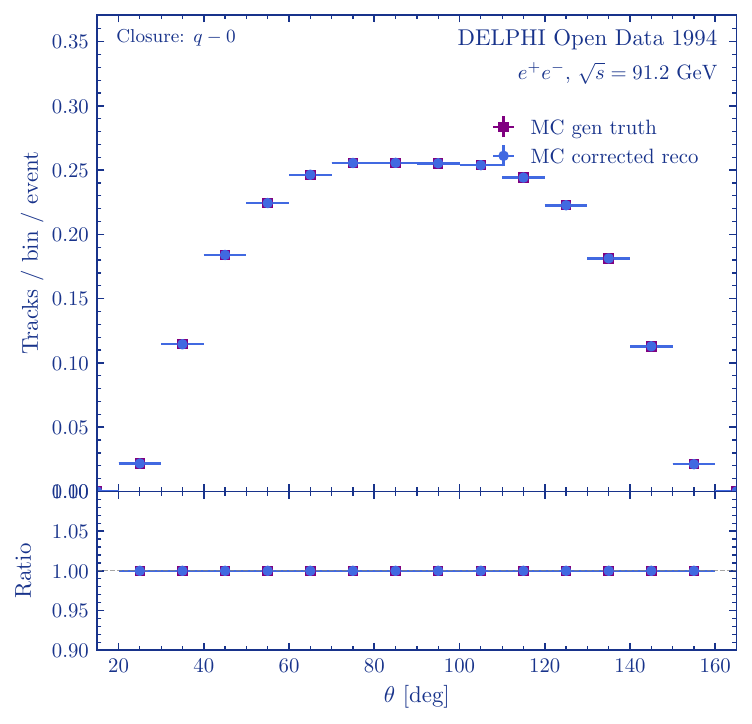}
    \caption{Closure test of the MC-based correction for positively charged (left) and negatively charged (right) tracks. The MC generator-level truth (purple squares) is compared to the corrected reconstructed MC (blue circles) as a function of the polar angle $\theta$. The lower panels show the ratio, which is consistent with unity.}
    \label{fig:closure}
\end{figure}

A closure test is performed by applying the full correction chain (fake subtraction followed by efficiency correction) to the reconstructed MC distribution and comparing the result to the generator-level truth.
The results are shown in Figure~\ref{fig:closure} for positively and negatively charged tracks separately.
The corrected reconstructed MC agrees with the generator-level truth across the full $\theta$ range, with the ratio consistent with unity in the lower panels, validating the MC-based correction procedure.

\begin{figure}[ht!]
    \centering
    \includegraphics[width=0.45\textwidth]{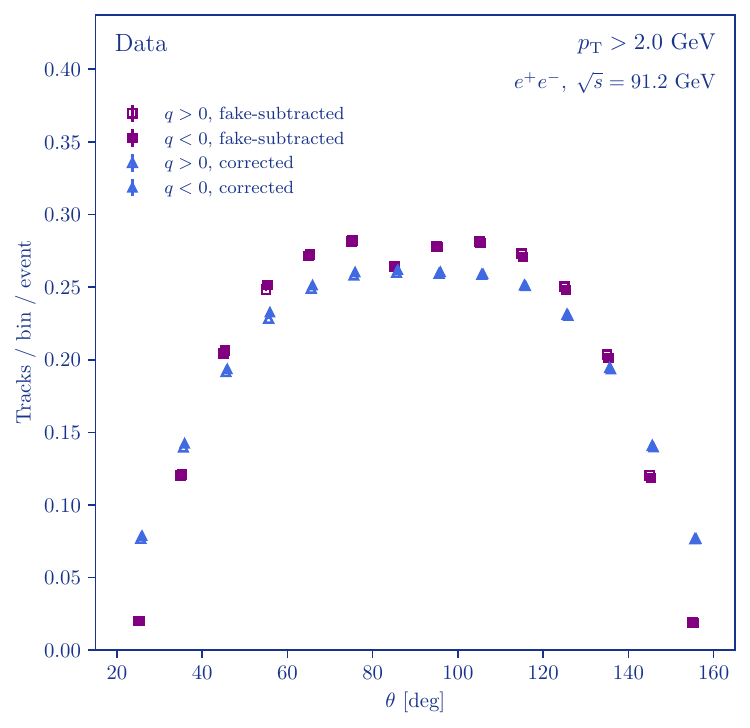}
    \includegraphics[width=0.45\textwidth]{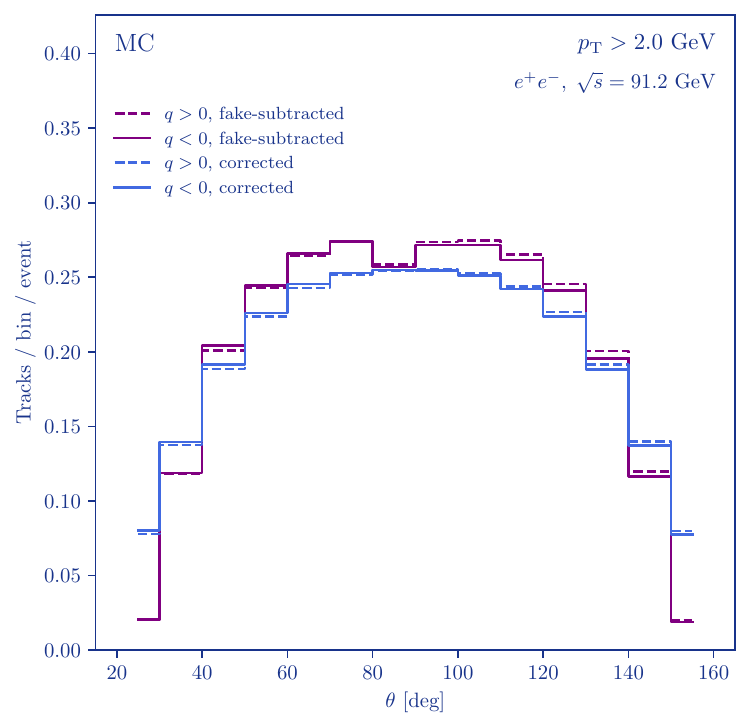}
    \caption{Correction chain for data (left) and reconstructed MC (right) for the $p_{\rm T} > 2$~GeV selection. In each panel, positively (open markers / dashed lines) and negatively (filled markers / solid lines) charged tracks are overlaid at the fake-subtracted (purple) and fully corrected (blue) stages, illustrating the residual charge balance after each correction step.}
    \label{fig:corrected_data}
\end{figure}

The full correction chain applied to the data is shown in Figure~\ref{fig:corrected_data} for the $p_{\rm T} > 2$~GeV selection, with data on the left and reconstructed MC on the right. In each panel, the positively and negatively charged tracks are overlaid at the fake-subtracted (purple) and fully corrected (blue) stages. After the MC-based correction, a clear global charge imbalance remains in the corrected data, indicating a residual data--MC difference in the charge-reconstruction asymmetry that the MC-based correction does not capture. This residual is addressed by the data-driven correction described in the next section.

\begin{figure}[htbp]
    \centering
    \includegraphics[width=0.55\textwidth]{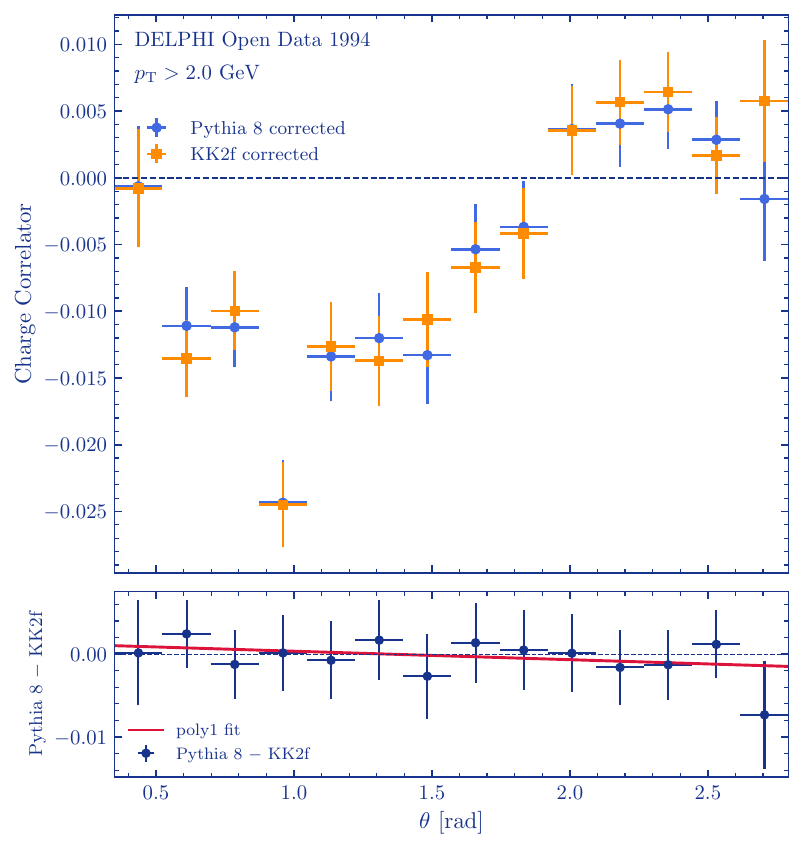}%
    \caption{MC-corrected one-point charge correlator obtained with \textsc{PYTHIA}~8 and
    \textsc{KK}2f as alternative MC samples in the analysis fiducial phase space of $p_{\rm T} > 2$~GeV.
    The lower panel shows the per-bin
    difference (\textsc{PYTHIA}~8 $-$ \textsc{KK}2f) with a first-order polynomial fit overlaid.}
    \label{fig:bruteforce_kk2f}
\end{figure}

The MC-based efficiency correction depends on the choice of generator used to derive it. This dependence is assessed by repeating the full correction with \textsc{KK}2f as an alternative generator in place of the nominal \textsc{PYTHIA}~8 sample and comparing the two results, as shown in Figure~\ref{fig:bruteforce_kk2f}. The difference between the two quantifies the sensitivity of the efficiency correction to the modeled track kinematics, $K/\pi$ abundance, and event environment, and is propagated as a generator-model systematic uncertainty (Section~\ref{sec:systematicsSource}).

\clearpage
\section{Scale factor from single-track charge misreconstruction measurement}
\label{sec:data_driven}

The MC-based correction described in Section~\ref{sec:mc_correction} relies on the simulation accurately reproducing the charge reconstruction performance of the detector.
Any mismodeling of the charge misreconstruction rate in the MC would introduce a bias in the corrected charge correlator.
Since the one-point charge correlator is constructed from the difference between positive- and negative-particle yields, it is particularly sensitive to charge misreconstruction. Even a small asymmetry in the rate at which positive tracks are reconstructed as negative, or vice versa, can shift the measured correlator.
A data-driven determination of the charge-reconstruction difference between data and MC is therefore essential to control this source of bias and to reduce the reliance on MC modeling of the detector response.

Whereas previous LEP and SLD experiments calibrated the detector response using track-level sagitta measurements and characterized the detector material through photon conversions, in this first measurement of the charge correlator, we bound charge-misidentification effects directly from the observed rate in a dedicated control region. Detector-level calibrations in DELPHI Open Data will be the natural next step to explore for future precision measurements. 

Measuring the charge misreconstruction rate requires a sample of tracks for which the true charge can be inferred with high purity independently of the reconstructed charge of the probe track. At the track level, the process $e^+e^- \to \tau^+\tau^-$ provides a particularly useful control region. Because $\tau$ leptons decay predominantly into low-multiplicity topologies, charge conservation and topology selections allow the true charge of selected probe tracks to be inferred with high purity, enabling a data-driven determination of charge-misreconstruction rates.

\begin{figure}[ht!]
    \centering
    \includegraphics[width=0.48\textwidth]{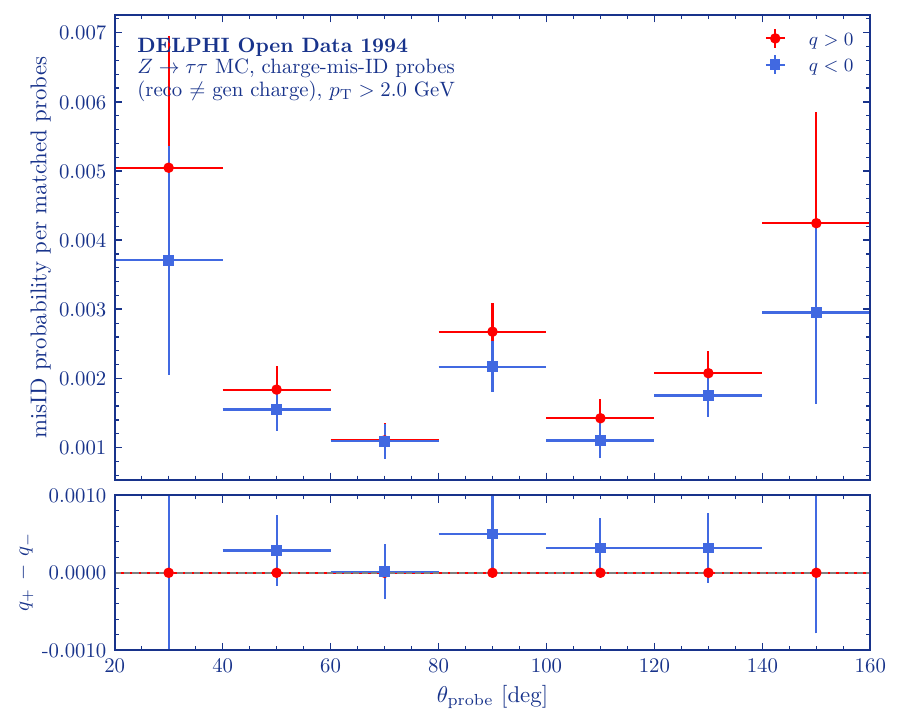}
    \includegraphics[width=0.48\textwidth]{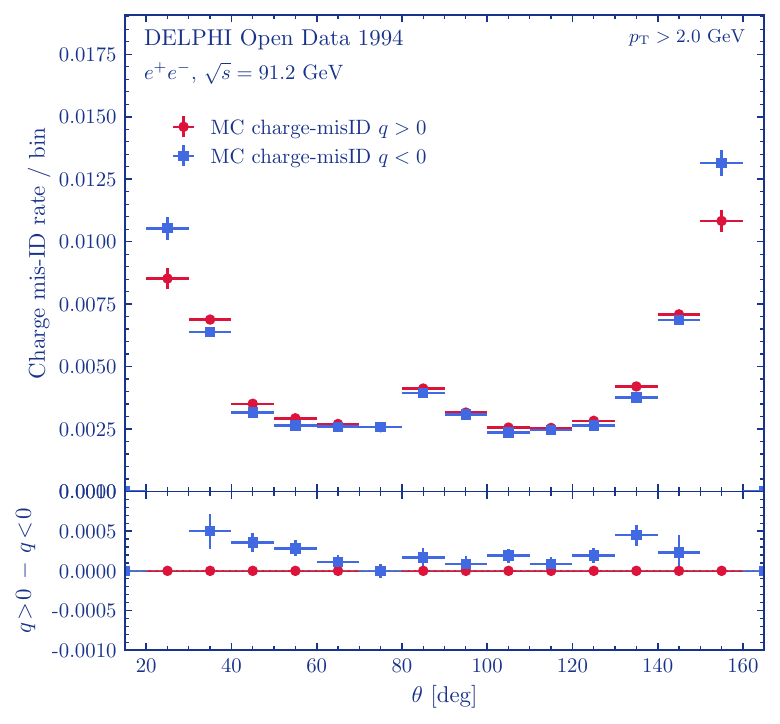}
    \caption{Per-track charge-misreconstruction rate obtained from truth matching in simulation, defined as the fraction of reconstructed tracks of a given charge whose matched generator-level partner carries the opposite charge, shown separately for the two reconstructed charges ($q>0$ and $q<0$), in the $\tau$ control-region MC (left) and the inclusive hadronic MC (right). In both environments, for most of the bins, the rate is larger for $q>0$, corresponding to a net negative-to-positive charge migration. The absolute rates are larger in the hadronic sample, while the difference between the rates for the two charges is consistent between the samples within the statistical precision. More details on the tag-and-probe selection are given below.}
    \label{fig:tau_misid_rate_mc}
\end{figure}

The event environment in $\tau^+\tau^-$ decays differs from that of hadronic $Z$ decays. The track multiplicity is much lower, the tracks are more isolated, and the kinematic distributions differ. The transferability of the correction is examined with generator-level studies and truth matching. Figure~\ref{fig:tau_misid_rate_mc} compares the per-track charge-misreconstruction rates, evaluated separately for positive and negative tracks, in the $\tau^+\tau^-$ and inclusive hadronic samples. The absolute rates are larger in the hadronic sample, as expected from the higher track density. The difference between the negative- and positive-track rates, which drives the correction, is instead consistent between the two samples in magnitude and $\theta$ dependence within the statistical precision. This behavior matches the expectation that environment effects, such as track-finding confusion in dense events, are largely charge-symmetric, while the charge-asymmetric component arises from track-level interactions with the detector material and from residual distortions of the track geometry. This picture is also independently supported by the decomposition of the simulated charge imbalance by primary quark flavor and by final-state hadron species (Figures~\ref{fig:flavor_universality} and~\ref{fig:species_universality}). Importantly, the correction does not transfer the absolute misreconstruction rate from $\tau^+\tau^-$ to hadronic events. It measures, and corrects for, the data--MC difference of the rate, which is the component arising from mismodeling of the track-level detector response and is common to the two event environments. Residual differences in the absolute rates between the environments do not enter the correction directly, since they are present in both data and simulation and cancel in the data--MC difference. The environment-sensitive component is moreover largely charge-symmetric and can therefore contribute only through the dilution term, whose impact is below the percent level (Section~\ref{sec:systematics}).

Alternative control samples based on $K_S^0 \to \pi^+\pi^-$ and $\phi \to K^+K^-$ decays were also considered but found to be unsuitable due to large combinatorial backgrounds. A cross-calibration using independent detector subsystems is left to future work. This section presents results for the 1994 sample; the corresponding 1995 distributions are shown in Appendix~\ref{app:1995_check}.

\subsection{$Z \to \tau^+\tau^-$ control region}
\label{sec:tau_control}

The $\tau$ control sample is selected by requiring events consistent with $\tau$-pair production.
Charged tracks are required to satisfy the same baseline quality criteria used for the main measurement (Table~\ref{tab:SelectionSummary}).
The event must contain between 2 and 4 such tracks, with a visible energy fraction $0.10 < E_{\rm vis}/E_{\rm cm} < 0.80$.
The low multiplicity requirement rejects hadronic $Z$ decays.
The lower bound on visible energy suppresses low-visible-energy backgrounds, such as two-photon events, while the upper bound suppresses high-visible-energy dilepton backgrounds, in particular Bhabha scattering and $Z \to \mu^+\mu^-$ events.
The event thrust is required to exceed 0.985, and the acollinearity between the two hemispheres, defined as 
$\theta_{\rm acoll} = 180^\circ - \theta(\vec{p}_1, \vec{p}_2)$, the 
deviation from exact back-to-back topology, must be less than $20^\circ$.
The prong configuration is required to be $1{+}1$, corresponding to the dominant single-prong $\tau$ decay topology.
Events containing any identified electron or muon are vetoed. This suppresses residual contamination from $Z \to e^+e^-$ and $Z \to \mu^+\mu^-$ events, and it also removes the leptonic $\tau$ decays, enriching the selected prongs in charged hadrons, predominantly pions.
The total visible mass is required to satisfy $15 < M_{\rm vis} < 55$~GeV.
The selection criteria are summarized in Table~\ref{tab:TauSelection}.

\begin{table}[ht]
\centering
\begin{tabularx}{\textwidth}{l|l}
\hline\hline
\multicolumn{2}{l}{$\tau$-pair event selection}  \\
\hline
Track quality              & same as Table~\ref{tab:SelectionSummary} \\
Charged track multiplicity & $2 \le N_{\rm ch} \le 4$ \\
Visible energy fraction    & $0.10 < E_{\rm vis}/E_{\rm cm} < 0.80$ \\
Thrust                     & $T > 0.985$ \\
Acollinearity              & $< 20^\circ$ \\
Lepton veto                & veto events with any identified $e$ or $\mu$\\
Visible mass               & $15 < M_{\rm vis} < 55$~GeV \\
\hline\hline
\end{tabularx}
\caption{Selection criteria for the $\tau$ control sample used in the charge misreconstruction rate measurement.}
\label{tab:TauSelection}
\end{table}

\begin{figure}
    \centering
    \includegraphics[width=0.6\textwidth]{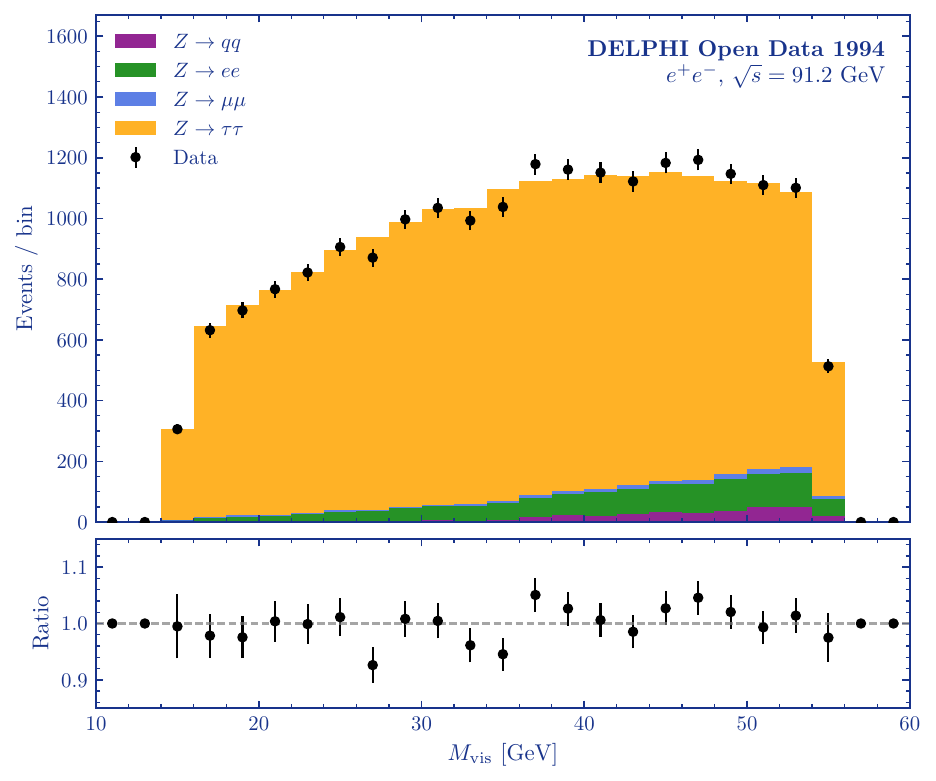}
    \caption{Visible mass $M_{\rm vis}$ distribution of the $\tau$ control sample. Data (markers) are compared to stacked MC contributions from $Z \to \tau^+\tau^-$ (orange), $Z \to \mu^+\mu^-$ (blue), $Z \to e^+e^-$ (green), and $Z \to q\bar{q}$ (purple). The lower panel shows the data/MC ratio.}
    \label{fig:tau_control_purity}
\end{figure}

The purity of the $\tau$ control sample is validated by comparing the visible mass distribution between data and MC, as shown in Figure~\ref{fig:tau_control_purity}.
The MC is broken down into $Z \to \tau^+\tau^-$ (dominant), $Z \to \mu^+\mu^-$, $Z \to e^+e^-$, and $Z \to q\bar{q}$ components.
The $\tau$-pair purity ($N_{\tau\tau}/N_{\rm sel}$) exceeds 96\% across the selected mass range, and the total leptonic purity ($1 - N_{q\bar{q}}/N_{\rm sel}$) exceeds 99\%.

\clearpage
\subsection{Tag and probe method}
\label{sec:tnp_method}

Within the selected $\tau$ sample, the charge misreconstruction rate is estimated using a tag-and-probe (TNP) method.
The event is required to contain a pair of opposite-hemisphere tracks with one track falling in the central region $40^\circ \le \theta \le 140^\circ$; given the acollinearity requirement, the other then lies in the wider range $20^\circ \le \theta \le 160^\circ$.
The prong topology selection purity, estimated using Monte Carlo simulation, is above 98\%.  
The cut flow table, including those from the control region event selection described above, can be seen in Appendix~\ref{app:tnp_check}. 
Of the two tracks, the one with the smaller relative momentum uncertainty $\Delta p/p$, i.e.\ the higher-quality momentum measurement, is chosen as the tag, and the other serves as the probe.
The charge misreconstruction rate of the tag track, estimated from Monte Carlo simulation, is below 0.08\%, while that of the probe track is around 0.2\%, both averaged over the phase space of this analysis. Differentially in $\theta$ the tag rate lies a factor of two to four below the probe rate over the bulk of the acceptance, with the separation shrinking in the outermost bins at either end, as shown in Appendix~\ref{app:tnp_check}.
Because the $\tau^+\tau^-$ pair is produced with zero net charge, the tag track provides a high-purity reference for the expected charge of the probe: a negatively charged tag implies a positively charged probe, and vice versa.
The selection biases in both tag-track quality and prong topology will be discussed in Appendix~\ref{app:tnp_check} and treated as systematic uncertainties. 

The charge misreconstruction rate is defined as the fraction of probe tracks whose reconstructed charge has the same sign as the tag, i.e.\ the same-sign fraction:
\begin{equation}\label{eq:misid}
    f^{\pm}(\theta) = \frac{N^{\rm SS}_{\pm}(\theta)}{N^{\rm tag}_{\pm}(\theta)},
\end{equation}
where $N^{\rm SS}_{\pm}(\theta)$ is the number of same-sign pairs and $N^{\rm tag}_{\pm}(\theta)$ the total number of tag entries, both as a function of the probe polar angle $\theta$ and binned by the true charge of the probe, which is fixed to be opposite the reconstructed charge of the tag, so that $f^{\pm} = P(\text{reco}\,{\mp}\,|\,\text{true}\,{\pm})$. Because the $\tau^+\tau^-$ pair carries zero net charge, the probe's true charge is fixed, opposite to that of the tag, by charge conservation alone, independently of the simulation. The track-quality requirement on the tag ensures that its reconstructed charge is correct to high purity, so the same-sign fraction of Equation~\ref{eq:misid} provides a direct, model-independent measurement of the probe charge-misreconstruction rate in both data and simulation, up to the small residual tag misreconstruction and multi-prong topology leakage effects that are taken into account as systematic uncertainties.

\begin{figure}[ht!]
    \centering
    \includegraphics[width=0.48\textwidth]{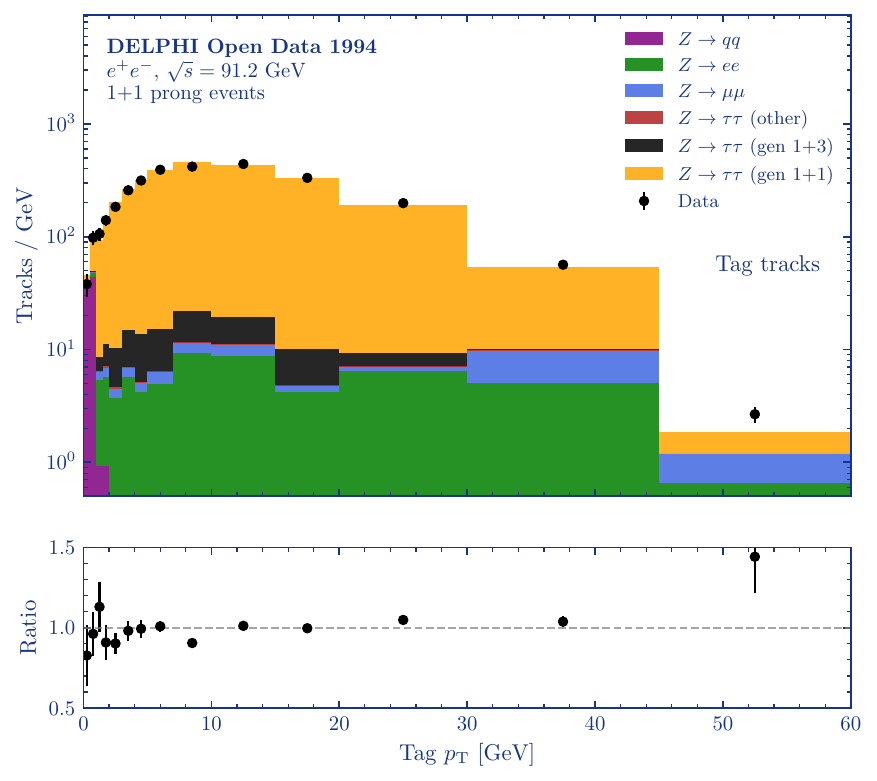}
    \includegraphics[width=0.48\textwidth]{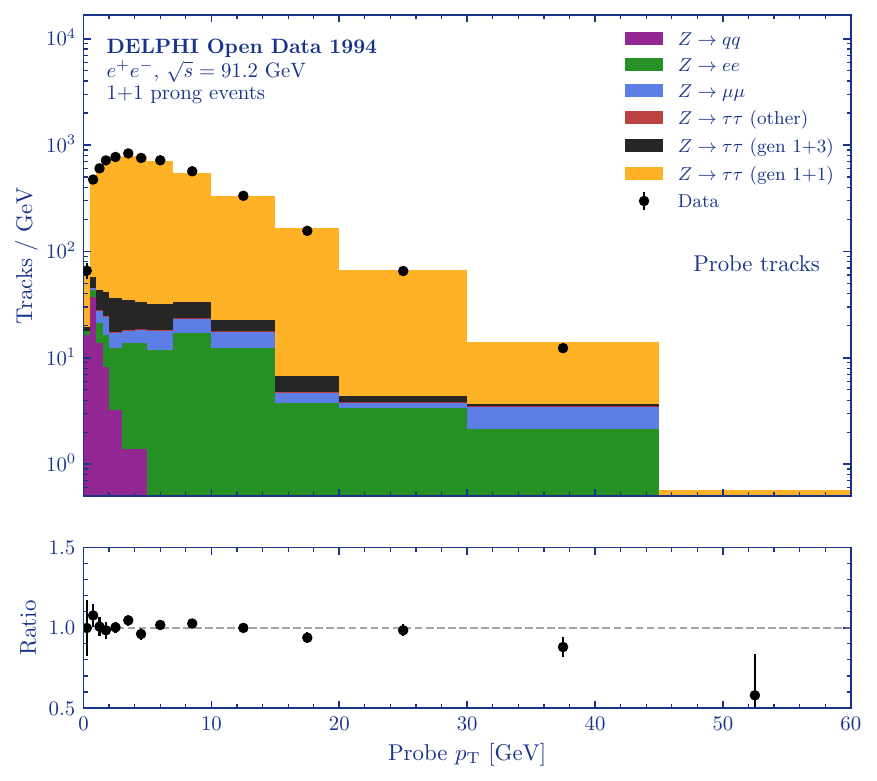}\\
    \includegraphics[width=0.48\textwidth]{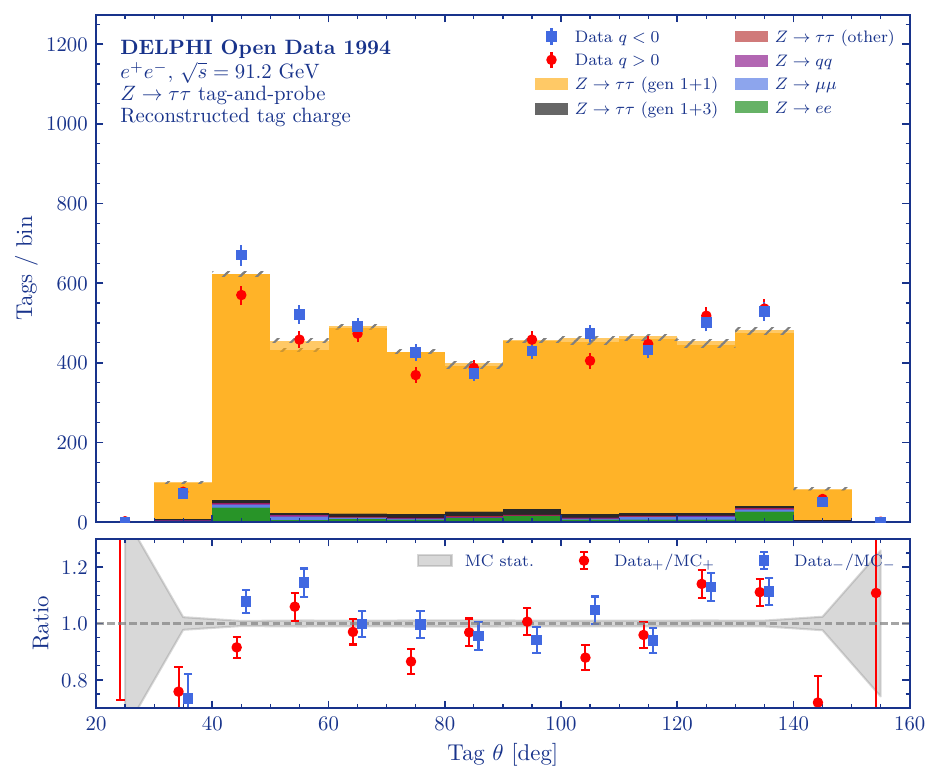}
    \includegraphics[width=0.48\textwidth]{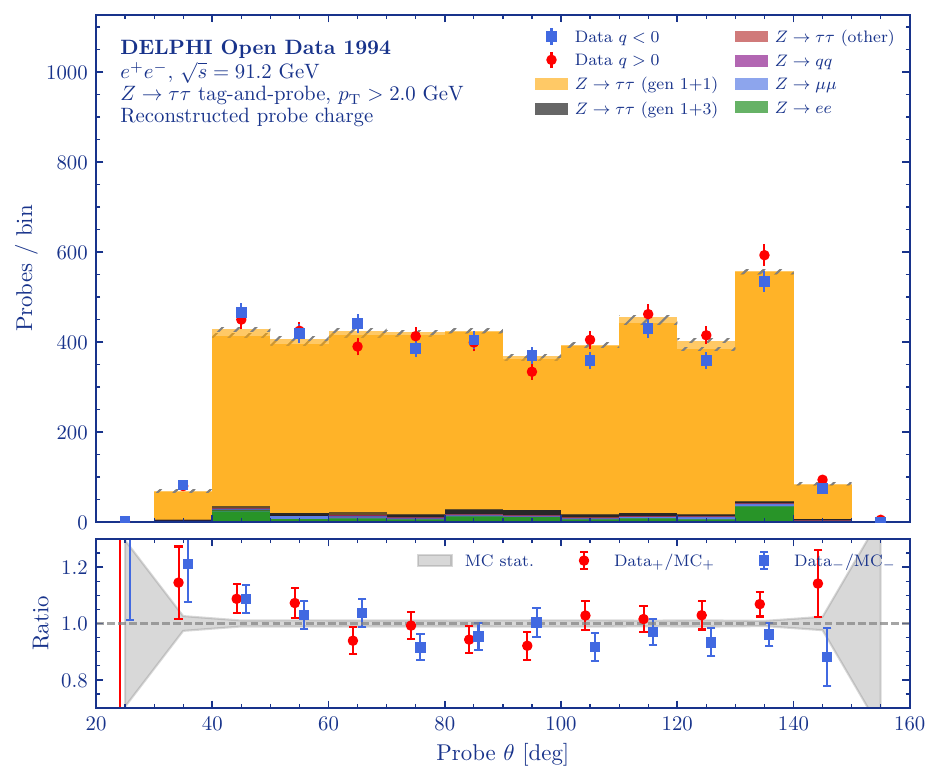}
    \caption{Kinematic distributions of the tag and probe tracks in the $\tau$ control sample. Top row: track $p_{\rm T}$ distribution of the tag (left) and probe (right), with stacked MC contributions from $Z \to \tau^+\tau^-$, $Z \to \mu^+\mu^-$, $Z \to e^+e^-$, and $Z \to q\bar{q}$. Bottom row: polar-angle distribution of the tag (left) and probe (right) tracks separated by reconstructed charge, $q > 0$ (red circles) and $q < 0$ (blue squares). Lower panels show the data/MC ratio.}
    \label{fig:tau_tnp_kin}
\end{figure}

The kinematic distributions of the tag and probe tracks, separated by reconstructed charge and track momentum, are shown in Figure~\ref{fig:tau_tnp_kin}. The top row shows the transverse-momentum distributions of the tag (left) and probe (right) tracks, split by charge with the stacked MC contributions overlaid, and the bottom row shows the corresponding polar-angle distributions. Good data--MC agreement is observed across the full kinematic range for both samples. An interesting feature of the $\theta$ distributions is that the tag, selected as the track with the smaller $\Delta p/p$, lies predominantly in the forward hemisphere, indicating that forward tracks systematically have higher momentum-measurement quality in the DELPHI 1994 detector configuration. This is worth noting but does not bias the correction, as the rate is constructed as a ratio (Equation~\ref{eq:misid}) that depends only on the probe charge distribution at a given probe $\theta$ and is insensitive to the absolute populations induced by the tag-quality asymmetry.

The charge misreconstruction rate is measured in exclusive $p_{\rm T}$ bins of the probe track.
Five coarse bins are used: $1.0 < p_{\rm T} < 2.0$, $2.0 < p_{\rm T} < 4.0$, $4.0 < p_{\rm T} < 6.0$, $6.0 < p_{\rm T} < 8.0$, and $p_{\rm T} > 8.0$~GeV.
The $1.0 < p_{\rm T} < 2.0$~GeV bin is included as an underflow bin: it does not enter the nominal $p_{\rm T} > 2$~GeV correction, and serves selections whose acceptance extends below $2$~GeV, such as the $|p| > 2$~GeV cross-check of Appendix~\ref{app:fiducial_check}.
In each bin, the data--MC residuals
$\delta^i_{\rm asym}(\theta) = (f^- - f^+)^{\rm data}_i - (f^- - f^+)^{\rm MC}_i$ and
$\delta^i_{\rm sum}(\theta) = (f^+ + f^-)^{\rm data}_i - (f^+ + f^-)^{\rm MC}_i$
are extracted as a function of $\theta$.
Inspired by Figure~\ref{fig:mc_fake_match} (bottom), the charge misreconstruction residuals are parametrized by a first-order polynomial $\chi^2$ fit in the region $20^\circ \le \theta \le 160^\circ$, excluding the band $80^\circ \le \theta \le 100^\circ$ where a known TPC cathode-plane crack distorts the local charge response.
The full $2 \times 2$ covariance matrix of the linear-fit parameters is retained in each probe-$p_{\rm T}$ bin and propagated downstream as the scale-factor uncertainty (Section~\ref{sec:systematics}). A zeroth-order polynomial fit is overlaid for visual comparison. 

\begin{figure}[ht!]
    \centering
    \includegraphics[width=0.9\textwidth]{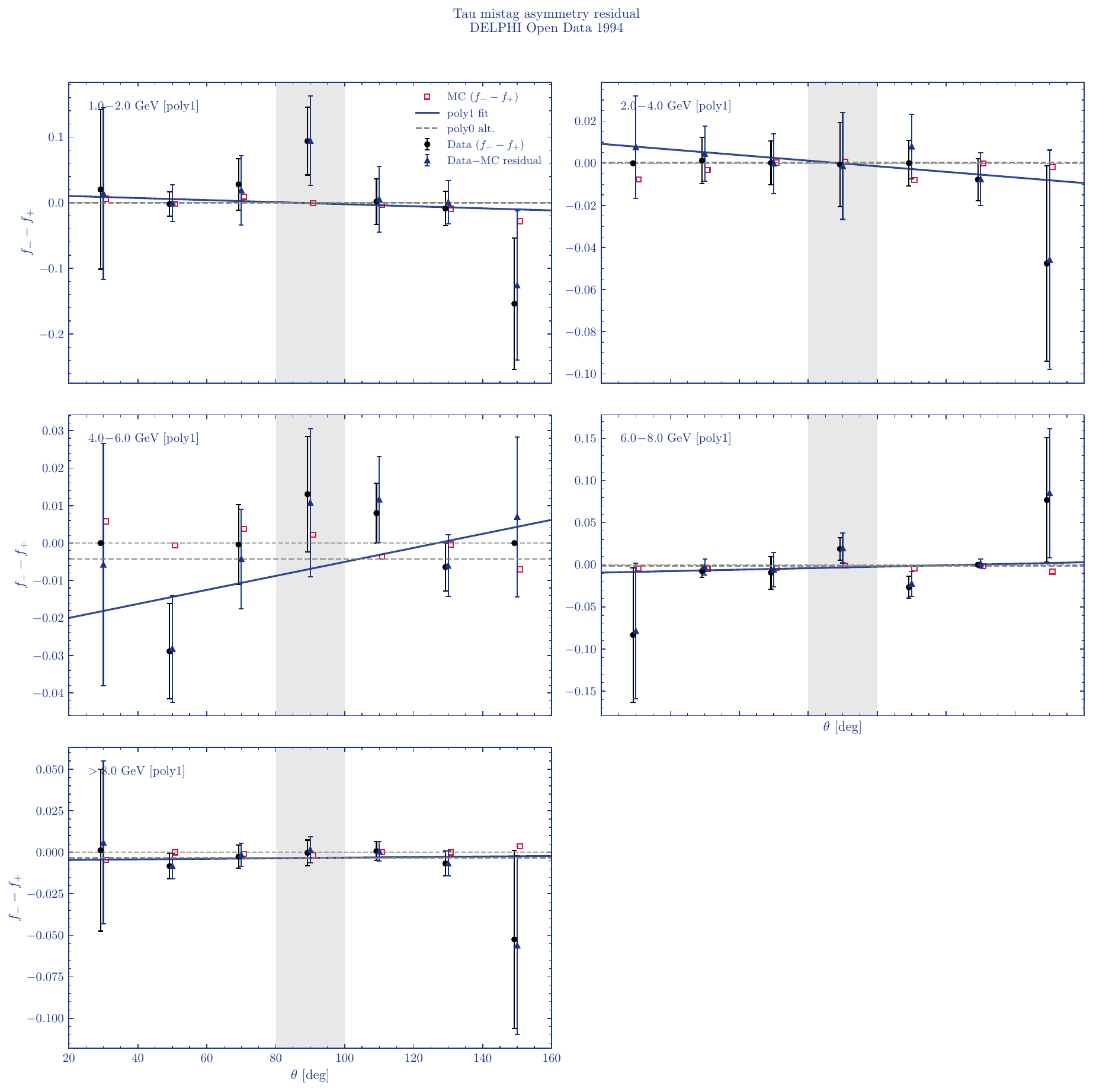}
    \caption{Data--MC residual of the charge misreconstruction asymmetry, $\delta^i_{\rm asym}(\theta) = \delta(f^- - f^+)$, as a function of probe $\theta$ in five exclusive probe $p_{\rm T}$ bins ($1$--$2$, $2$--$4$, $4$--$6$, $6$--$8$, and $>8$~GeV). Data (triangles) are shown with combined data and MC statistical uncertainties, and MC statistical uncertainties are not shown. The solid line shows the nominal first-order polynomial fit in $20^\circ$--$160^\circ$ (excluding the $80^\circ$--$100^\circ$ region indicated by the gray band, where a known TPC cathode-plane crack distorts the local charge response) used as the data-driven correction; the dashed line shows the zeroth-order alternative, overlaid for visual comparison only. The systematic uncertainty on the scale factor is obtained from the full statistical-uncertainty propagation of the linear-fit parameters (Section~\ref{sec:systematics}).}
    \label{fig:misid_asymmetry}
\end{figure}

Figure~\ref{fig:misid_asymmetry} shows the data--MC residual of the charge misreconstruction asymmetry, $\delta^i_{\rm asym}(\theta) = \delta(f^- - f^+)$, as a function of probe $\theta$ in the five probe-$p_{\rm T}$ bins.
A non-zero residual is visible in some bins, with a $\theta$-dependence that can be captured by the first-order polynomial parametrization. The zeroth-order alternative shown in dashed lines is overlaid for visual comparison. The slope of the first-order polynomial is statistically consistent with zero, within one standard deviation for the bins with $p_{\rm T}>6$~GeV and within two standard deviations for lower $p_{\rm T}$ bins. The fitted slopes and constants are shown in Table~\ref{tab:misid_fits_1994}.  The systematic uncertainty associated with the scale factor is obtained from the full statistical-uncertainty propagation of the linear-fit parameters (Section~\ref{sec:systematics}).

\begin{table}[htbp]
\centering
\begin{tabular}{lcccc}
\hline\hline
 & & \multicolumn{2}{c}{linear fit} & constant fit \\
\cline{3-4}\cline{5-5}
$p_T$ [GeV] & $N_{\mathrm{tag}}$ & $b$ [$10^{-4}$/deg] & $\chi^2/\mathrm{ndf}$ & $c$ [$10^{-3}$] ~~ $\chi^2/\mathrm{ndf}$ \\
\hline
2.0--4.0 & 1607 & $-1.50 \pm 1.20$ & $0.7/4$ & $+1.35 \pm 4.67$ ~~ $2.3/5$ \\
4.0--6.0 & 1497 & $+1.56 \pm 0.96$ & $4.7/4$ & $-2.47 \pm 3.68$ ~~ $7.4/5$ \\
6.0--8.0 & 1320 & $+0.92 \pm 0.96$ & $4.9/4$ & $+0.97 \pm 1.79$ ~~ $5.8/5$ \\
$>$8.0 & 4371 & $+0.17 \pm 1.14$ & $2.1/4$ & $-3.39 \pm 3.43$ ~~ $2.1/5$ \\
\hline\hline
\end{tabular}
\caption{Charge mis-identification asymmetry residual $\delta(f_--f_+)$ for the 1994 data set. The $\tau$ tag-and-probe data$-$MC residual is fitted in each exclusive $p_T$ bin over $20$--$160^\circ$ excluding the $80$--$100^\circ$ crack (6 bins). The nominal model is linear in $\theta$; its slope $b$ is quoted. The constant-only fit $\delta(\theta)=c$ is shown for comparison. No fitted slope differs from zero by more than $1.7\sigma$ in either data set.}
\label{tab:misid_fits_1994}
\end{table}

\begin{figure}[ht!]
    \centering
    \includegraphics[width=0.9\textwidth]{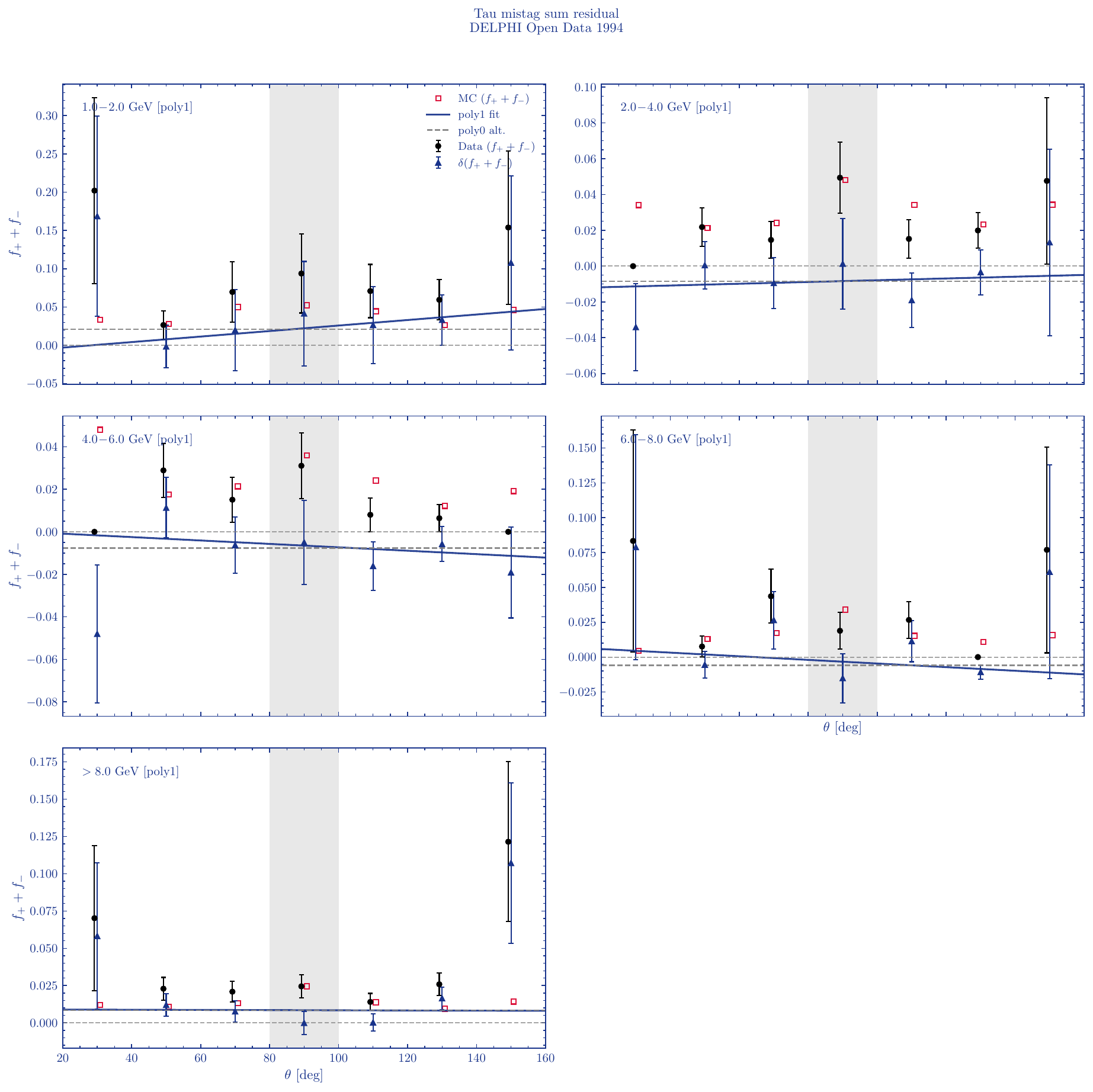}
    \caption{Data--MC residual of the charge misreconstruction sum, $\delta^i_{\rm sum}(\theta) = \delta(f^+ + f^-)$, as a function of probe $\theta$ in the same five exclusive probe $p_{\rm T}$ bins as Figure~\ref{fig:misid_asymmetry}. Data (triangles) are shown with combined data and MC statistical uncertainties, and MC statistical uncertainties are not shown. The solid curve shows the nominal first-order polynomial fit in $20^\circ$--$160^\circ$ (excluding the $80^\circ$--$100^\circ$ region indicated by the gray band, where a known TPC cathode-plane crack distorts the local charge response) used as the data-driven correction; the dashed curve shows the zeroth-order alternative, overlaid for visual comparison only. The systematic uncertainty on the scale factor is obtained from the full statistical-uncertainty propagation of the linear-fit parameters (Section~\ref{sec:systematics}). Because the sum residual is applied as a dilution on $\mathcal{Q} = N^+ - N^-$ rather than on the much larger total yield, its impact on the corrected charge correlator is small compared to that of the asymmetry residual.}
    \label{fig:misid_sum}
\end{figure}

Figure~\ref{fig:misid_sum} shows the corresponding residual of the misreconstruction sum, $\delta^i_{\rm sum}(\theta) = \delta(f^+ + f^-)$, in the same exclusive bins. The asymmetry residual ultimately drives the bulk of the data-driven correction: as discussed in Section~\ref{sec:misid_correction}, $\delta_{\rm asym}$ enters the corrected charge correlator multiplied by the total track yield $N_{\rm tot}(\theta) = N^+ + N^-$, whereas $\delta_{\rm sum}$ acts as a dilution on the much smaller difference $\mathcal{Q} = N^+ - N^-$, so even a comparable residual size translates into a much smaller effect on $\mathcal{Q}$.

\subsection{Application of the charge misreconstruction correction}
\label{sec:misid_correction}
Charge misreconstruction both dilutes and biases the measured charge correlator. Let $f^{+}(\theta)$ and $f^{-}(\theta)$ be the per-track sign-flip probabilities for tracks of true positive and true negative charge, $f^{\pm} = P(\text{reco}\,{\mp}\,|\,\text{true}\,{\pm})$, and let $\Sigma = f^+ + f^-$ and $\Delta = f^- - f^+$ denote their symmetric and asymmetric combinations. The reconstructed and true correlators are then related by
\begin{equation}\label{eq:dilution}
    \mathcal{Q}_{\rm rec} = \mathcal{Q}_{\rm true}\,(1-\Sigma) + N_{\rm tot}\,\Delta,
\end{equation}
where the first term is a multiplicative dilution and the second an additive bias proportional to the total track yield. The MC-based correction of Section~\ref{sec:mc_correction} removes the charge migration predicted by the simulation, leaving a residual effect governed by the residual data--MC differences $\delta_{\rm sum} = \Sigma_{\rm data}-\Sigma_{\rm MC}$ and $\delta_{\rm asym} = \Delta_{\rm data}-\Delta_{\rm MC}$. The same inversion as Equation~\ref{eq:dilution} then applies with $\Sigma$ and $\Delta$ replaced by these residuals. 
The data-driven correction reads
\begin{equation}\label{eq:full_correction}
    \mathcal{Q}_{\rm corr}(\theta) =
    \frac{\mathcal{Q}_{\rm eff}(\theta) - \delta_{\rm asym}(\theta)\,N_{\rm tot}(\theta)}
         {1 - \delta_{\rm sum}(\theta)},
\end{equation}
where $\mathcal{Q}_{\rm eff}$ is the MC-corrected correlator of Section~\ref{sec:mc_correction}.

The residuals $\delta_{\rm asym}(\theta)$ and $\delta_{\rm sum}(\theta)$ are obtained from the per-$p_{\rm T}$-bin $\tau$ measurements of the previous section, and then reweighted to the hadronic-event $p_{\rm T}$ spectrum at each polar angle
\begin{equation}\label{eq:pt_reweight}
    \delta_X(\theta) = \sum_i w^{\rm had}_i(\theta)\,\delta_X^i(\theta),
    \qquad X \in \{\rm asym, sum\},
\end{equation}
with per-angle weights
\begin{equation}\label{eq:had_weights}
    w^{\rm had}_i(\theta) =
    \frac{N^{\rm had}(p_{\rm T}\!\in\! i,\,\theta)}{N^{\rm had}(\theta)}
\end{equation}
given by the fractional population of hadronic-event tracks in each exclusive $p_{\rm T}$ bin. This folds the $\tau$-derived per-bin residual through the hadronic momentum spectrum, whose shape differs from that of the $\tau$ probe spectrum. The $\delta_{\rm sum}$ contribution is small, as it enters only as a dilution of the already-small signal $\mathcal{Q}$.
\textit{\textbf{The dominant correction comes from the asymmetric residual $\bm{\delta_{\rm asym}}$, amplified by the much larger total track yield $\bm{N_{\rm tot}}$.}}

\begin{figure}[ht!]
    \centering
    \includegraphics[width=0.5\textwidth]{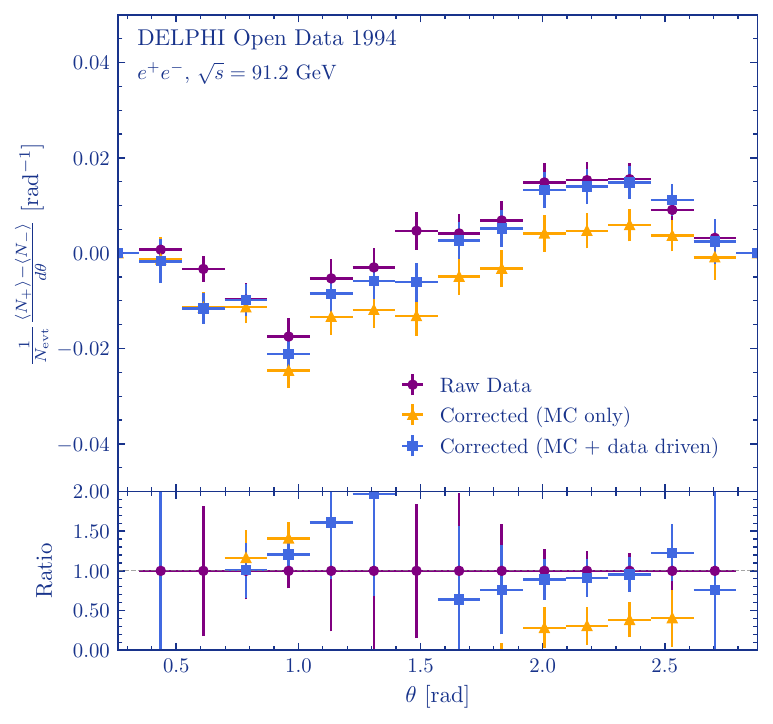}
    \caption{One-point charge correlator at three correction stages for tracks with $p_{\rm T} > 2$~GeV. The raw data (purple), MC-only corrected (orange), and fully corrected including the data-driven charge misreconstruction correction (blue) are shown. The lower panel displays the ratio to the raw data.}
    \label{fig:misid_corr}
\end{figure}

The procedure is validated end-to-end by the bias-injection closure test of Section~\ref{sec:misid_closure}. The per-$p_{\rm T}$-bin extraction recovers the injected charge asymmetry (Figure~\ref{fig:misid_closure_extraction}), and the fully corrected distribution agrees with the un-injected truth (Figure~\ref{fig:misid_closure_application}) to within a small residual that is assigned as a systematic uncertainty. The full correction chain is shown in Figure~\ref{fig:misid_corr} for the nominal $p_{\rm T} > 2$~GeV selection, comparing the raw, MC-corrected, and fully corrected distributions.

\subsection{Data-driven correction closure: bias injection test}
\label{sec:misid_closure}
The data-driven correction is validated with a bias-injection closure test. A known charge-misreconstruction asymmetry, prescribed by the per-$p_{\rm T}$- and $\theta$-bin parametrized residuals $\delta = f^- - f^+$ measured on the 1994 sample (Section~\ref{sec:tnp_method}), is injected at the generator level by randomly flipping track charges in both the $Z \to \tau^+\tau^-$ control sample and the hadronic sample. Although the absolute charge-misreconstruction rates may differ between the two environments, only the asymmetric component $\delta$ enters the data-driven correction; injecting a rate that reproduces $\delta$ alone is therefore sufficient to probe everything the procedure is sensitive to. The corrected charge correlator obtained after the full data-driven correction chain is then compared against the un-injected generator-level truth. The test is run in the high-statistics limit, using all available generator-level MC events without luminosity scaling, so as to suppress statistical fluctuations and isolate any residual non-closure of the procedure.

\begin{figure}[ht!]
    \centering
    \includegraphics[width=0.9\textwidth]{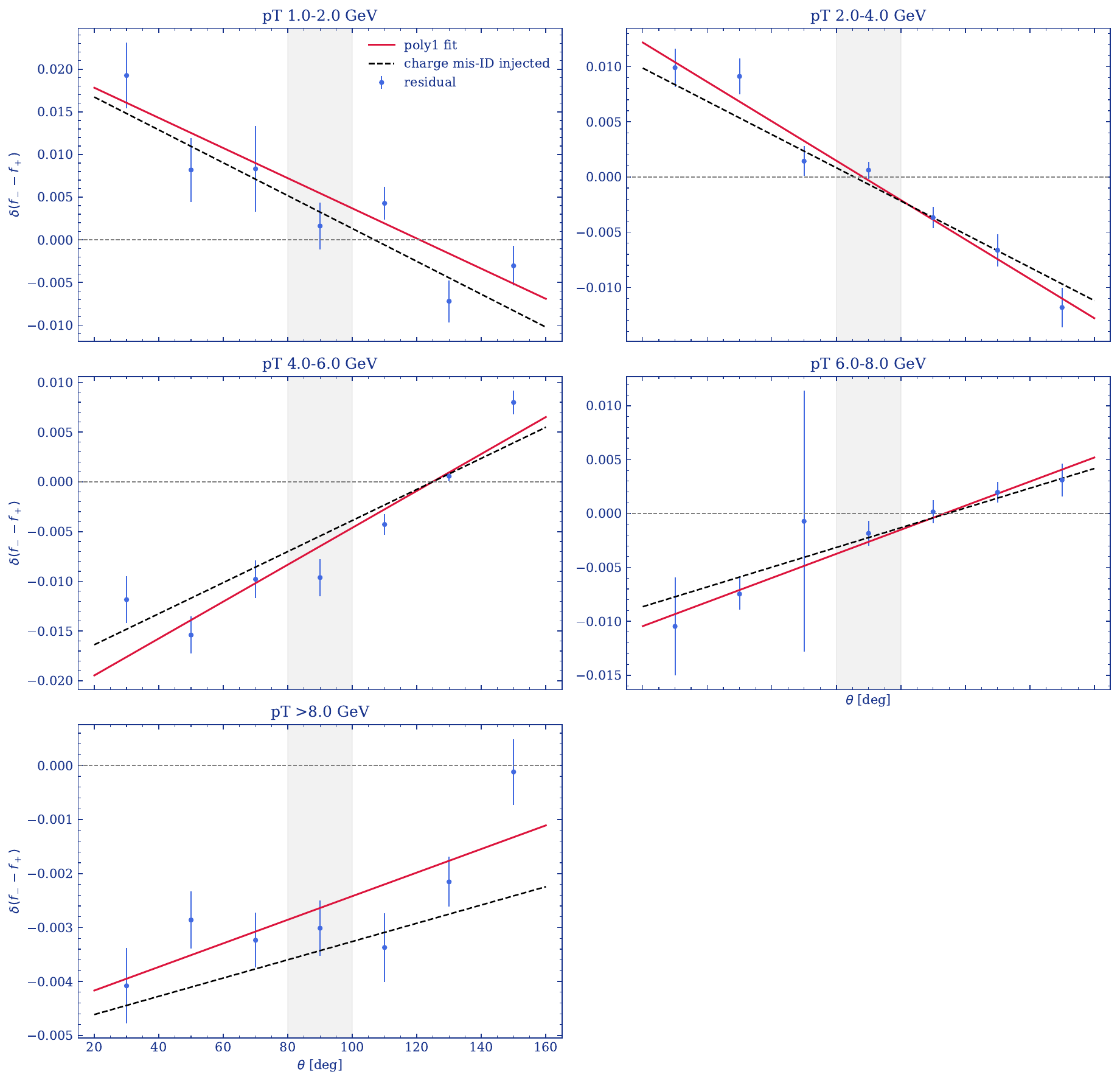}
    \caption{Closure test of the data-driven correction: extracted versus injected $\delta_{\rm asym}^i(\theta)$ residuals in five exclusive probe-$p_{\rm T}$ bins, obtained by running the $\tau$ TNP procedure on a generator-level MC sample restricted to truth $1{+}1$-prong events with a known charge-misreconstruction rate injected.}
    \label{fig:misid_closure_extraction}
\end{figure}

\begin{figure}[ht!]
    \centering
    \includegraphics[width=0.49\textwidth]{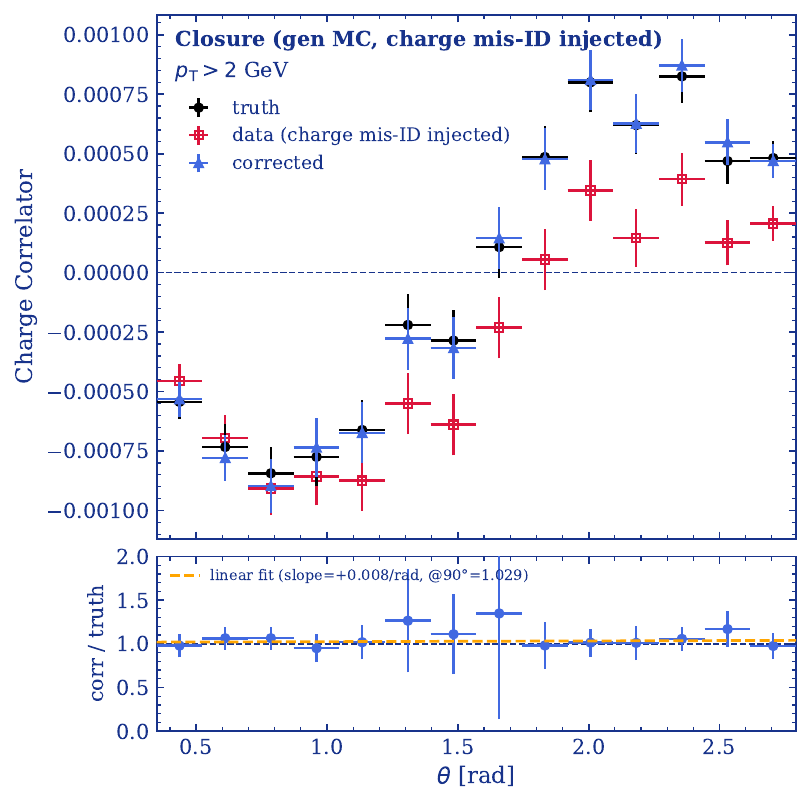}
    \includegraphics[width=0.49\textwidth]{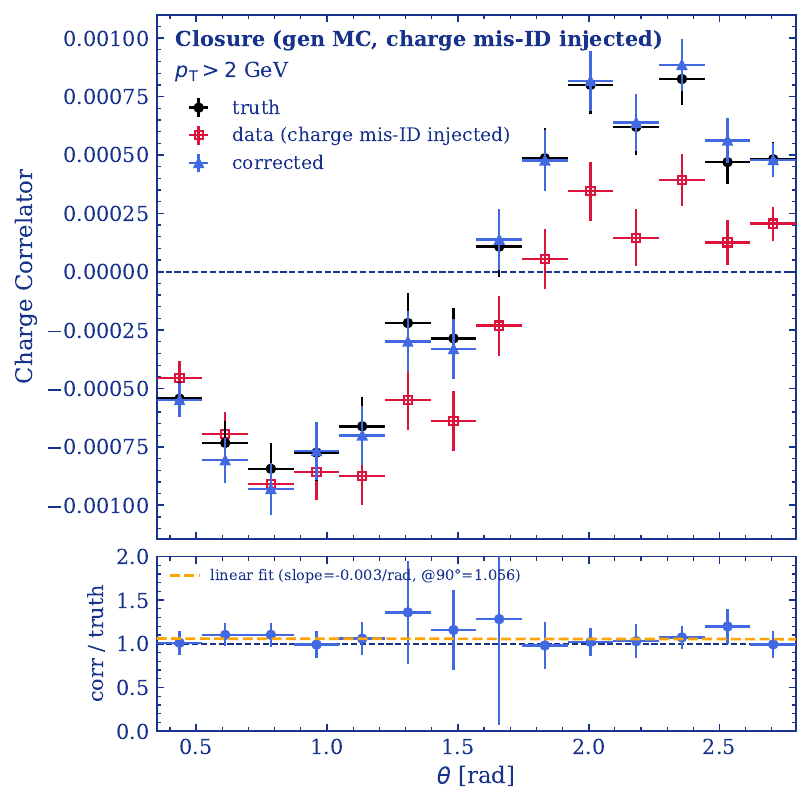}
    \caption{Closure test of the data-driven correction: the corrected hadronic charge correlator (blue) compared to the un-injected generator-level truth (black) and the raw injected correlator before correction (red). Left: idealized configuration, with injection and extraction restricted to truth-level $1{+}1$-prong $Z \to \tau^+\tau^-$ events. Right: realistic configuration, including $1{+}3$-prong leakage into the $1{+}1$ selection. The corrected output agrees with the truth across the full $\theta$ range; the residual $\sim 5\%$ shape shift in the realistic configuration is assigned as a systematic uncertainty.}
    \label{fig:misid_closure_application}
\end{figure}

The injected $\tau$ sample is processed through the TNP extraction of Section~\ref{sec:tnp_method} in two configurations. In the first, idealized configuration, the injection and the TNP extraction are restricted to truth-level $1{+}1$-prong $Z \to \tau^+\tau^-$ events. The resulting per-$p_{\rm T}$-bin residuals, shown in Figure~\ref{fig:misid_closure_extraction} overlaid with the injected truth, are recovered by the linear parametrization within the statistical precision. The extracted residual is then propagated with the hadronic-event $p_{\rm T}$ spectrum (Equations~\ref{eq:pt_reweight}--\ref{eq:had_weights}) and applied to the injected hadronic sample charge correlator distribution via Equation~\ref{eq:full_correction}. Figure~\ref{fig:misid_closure_application} (left) compares the corrected output to the un-injected generator-level truth. The two agree across the full $\theta$ range within the statistical precision, while the raw injected correlator (also overlaid) exhibits the clear bias that the procedure removes. This configuration validates the algebraic inversion of the correction formula, Equation~\ref{eq:full_correction}.

In the second, realistic configuration, the injection and extraction are performed on all selected $1{+}1$-topology $Z \to \tau^+\tau^-$ events, which include leakage from $1{+}3$-prong decays in which two of the three prongs are not reconstructed. This reproduces the composition of the real control sample. The corrected output, the un-injected truth, and the raw injected correlator are shown in Figure~\ref{fig:misid_closure_application} (right). A residual shape shift of $\sim 5\%$ remains after correction and is assigned as a systematic uncertainty on the data-driven procedure. A complementary, hadronic-event--based correction that does not rely on the $\tau$ control region is presented in Appendix~\ref{sec:hadronic_correction}.

\clearpage
\section{Statistical and systematic uncertainties}
\label{sec:systematics}

The measurement of the one-point charge correlator is statistically limited, both in the final result and in the data-driven correction procedure. This section discusses the treatment of statistical and systematic uncertainties.

\subsection{Statistical uncertainty}
The statistical uncertainties are evaluated from the full event-by-event covariance. For each pair of angular bins $(a,b)$, with the per-event net charge $q_j^{(a)} = \sum_{i \in \mathrm{bin}\,a} q_i$, the event-by-event covariance is defined as
\begin{equation}\label{eq:stat_canonical}
\mathrm{Cov}(\mathcal{Q}_a, \mathcal{Q}_b)
= \frac{1}{N_{\rm evt}\,(N_{\rm evt}-1)\,\Delta\theta_a\,\Delta\theta_b}
  \sum_{j=1}^{N_{\rm evt}}
  \bigl(q_j^{(a)} - \bar{q}^{(a)}\bigr)
  \bigl(q_j^{(b)} - \bar{q}^{(b)}\bigr),
\end{equation}
where $\bar{q}^{(a)}$ is the sample average charge in bin $a$.
This captures both the within-event track correlations that inflate the variance relative to Poisson and the charge-balance anti-correlation across bins that reduces it.

\begin{figure}[ht!]
    \centering
    \includegraphics[width=0.55\textwidth]{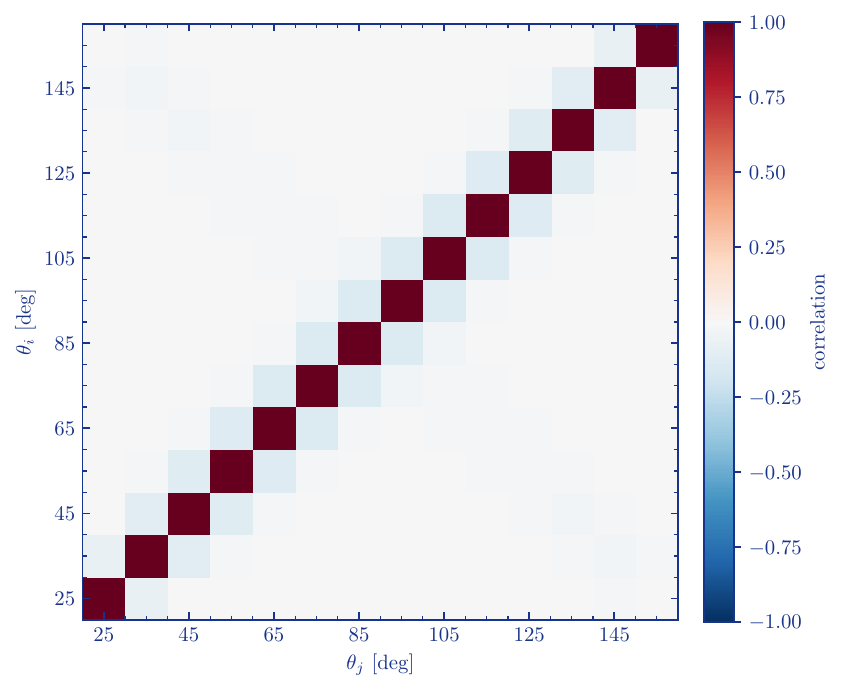}
    \caption{Statistical correlation matrix of $\mathcal{Q}(\theta)$ between angular bins, derived from the event-by-event covariance.}
    \label{fig:stat_cov}
\end{figure}

Because the correction is applied per charge with $\varepsilon^+ \neq \varepsilon^-$, Equation~\ref{eq:stat_canonical} cannot be propagated as a single net-charge object. Three per-charge covariances are therefore accumulated,
\begin{equation}\label{eq:stat_percharge}
C_{++}(a,b) = \mathrm{Cov}(Y^+_a, Y^+_b), \qquad
C_{--}(a,b) = \mathrm{Cov}(Y^-_a, Y^-_b), \qquad
C_{+-}(a,b) = \mathrm{Cov}(Y^+_a, Y^-_b),
\end{equation}
which carry the same event-by-event information as Equation~\ref{eq:stat_canonical}, with $Y^{\pm}_a$ the raw track yields of Equation~\ref{eq:charged_densities}. The corrected correlator is
\begin{equation}\label{eq:stat_qcorr}
\mathcal{Q}_{\rm corr}(a) = \frac{Y^+_a - n^+_{{\rm fake},a}}{s^+(a)} - \frac{Y^-_a - n^-_{{\rm fake},a}}{s^-(a)},
\end{equation}
with per-charge scale factors $s^\pm(a) = N_{\rm evt}\,\varepsilon^\pm(a)\,d(a)\,\Delta\theta_a$ built from the number of events $N_{\rm evt}$, the per-charge efficiency $\varepsilon^\pm$, the residual misreconstruction dilution $d(a) = 1 - \delta_{\rm sum}(a)$, and the bin width $\Delta\theta_a$. Here $n^\pm_{{\rm fake},a}$ is the simulation-derived subtraction of fake and charge-misidentified tracks. The correction factors are held at their central values in the propagation, and their statistical uncertainties are accounted for separately as systematic uncertainties (Section~\ref{sec:systematicsSource}). The statistical covariance is then obtained by propagating the three per-charge covariances through the scale factors,
\begin{equation}
V^{\rm stat}_{ab} = \frac{C_{++}(a,b)}{s^+(a)\,s^+(b)} + \frac{C_{--}(a,b)}{s^-(a)\,s^-(b)} - \frac{C_{+-}(a,b)}{s^+(a)\,s^-(b)} - \frac{C_{+-}(b,a)}{s^-(a)\,s^+(b)}.
\end{equation}
The two cross terms are distinct because $C_{+-}$ is in general asymmetric. By construction, it reduces to Equation~\ref{eq:stat_canonical} when $s^+ = s^-$. The resulting correlation matrix, shown in Figure~\ref{fig:stat_cov}, exhibits the near-diagonal anti-correlation characteristic of the event-by-event charge balance.

The 1994 and 1995 correlators are corrected independently and combined with weights given by the number of selected events in each year. The combined statistical covariance is added to the systematic covariance of Section~\ref{sec:syst_profile} to form the total covariance used in the comparison with the prediction (Section~\ref{sec:results}). Its contraction with the bin widths gives the statistical uncertainty on the integrated charge used in Section~\ref{sec:syst_profile}.



\subsection{Systematic uncertainty sources}
\label{sec:systematicsSource}

\paragraph{High-momentum fake tracks}
A significant discrepancy exists between data and simulation in the high-$p_{\rm T}$ track spectrum, as visible in Figure~\ref{fig:data_mc_theta}. This systematic addresses the possibility that the data contains a source of high-$p_{\rm T}$ background tracks not modeled in the simulation. The potential impact is estimated by removing a fraction of these tracks from the data. Specifically, a $p_{\rm T}$-dependent fraction, ranging from 10\% to 40\% of charged tracks with $p_{\rm T} > 30$~GeV, is randomly removed from the data sample. The full difference between the nominal result and the result from this modified data sample, evaluated at the reconstructed level, is less than 1\%. 

\paragraph{Charged track momentum scale}
The momentum scale of charged tracks is varied by a $p_{\rm T}$-dependent factor (from 0.1\% at low $p_{\rm T}$ to 2.5\% at high $p_{\rm T}$), following the procedure of Ref.~\cite{DELPHI:2000uri}.
The analysis is repeated with the rescaled momenta; the full effect on the charge correlator is below 1\%.

\paragraph{MC model dependence of the MC-based correction}
The MC-based correction depends on the choice of generator used to derive it. This dependence is assessed by repeating the full correction with \textsc{KK}2f as an alternative generator in place of the nominal \textsc{PYTHIA}~8 sample and comparing the two results; the difference between the two quantifies the dependence of the efficiency correction on the modeled track kinematics, $K/\pi$ abundance, and event environment, and is propagated as a generator-model systematic uncertainty (Figure~\ref{fig:bruteforce_kk2f}).

\paragraph{Track finding efficiency}
The DELPHI collaboration performed detailed calibrations of the track reconstruction, tuning the simulated hit efficiencies to match data~\cite{Elsing:2000fv, Osterberg:1998xxx, DELPHI:1995dsm}. A previous DELPHI analysis~\cite{DELPHI:2000uri} estimated the residual efficiency uncertainty by randomly dropping reconstructed tracks at the 2\% level in the simulation. We adopt the same random-drop technique, applied separately to positive and negative tracks, and propagate the per-charge differences with respect to the nominal as independent systematic shifts on $N^+$ and $N^-$. Because the two charges are varied independently, the resulting shift follows the total charged-particle density $N_{\rm tot}(\theta)$ rather than the much smaller charge correlator, and is therefore sizable and, to good approximation, symmetric under $\theta \to 180^\circ\!-\theta$. The charge-conservation constraint of Section~\ref{sec:syst_profile} bounds its angular integral but not its full angular shape, and a residual at the $10\%$ level remains, making this the second largest systematic source. Any kinematic dependence of the track finding efficiency on event activity or track multiplicity is additionally covered by the MC model-dependence systematic. 

\begin{figure}[H]
    \centering
    \includegraphics[width=0.95\textwidth]{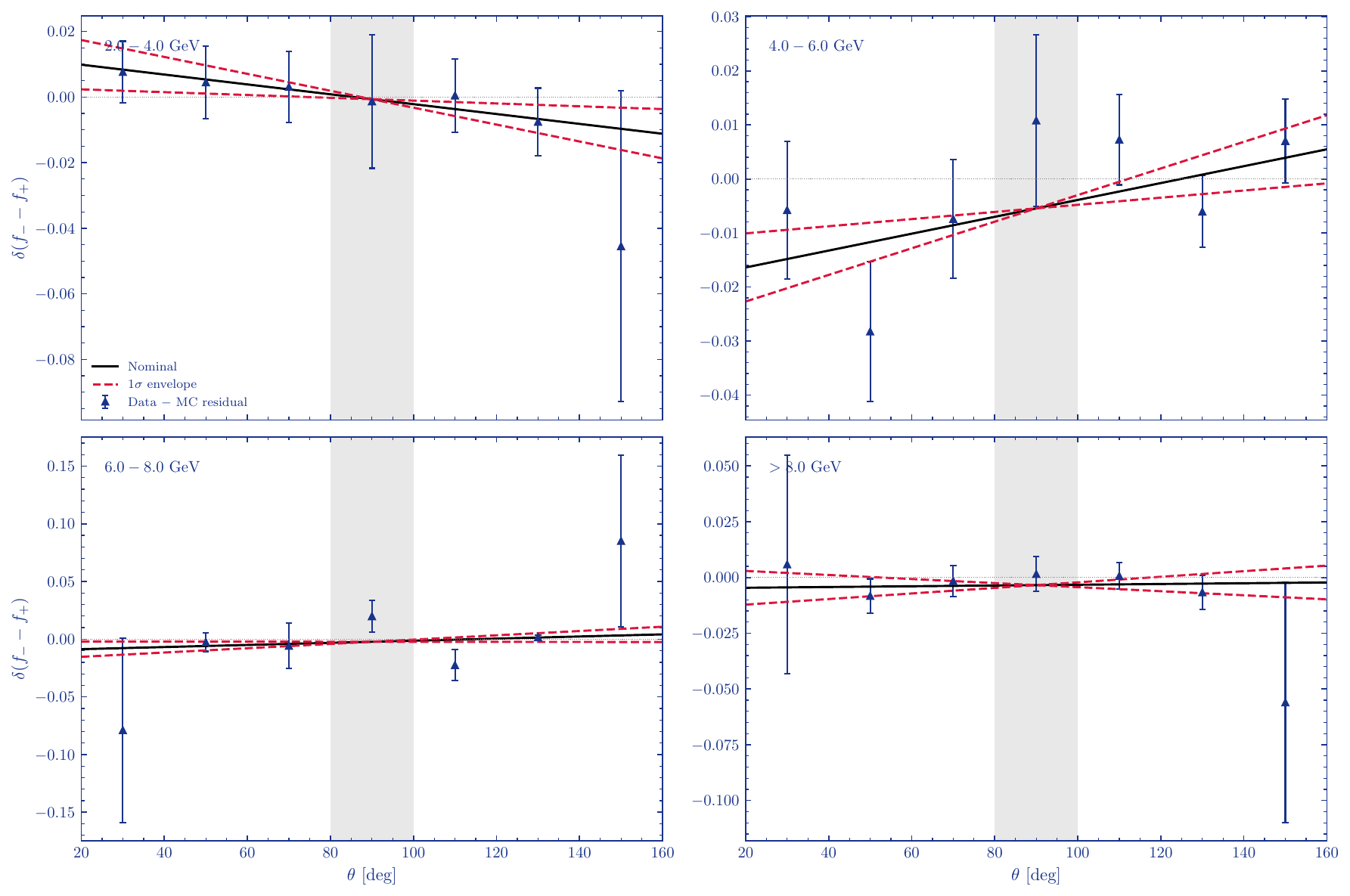}\\[0.5em]
    \includegraphics[width=0.95\textwidth]{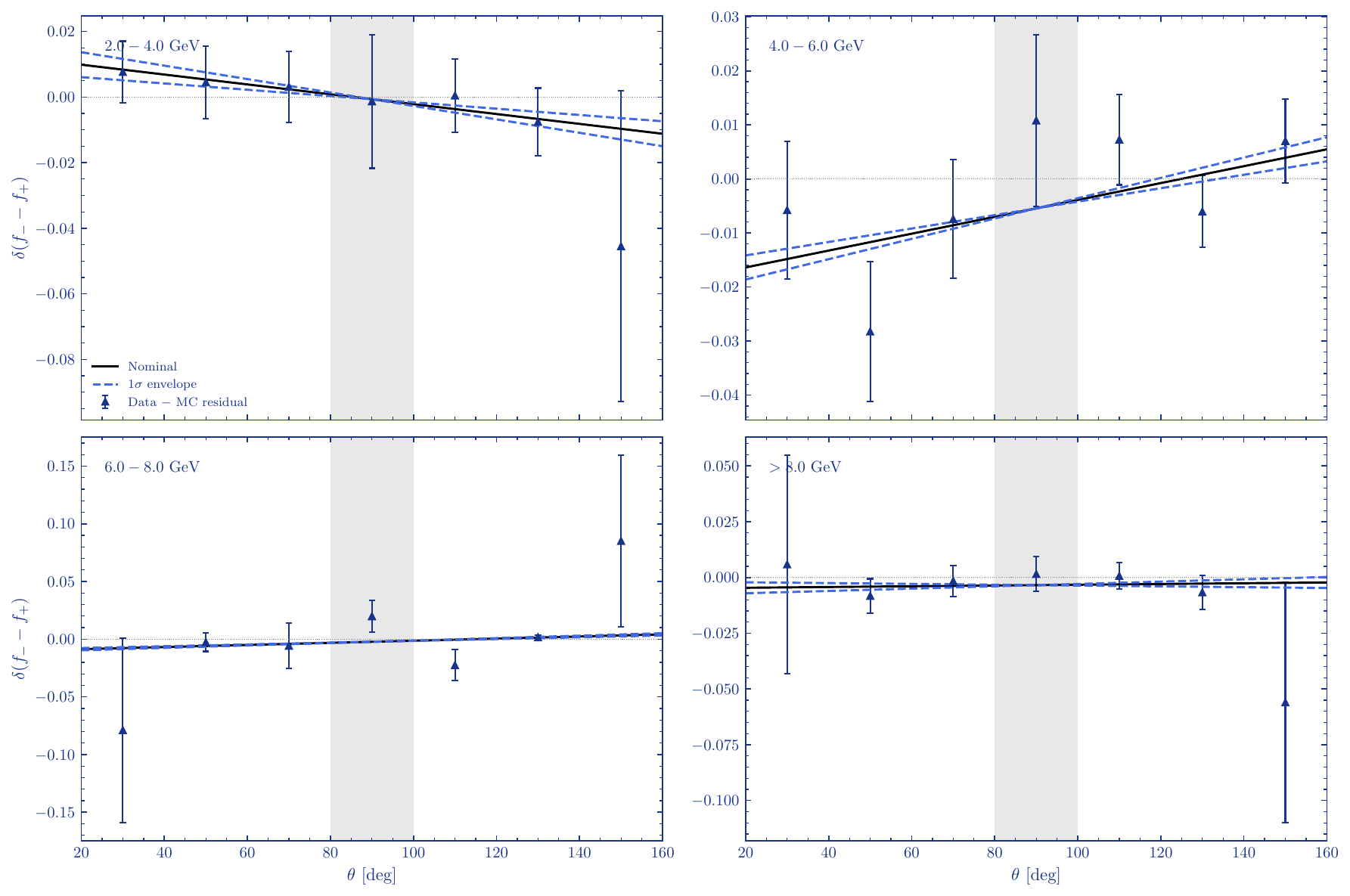}
    \caption{$\pm 1\sigma$ envelopes of the two eigenmodes ($\lambda_+$, top; $\lambda_-$, bottom) of the covariance matrix of the linear parametrization of $\delta_{\rm asym}(\theta)$, summed coherently across the four probe-$p_{\rm T}$ bins entering the correction and propagated as independent shift vectors representing the full statistical uncertainty of the data-driven charge-misreconstruction bias scale factor. The shaded band between $80^\circ$ and $100^\circ$ marks the bins containing the cathode plane of the DELPHI TPC.}
    \label{fig:misid_eigenmodes}
\end{figure}

\paragraph{Charge misreconstruction bias scale factor}
The asymmetric component of the charge misreconstruction, $\delta_{\rm asym}(\theta) = (f^- - f^+)_{\rm data} - (f^- - f^+)_{\rm MC}$, directly biases the charge correlator by introducing a spurious difference between the positive and negative track rates.
The data-driven correction (Section~\ref{sec:data_driven}) is obtained from a linear parametrization of the per-$p_{\rm T}$-bin residual in the range $20^\circ$--$160^\circ$, excluding the band $80^\circ$--$100^\circ$ affected by the TPC cathode-plane crack.
The statistical uncertainty of the scale factor is fully propagated to the corrected charge correlator via the covariance matrix of the linear-fit parameters. In each of the four probe-$p_{\rm T}$ bins entering the correction ($2$--$4$, $4$--$6$, $6$--$8$, and $>8$~GeV), the $2 \times 2$ covariance is diagonalized into two orthogonal eigenmodes, $\lambda_{+}$ and $\lambda_{-}$, yielding $2 \times 4 = 8$ independent shift vectors. These are treated as fully uncorrelated across both the eigenmode index and the probe-$p_{\rm T}$ index. The $\pm 1\sigma$ envelopes on the corrected charge correlator from the $\lambda_{+}$ and $\lambda_{-}$ modes, summed coherently across the four probe-$p_{\rm T}$ bins, are shown in Figure~\ref{fig:misid_eigenmodes}. This is the dominant systematic source.

\paragraph{Charge misreconstruction dilution scale factor}
The symmetric component, $\delta_{\rm sum}(\theta) = (f^+ + f^-)_{\rm data} - (f^+ + f^-)_{\rm MC}$, dilutes the measured charge correlator by modifying the overall misreconstruction rate.
The statistical uncertainty of this scale factor is propagated in exactly the same way as for the bias scale factor: in each of the four probe-$p_{\rm T}$ bins, the $2\times 2$ covariance of the linear fit to $\delta_{\rm sum}^i(\theta)$ is diagonalized into two orthogonal eigenmodes, giving an additional $2\times 4 = 8$ independent shift vectors that are treated as fully uncorrelated across both the eigenmode index and the probe-$p_{\rm T}$ index.
The impact on the corrected charge correlator is much smaller than that of the bias term (below 1\%), because $\delta_{\rm sum}$ enters as a dilution on the small difference $\mathcal{Q} = N^+ - N^-$ rather than being amplified by the much larger total track yield $N_{\rm tot} = N^+ + N^-$ as $\delta_{\rm asym}$ is. 

\paragraph{Tag-and-probe selection bias}
The TNP procedure for measuring the per-track charge misreconstruction rate carries residual selection biases from limited tag purity, studied in Appendix~\ref{app:tnp_check}. 
A comparison between the TNP-extracted asymmetric residual $\delta_{\rm asym}^{\rm TNP}$ and the corresponding generator-level truth in MC reveals a $0.1\%$ discrepancy on the misreconstruction rate. 
Propagated through the data-driven correction, this absolute discrepancy is amplified by the total track yield $N_{\rm tot}$ and translates into a relative impact on the corrected charge correlator of $7.0\%$ at $\theta = 55^\circ$ and $9.7\%$ at $\theta = 125^\circ$ in the 1994 dataset. The full $\theta$-dependent shift is shown in Figure~\ref{fig:tnp_nonclosure}.

\paragraph{Tag-and-probe sample contamination modeling}
A decomposition of $\delta_{\rm asym}^{\rm TNP}$ by generator-level event topology (Appendix~\ref{app:tnp_check}) reveals that the dominant 
contribution to the measured residual comes from $1{+}3$ prong leakage into the nominal $1{+}1$ reco sample. 
Although the leakage fraction after the tag-and-probe selection is only $1.7\%$, the leaked events carry an intrinsically charge-asymmetric structure that survives in the data--MC residual unless the topology is modeled exactly in simulation. 
To account for imperfect modeling, a $10\%$ uncertainty is assigned to the $1{+}3$ topology contribution and propagated as a fully correlated shift across $\theta$ bins, contributing at or below the percent level to the corrected charge correlator.

\paragraph{Tag-and-probe sample contamination bias}
The TNP measurement can be biased by genuine same-sign track pairs from $1{+}3$-prong $\tau$ decays in which two of the three prongs are not reconstructed; such events leak into the $1{+}1$-prong control sample and inflate the measured same-sign fraction. This bias is estimated with the bias-injection closure test on generator-level MC described in Section~\ref{sec:misid_closure}, from the difference between the idealized and realistic configurations. The resulting residual non-closure, shown in Figure~\ref{fig:misid_closure_application} and amounting to about $5\%$, is assigned as a systematic uncertainty for this effect.

\subsection{Application of the global charge-conservation constraint}
\label{sec:syst_profile}
We exploit global electric-charge conservation to constrain the systematic uncertainties. The true charge correlator is odd under $\theta \to 180^\circ\!-\theta$, and the fiducial region is symmetric about $\theta = 90^\circ$, so its angular integral over the measured range vanishes exactly. The deviation from zero observed in data therefore arises only from statistical fluctuations, of size $\sigma_{\rm stat}$, and from the parity-even component of the residual detector bias. Each of the systematic sources is instead fully correlated across $\theta$, and the parity-even component would shift the integrated charge well away from zero. The integrated charge measured in data therefore effectively bounds the normalization of each correlated systematic.

Let the corrected one-point charge correlator be represented as a vector $\mathbf{Q}$ of its measured values in the $N_\theta = 14$ bins of polar angle $\theta$. Each systematic source $i$, $i = 1,\dots,K$, is propagated as a coherent $1\sigma$ shift of $\mathbf{Q}$, encoded as a length-$N_\theta$ vector $\mathbf{d}_i$ giving the bin-by-bin response of $\mathbf{Q}$ to a one-sigma variation of that source. The $\theta$ dependence of each shift vector is parametrized by a cubic function, applied uniformly to all sources to suppress bin-to-bin fluctuations while retaining the leading and next-to-leading terms in each parity sector. The parametrization acts primarily on the shifts derived from differences of finite simulated samples, as intended, and leaves the dominant charge-misreconstruction source unchanged to within about $10\%$ across the bulk of the acceptance. The correlated systematic offset $\boldsymbol{\beta}$ of the measured spectrum then has the covariance built source-by-source as
\begin{equation}
V_{\rm syst} = \sum_{i=1}^{K} \mathbf{d}_i\,\mathbf{d}_i^{\!\top}, 
\label{eq:Vsyst-bare}
\end{equation}
an $N_\theta \times N_\theta$ matrix giving the systematic covariance between $\theta$ bins. The global charge-conservation constraint derived below modifies this matrix in place.

\begin{figure}[ht!]
    \centering
    \includegraphics[width=0.6\textwidth]{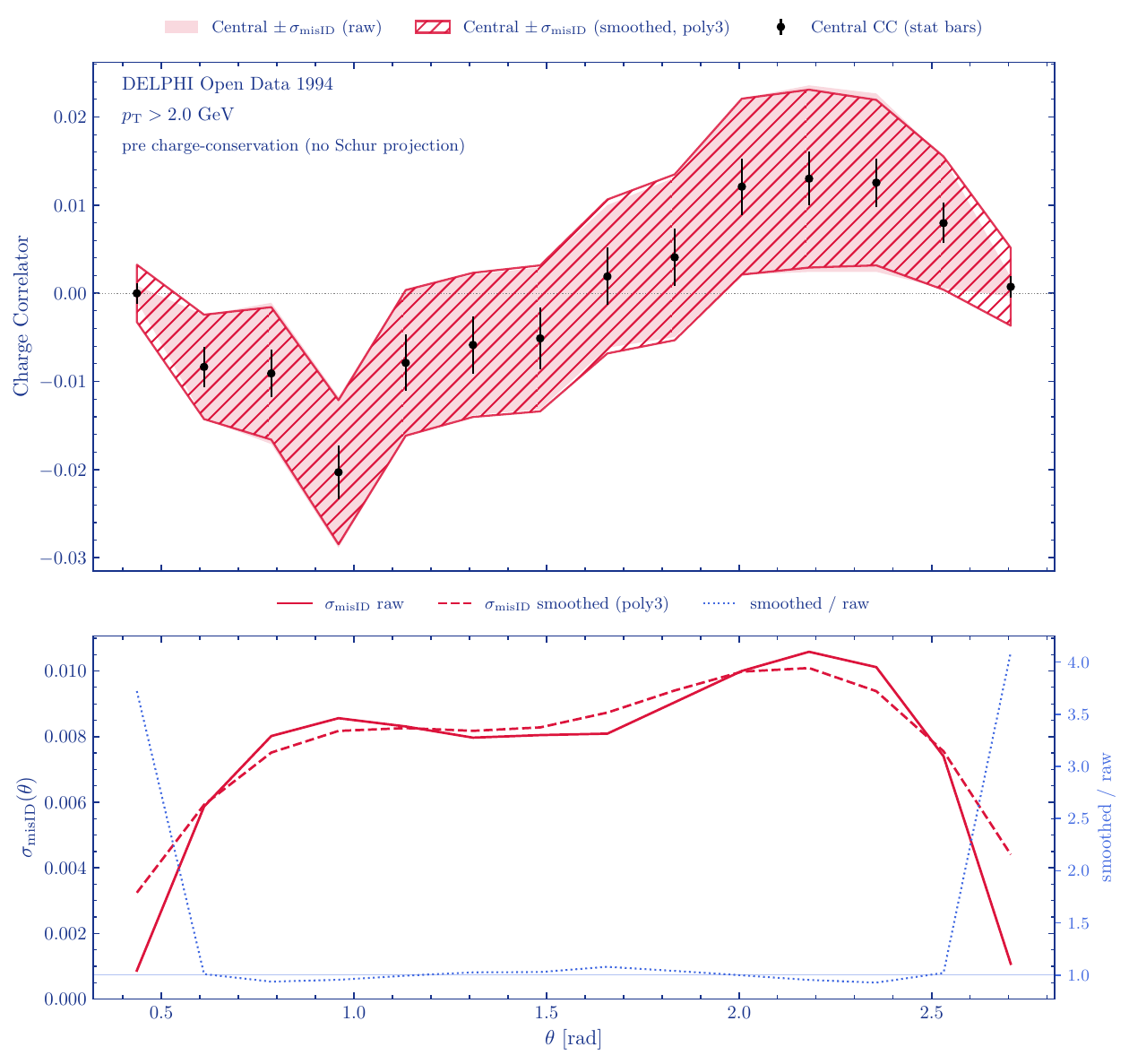}
    \caption{Charge-misreconstruction bias scale-factor uncertainty for the 1994 dataset, before the charge-conservation constraint is applied. The top panel shows the measured charge correlator with the raw (solid band) and cubic-parametrized (hatched band) $\pm 1\sigma$ envelopes of this source. The bottom panel shows the corresponding raw and parametrized uncertainties together with their ratio. The two agree to within about $10\%$ across the bulk of the acceptance, while in the outermost bins, where the raw estimate drops sharply, the parametrization yields a conservatively larger uncertainty. In the plot labels, $\sigma_{\rm misID}$ denotes the total $\pm 1\sigma$ envelope of this source, summed over its eigenmode components, \texttt{poly3} denotes the cubic parametrization of the shift vectors, and pre charge-conservation (no Schur projection) indicates that the covariance update of Equation~\ref{eq:schur} has not yet been applied. The black points show the corrected charge correlator with statistical error bars.}
    \label{fig:misid_smooth_band}
\end{figure}

The effect of the parametrization on the dominant source, the charge-misreconstruction bias scale factor, is shown in Figure~\ref{fig:misid_smooth_band}, which compares the raw and parametrized uncertainty envelopes around the measured correlator, prior to the application of the charge-conservation constraint. The parametrized uncertainty agrees with the raw one to within about $10\%$ across the bulk of the acceptance. In the outermost bins, where the raw estimate itself drops sharply, the parametrization yields a larger uncertainty, by up to a factor of four, which is the conservative direction.

The integrated charge correlator is the scalar
\begin{equation}
I \equiv \mathbf{c}^{\!\top}\mathbf{Q},
\end{equation}
where $\mathbf{c}$ is the vector of bin widths that turns the $\theta$-binned correlator into the angular integral $\int \mathcal{Q}(\theta)\,\mathrm{d}\theta$, so that $I$ measures the residual net charge in the fiducial region. Because charge conservation fixes the integral of the true charge correlator to zero within fluctuation, the integrated net charge observed in data, $I_{\rm obs}$, receives contributions only from the systematic offsets and from the fluctuation $\varepsilon$ of the integral,
\begin{equation}
\label{eq:cc_constraint}
I_{\rm obs} \;=\; \mathbf{c}^{\!\top}\boldsymbol{\beta} \;+\; \varepsilon ,
\qquad \varepsilon \sim \mathcal{N}\!\left(0,\,\sigma_{\rm stat}^{2}\right),
\end{equation}
where $\sigma_{\rm stat}$ is the statistical uncertainty assigned to the measured integral. The constraint is imposed by minimizing
\begin{equation}
\label{eq:cc_chi2}
\chi^{2}(\boldsymbol{\beta},\varepsilon)
 \;=\; \boldsymbol{\beta}^{\!\top} V_{\rm syst}^{-1}\,\boldsymbol{\beta}
 \;+\; \frac{\varepsilon^{2}}{\sigma_{\rm stat}^{2}}
\end{equation}
subject to Equation~\ref{eq:cc_constraint}, using a Lagrange multiplier for the constraint. The minimization is analytic. Writing $s \equiv \mathbf{c}^{\!\top} V_{\rm syst}\,\mathbf{c} + \sigma_{\rm stat}^{2}$, the best-fit solution
\begin{equation}
\label{eq:cc_pull}
\hat{\boldsymbol{\beta}} \;=\; \frac{V_{\rm syst}\,\mathbf{c}}{s}\; I_{\rm obs},
\qquad
\hat{\varepsilon} \;=\; \frac{\sigma_{\rm stat}^{2}}{s}\; I_{\rm obs}
\end{equation}
divides the observed imbalance between the correlated systematics and the statistical fluctuation in proportion to their prior variances, and propagation of the fitted offset yields the constrained systematic covariance
\begin{equation}
V_{\rm syst} \;\longrightarrow\; V_{\rm syst} \;-\; \frac{V_{\rm syst}\,\mathbf{c}\,\mathbf{c}^{\!\top}\,V_{\rm syst}}{\mathbf{c}^{\!\top} V_{\rm syst}\,\mathbf{c} \;+\; (\sigma_{\rm stat})^{2}}.
\label{eq:schur}
\end{equation}
This is the standard covariance update of a least-squares fit with a linear constraint. The constraint acts along a single direction in the space of $\theta$ bins. Because the integrated charge is one scalar, constraining it reduces the uncertainty only along the combination $V_{\rm syst}\,\mathbf{c}$ that the constraint resolves, while the uncertainty in all directions orthogonal to it is unaffected. The limit $\sigma_{\rm stat}\to 0$ corresponds to an exact constraint, while $\sigma_{\rm stat}\to\infty$ removes it. In the nominal result, $\sigma_{\rm stat}$ is set to the statistical uncertainty on the measured integrated charge. The central-value correction implied by Equation~\ref{eq:cc_pull} is proportional to $I_{\rm obs}$, which is consistent with zero within its statistical uncertainty. It is therefore negligible and not applied, and only the covariance update of Equation~\ref{eq:schur} is retained.

A simultaneous fit of the control and signal regions, with a full likelihood treatment of the systematic uncertainties, is deferred to a future dedicated precision measurement.

\begin{figure}[ht!]
    \centering
    \includegraphics[width=0.50\textwidth]{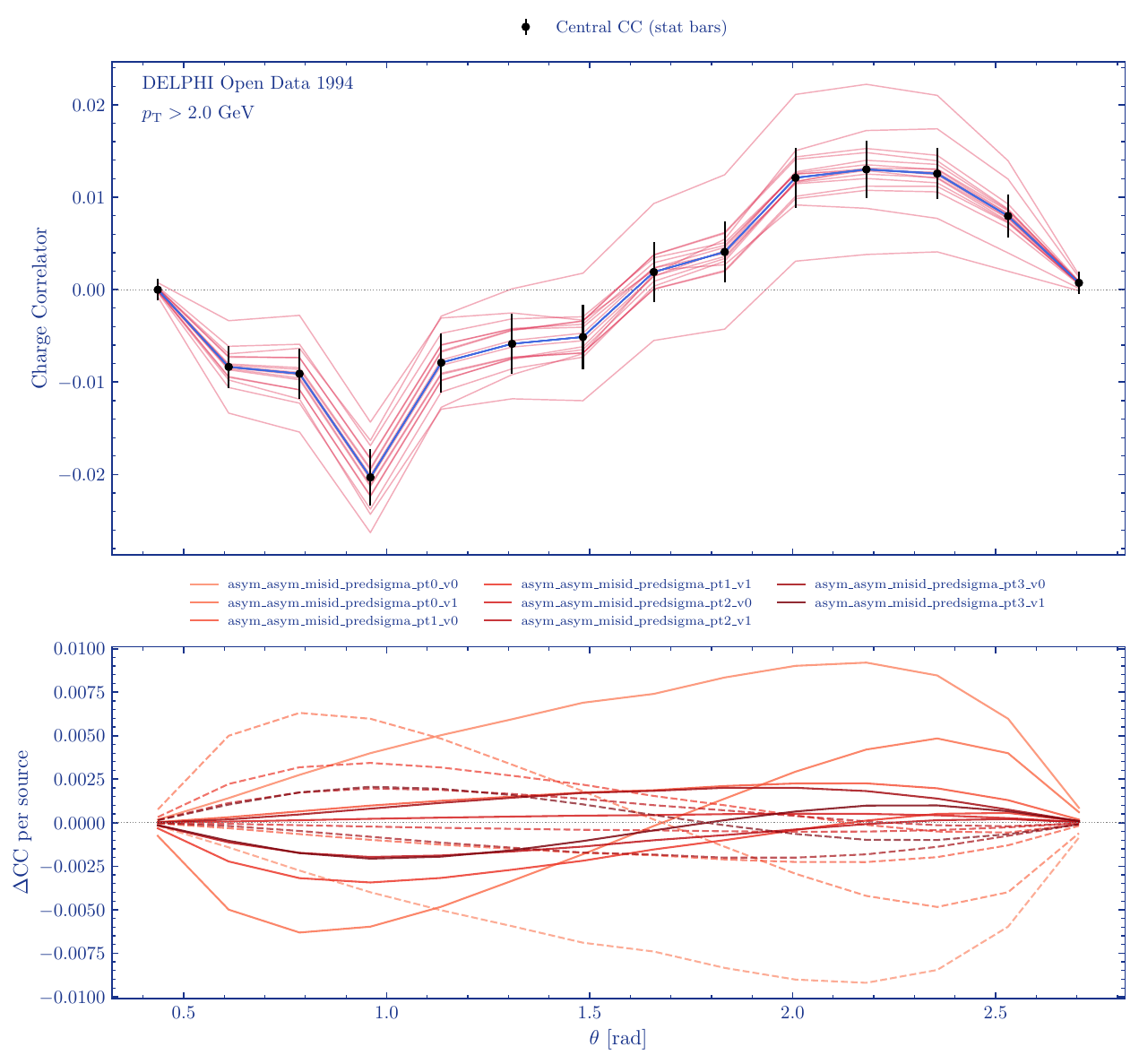}%
    \hfill
    \includegraphics[width=0.50\textwidth]{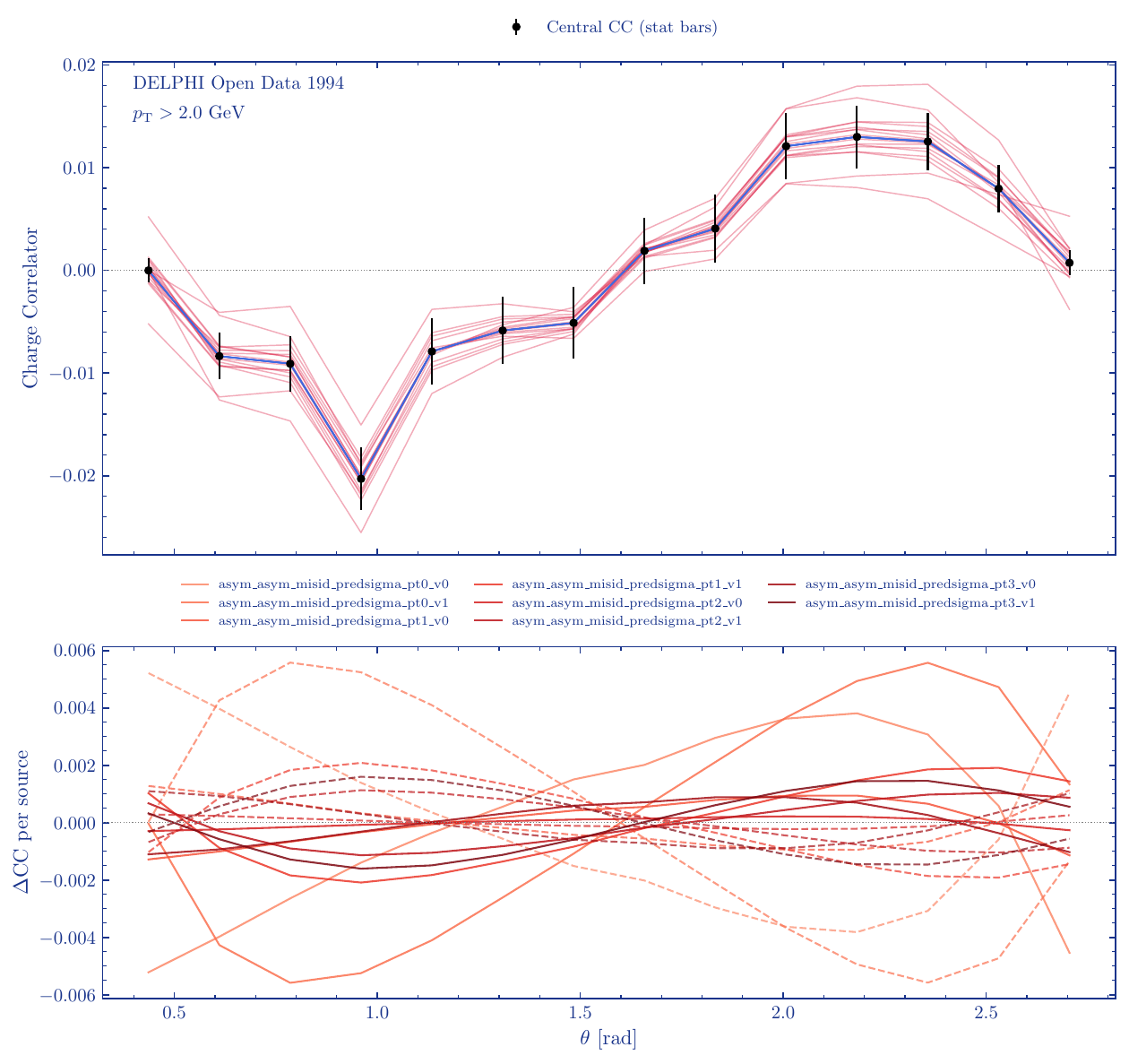}
    \caption{Size and shape of the leading systematic uncertainty, the charge misreconstruction asymmetry residual, before (left) and after (right) applying the global charge conservation constraint, for the 1994 dataset. In each panel, the top part shows the corrected charge correlator, labeled Central CC and drawn as black points with statistical error bars, overlaid with the correlator obtained after each individual systematic shift, and the bottom part shows the per-source shifts $\Delta$CC. The legend entries denote the eight eigenmode shift vectors of this source, where \texttt{pt0} through \texttt{pt3} label the four probe-$p_{\rm T}$ bins ($2$--$4$, $4$--$6$, $6$--$8$, and $>8$~GeV) and \texttt{v0} and \texttt{v1} label the two eigenmodes $\lambda_{+}$ and $\lambda_{-}$ of the linear-fit covariance in each bin (Figure~\ref{fig:misid_eigenmodes}). Solid and dashed curves show the up and down variations of each mode.}
    \label{fig:sys_shapes}
\end{figure}

\begin{figure}[ht!]
    \centering
    \includegraphics[width=0.50\textwidth]{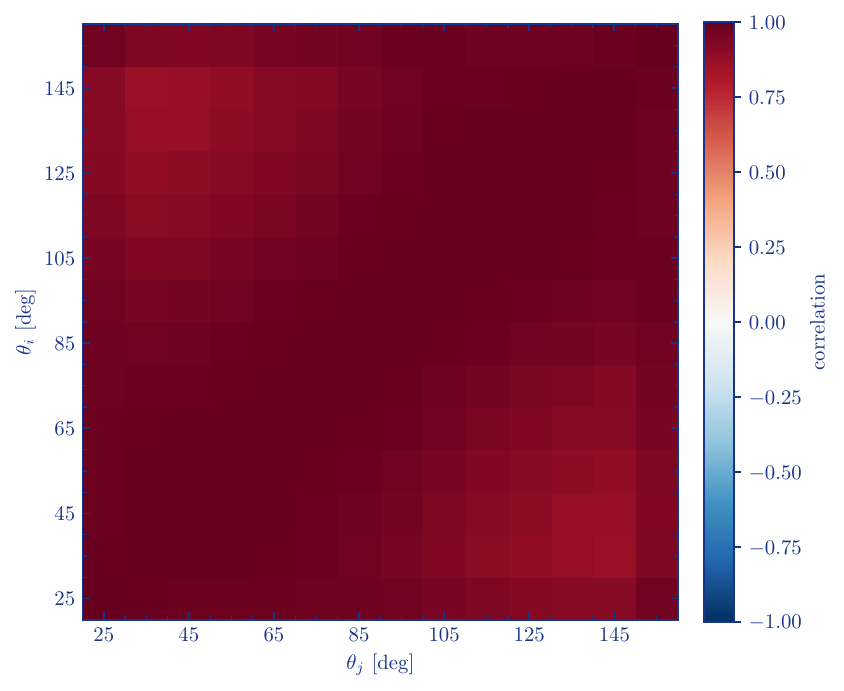}%
    \hfill
    \includegraphics[width=0.50\textwidth]{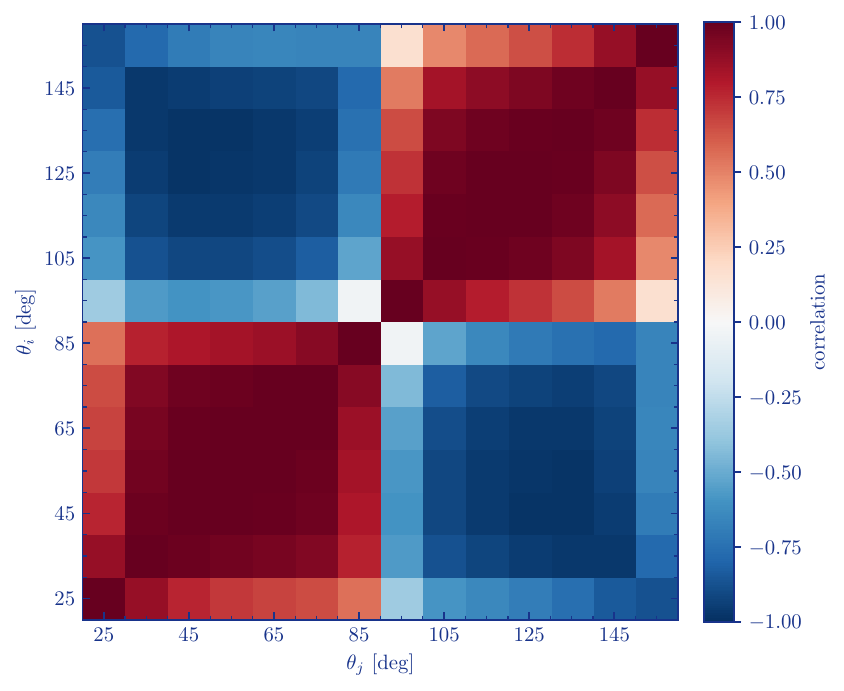}
    \caption{Systematic correlation matrix before (left) and after (right) applying the global charge conservation constraint. The constraint induces additional anti-correlation between angular bins.}
    \label{fig:sys_cov}
\end{figure}

\begin{figure}[ht!]
    \centering
    \includegraphics[width=0.75\textwidth]{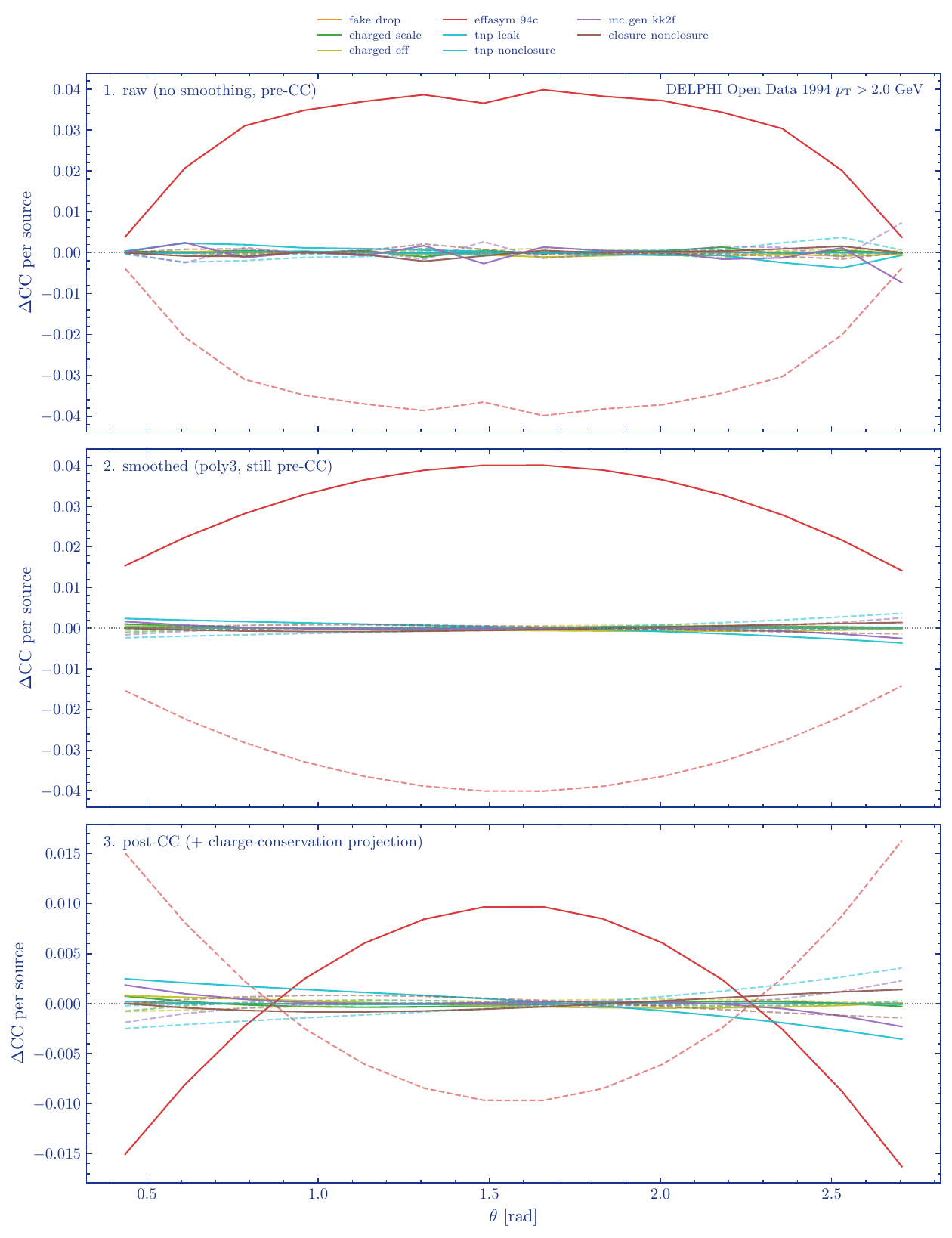}
    \caption{Per-source shifts of the corrected charge correlator for the systematic sources other than the charge-misreconstruction scale factor, for the 1994 dataset. The three panels show the raw shifts, their cubic parametrization, and the result of the charge-conservation constraint. The dominant curve is the per-charge track finding efficiency shift, which follows the total charged-particle density and is approximately symmetric under $\theta \to 180^\circ\!-\theta$. The constraint bounds its angular integral and strongly reduces it, leaving its residual shape component, while the remaining sources are small at every stage. The solid and dashed curves show the up and down variations of each source. In the legend, \texttt{fake\_drop} denotes the high-momentum fake-track removal, \texttt{charged\_scale} the track momentum scale, \texttt{charged\_eff} the track finding efficiency variation and \texttt{effasym\_94c} the charge-asymmetric efficiency scale factor for the $94\_\text{c}$ configuration, \texttt{tnp\_leak} the tag-and-probe sample contamination modeling from $1{+}3$-prong leakage, \texttt{tnp\_nonclosure} the tag-and-probe selection bias, \texttt{closure\_nonclosure} the tag-and-probe sample contamination bias from the bias-injection test, and \texttt{mc\_gen\_kk2f} the \textsc{PYTHIA}~8 versus \textsc{KK}2f generator-model difference.}
    \label{fig:other_syst_shapes}
\end{figure}

The constraint acts primarily on the normalization component of the systematic shifts. A coherent shift with a non-zero angular integral is directly bounded by the measured integrated charge, whereas a shape variation aligned with the parity-odd signal, which integrates to zero over the symmetric fiducial range, is left largely unchanged. A comparison of the leading systematic uncertainty, the charge misreconstruction asymmetry residual, before and after applying the constraint is shown in Figure~\ref{fig:sys_shapes}. For this source, the constraint reduces the shift to approximately $67\%$ of its unconstrained value in the bins where the $\sin 2\theta$ modulation is largest. The constraint also introduces anti-correlation structure between forward and backward angular bins of the systematic uncertainty. The corresponding correlation matrices before and after the constraint are shown in Figure~\ref{fig:sys_cov}. The remaining systematic sources are shown in Figure~\ref{fig:other_syst_shapes} at the three stages of the procedure, namely the raw shift vectors, their cubic parametrization, and the result of the charge-conservation constraint. The parametrization suppresses the bin-to-bin fluctuations of the shifts derived from differences of finite simulated samples. The largest of these sources is the per-charge track finding efficiency shift, which follows the total charged-particle density and is approximately symmetric under $\theta \to 180^\circ\!-\theta$. The constraint bounds its angular integral and reduces it strongly, leaving the residual shape component that remains the second largest source, while the smaller sources are essentially unchanged.

\paragraph{Summary of the size of systematic sources} The systematic uncertainty is dominated by the charge-misreconstruction asymmetry residual, which contributes roughly $30\%$ at $\theta = 55^\circ$ and $49\%$ at $\theta = 125^\circ$ in the 1994 dataset and is by far the leading source. The next-largest contribution is the charged-track finding efficiency, roughly $12\%$ and $18\%$, whose per-charge random-drop shift follows the total charged-particle density and is only partially removed by the charge-conservation constraint. It is followed by the tag-and-probe selection bias, roughly $7$ to $10\%$, and the tag-and-probe sample contamination bias, the residual non-closure of the bias-injection test, roughly $4$ to $5\%$. All remaining sources, namely the MC generator-model difference, the charge-misreconstruction dilution residual, and the tag-and-probe contamination modeling, are at or below the percent level in the 1994 dataset. Representative values at $\theta = 55^\circ$ and $\theta = 125^\circ$ are summarized in Table~\ref{tab:systematic_summary} for the 1994 dataset and in Table~\ref{tab:systematic_summary_95} of Appendix~\ref{app:1995_check} for 1995. Each per-source entry is obtained by imposing the charge-conservation constraint on that source alone, while the total imposes it on all sources simultaneously; the entries are therefore not additive. The total systematic uncertainty exceeds the statistical uncertainty in both years, by roughly a factor of two in the 1994 dataset. The leading systematic originates from the finite statistics of the $\tau$ control region, propagated to the corrected correlator through the covariance of the residual fit, so the analysis remains ultimately statistics-limited.
\textit{It is important to notice that a detector effect at the sub-percent level drives the precision of this measurement. The effect must itself be measured from data, and it is the statistical precision of this auxiliary measurement in the $\tau$ control region, rather than the statistics of the hadronic signal sample alone, that dominates the total uncertainty.}
The per-track charge-misreconstruction rate is of order a few per mille, and its residual data--MC asymmetry is statistically consistent with zero. Rather than being set to zero, the fitted residual, whose magnitude ranges from zero to below $2\%$ across the acceptance, is propagated in full together with its statistical covariance: amplified by the total track density $N_{\rm tot}$ (Equation~\ref{eq:full_correction}) relative to a signal that is itself only about $1\%$ of $N_{\rm tot}$, an $\mathcal{O}(10^{-3})$ uncertainty on the residual maps onto several tens of percent of the measured correlator.

\begin{table}[h]
\centering
\caption{Relative systematic uncertainties (\%) on the corrected one-point charge correlator at two representative $\theta$ bins ($p_{\mathrm{T}} > 2.0$ GeV, 1994). Each per-source entry is obtained by imposing the charge-conservation constraint on that source alone, whereas the total is obtained by imposing it on all sources simultaneously; the entries are therefore not additive.}
\label{tab:systematic_summary}
\begin{tabular}{lcc}
\hline\hline
Source & $\theta = 55^\circ$ & $\theta = 125^\circ$ \\
\hline
Charged-track finding efficiency & 12.20 & 18.24 \\
Charge misreconstruction asymmetry residual & 30.15 & 48.52 \\
Charge misreconstruction dilution residual & 0.18 & 0.27 \\
Tag-and-probe selection bias & 7.01 & 9.68 \\
Tag-and-probe sample contamination modeling & 0.24 & 0.22 \\
MC generator model (\textsc{PYTHIA}~8 vs \textsc{KK}2f) & 0.57 & 0.38 \\
Tag-and-probe sample contamination bias & 4.01 & 4.57 \\
\hline
Total systematic (post-constraint) & 31.85 & 49.08 \\
Statistical & 14.98 & 23.53 \\
\hline\hline
\end{tabular}
\end{table}

\paragraph{Combination of the systematic sources} Each source enters the systematic covariance of Equation~\ref{eq:Vsyst-bare} as a single coherent shift vector $\mathbf{d}_i$, fully correlated across $\theta$ bins, and the sources are taken to be mutually independent so that their contributions add in quadrature. For the combined 1994$+$1995 result, the modeling systematics common to both years, namely the \textsc{PYTHIA}-vs-\textsc{KK}2f generator-model difference and the non-closure of the bias-injection test, are propagated as fully correlated across years. The remaining sources are driven by year-specific detector conditions and control-region statistics, including the track finding efficiency, the momentum scale, the fake subtraction, and the data-driven charge-misreconstruction components. Because the DELPHI detector and reconstruction-software configurations differ between the two years, with the data processed under the separate $94\_\text{c}$ and $95\_\text{d}$ tags, these detector-related effects are genuinely independent from one year to the next. Following the recommendation of the DELPHI data-preservation team, they are evaluated independently for each year and treated as uncorrelated across years. The two years are combined with weights given by the number of selected events in each year. The shift vectors of the yearly correlated sources are combined with these weights before entering $V_{\rm syst}$, whereas the contributions of the yearly uncorrelated sources are added in quadrature with the same weights. Representative values at $\theta = 55^\circ$ and $\theta = 125^\circ$ are summarized in Table~\ref{tab:systematic_summary_9495}.

\begin{table}[h]
\centering
\caption{Relative systematic uncertainties (\%) on the corrected one-point charge correlator for the combined 1994$+$1995 data, quoted in the $\theta = 55^\circ$ and $\theta = 125^\circ$ bins. The quoted values are the square roots of the diagonal elements of the systematic covariance, expressed relative to the measured charge correlator in each bin. Each per-source entry is obtained by imposing the constraint on that source alone, whereas the total is obtained by imposing it on all sources simultaneously. The entries are therefore not additive. Their sum in quadrature does not reproduce the total. Sources contributing below the percent level are not listed individually but are included in the total. The charge misreconstruction asymmetry residual is the combination of 16 uncorrelated components, eight per data-taking year.}
\label{tab:systematic_summary_9495}
\begin{tabular}{lcc}
\hline\hline
Source & $\theta = 55^\circ$ & $\theta = 125^\circ$ \\
\hline
Charged-track finding efficiency & 10.34 & 12.99 \\
Charge misreconstruction asymmetry residual & 25.94 & 35.24 \\
Tag-and-probe selection bias & 5.67 & 6.58 \\
Tag-and-probe sample contamination bias & 4.55 & 4.24 \\
MC generator model (\textsc{PYTHIA}~8 vs \textsc{KK}2f) & 0.69 & 1.62 \\
\hline
Total systematic & 27.61 & 35.58 \\
Statistical & 13.85 & 18.15 \\
\hline\hline
\end{tabular}
\end{table}

\clearpage
\section{Results}
\label{sec:results}

The one-point charge correlator is measured with DELPHI Open Data collected at $\sqrt{s} = 91.2$~GeV, for tracks with $p_{\rm T} > 2.0$~GeV in the polar-angle range $20^\circ \le \theta \le 160^\circ$. It is corrected for detector effects with the simulation-based correction described in Section~\ref{sec:mc_correction} and the data-driven charge-misreconstruction correction of Section~\ref{sec:data_driven}, and for the shape change induced by the hadronic event selection described in Section~\ref{sec:data_samples}, derived from simulation. The 1994 and 1995 samples differ in detector configuration and are analyzed independently throughout, each with its own detector simulation and $\tau^+\tau^-$ control-region measurement.

\begin{figure}[ht!]
    \centering
    \includegraphics[width=0.55\textwidth]{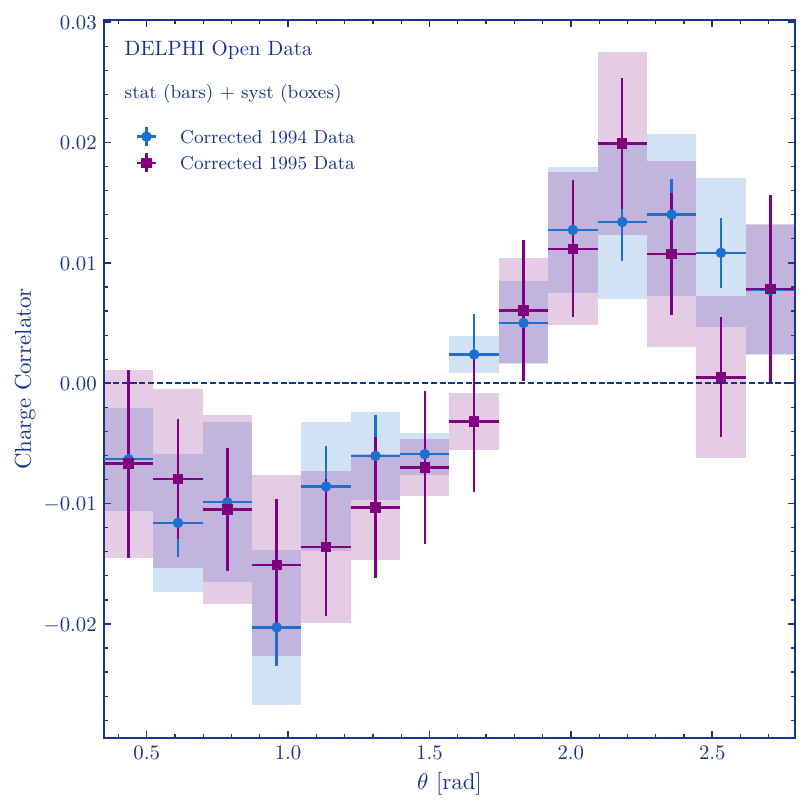}
    \caption{Fully corrected one-point charge correlator for tracks with $p_{\rm T} > 2.0$~GeV, comparing the DELPHI 1994 and 1995 samples. The two datasets are corrected independently, each with its own detector simulation and $\tau^+\tau^-$ control-region measurement. Data points are shown with statistical uncertainties (error bars) and systematic uncertainties (shaded boxes). The 1995 sample corresponds to 15~pb$^{-1}$ against 46~pb$^{-1}$ in 1994, and its statistical precision is correspondingly lower.}
    \label{fig:year_comparison}
\end{figure}

Figure~\ref{fig:year_comparison} compares the two corrected results. Both exhibit the characteristic $\sin(2\theta)$ modulation anticipated in the SM, with the same sign and comparable amplitude. The larger fluctuation of the 1995 sample reflects its smaller integrated luminosity, 15~pb$^{-1}$ against 46~pb$^{-1}$ in 1994. The two measurements are consistent within their uncorrelated uncertainties and are therefore combined. Because the two years are corrected independently, their agreement also tests the reproducibility of the data-driven correction across detector configurations and control samples.

\begin{figure}[ht!]
    \centering
    \includegraphics[width=0.55\textwidth]{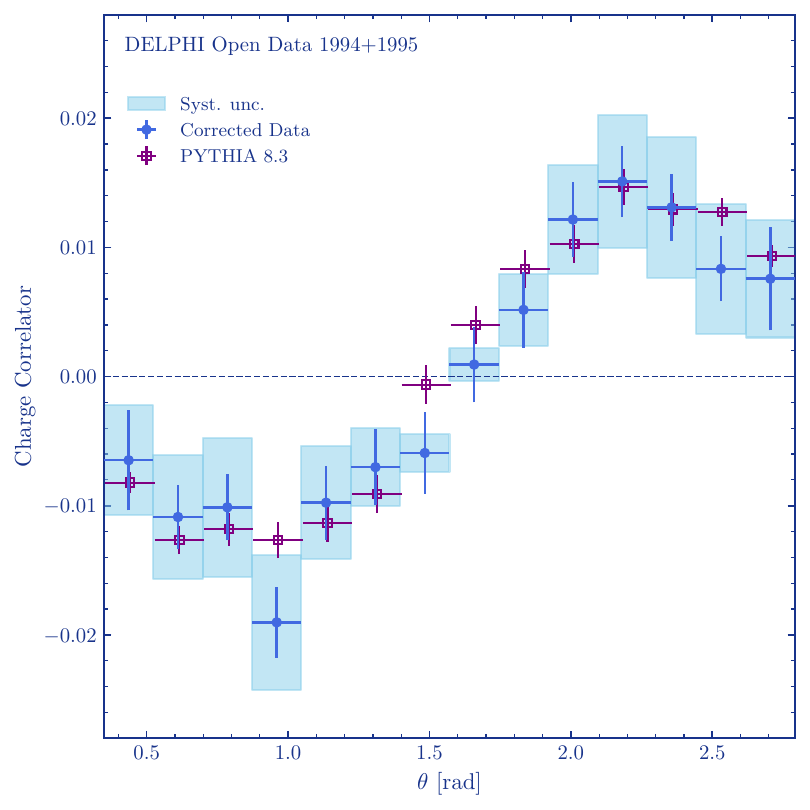}
    \caption{Fully corrected one-point charge correlator for tracks with $p_{\rm T} > 2.0$~GeV, combining the DELPHI 1994 and 1995 data, compared to the \textsc{PYTHIA}~8.3 generator-level prediction. Data points are shown with statistical uncertainties (error bars) and systematic uncertainties (shaded boxes). The data exhibit a clear $\sin(2\theta)$ modulation, consistent with the prediction within the evaluated uncertainties.}
    \label{fig:final_result_pythia}
\end{figure}

The combined result, weighted by the number of selected events in each year and with the between-year correlations described in Section~\ref{sec:syst_profile}, is shown in Figure~\ref{fig:final_result_pythia}, together with the \textsc{PYTHIA}~8.3 generator-level prediction evaluated with the same track momentum and polar-angle selections and including initial-state radiation. The one-point charge correlator is negative in the forward hemisphere and positive in the backward one, the sign expected from $\sum_q R_q Q_q A_{\rm FB}^q$, in which the down-type contributions dominate the sum, and it crosses zero at $\theta = 90^\circ$ as required by the symmetry. The measurement agrees with the prediction over the full angular range, within the evaluated uncertainties, and, for the first time, resolves the hadronic parity-violating charge flow as a function of polar angle. 
 
This measurement establishes the experimental viability of the observable rather than providing a precision test of the Standard Model. Precision extraction of $A_{\rm FB}$ is not yet possible, but the measurement motivates the experimental and theoretical developments that such an extraction would require.


\clearpage
\section{Summary and outlook}
\label{sec:summary}

We report the first measurement of the one-point charge correlator in $e^+e^-$ collisions at the $Z$ pole, using $61~\mathrm{pb}^{-1}$ of archival DELPHI Open Data recorded at $\sqrt{s} = 91.2$~GeV in 1994 and 1995. The observable is the normalized flux of electric charge as a function of the polar angle with respect to the $e^-$ beam direction. Unlike the jet-charge observables on which the previous hadronic asymmetry measurements were built, it involves no jets, no thrust hemispheres, and no event-by-event association of a reconstructed object with a primary-quark direction. The measurement is performed for tracks with $p_{\rm T} > 2$~GeV in $20^\circ \le \theta \le 160^\circ$.

The detector effects are corrected in two stages. First, a simulation-based correction is derived and applied separately for positive and negative charged tracks. Second, the residual data--simulation charge reconstruction differences are bounded by a tag-and-probe approach in the single-prong $e^+e^-\to Z\to\tau^+\tau^-$ sample, where charge conservation fixes the true charge of the probe track. The result, measured in bins of probe $p_{\rm T}$ and $\theta$, is transferred to the hadronic events. The parity-odd component of the residual, which is aligned with the angular structure of the signal, is statistically consistent with zero but is propagated in full. The data-driven correction therefore gives the dominant systematic uncertainty, whose size is set by the statistics of the control sample. The fully corrected one-point charge correlator exhibits the characteristic $\sin(2\theta)$ modulation and agrees with the \textsc{PYTHIA}~8.3 prediction.


While energy flux has been measured in $e^+e^-$ annihilation for decades, and the jet charge ever since Feynman and Field, the hadronic charge-flux had never been. This work demonstrates its measurement, and identifies what a precise measurement will take. As more archival $e^+e^-$ datasets become available and are being analyzed~\cite{Electron-PositronAlliance:2019cpi, Chen:2021uws, Chen:2023njr, Electron-PositronAlliance:2025fhk, Electron-PositronAlliance:2025hze, Zhang:2025delphiEEC, Defranchis:2026wyw, Cheng:2026zvl, Fan:2026ckf}, this work opens a broad program of measurements of charge flux observables, from energy-weighted charge correlators, to correlators of flavor currents~\cite{Lee:2026bcq}, to the two-point charge correlator, and to fluxes of other conserved quantum numbers such as baryon number and strangeness, which have since been proposed as probes of hadronization~\cite{Ke:2026zia}. 

\section*{Acknowledgements}

This work would not have been possible without the decades of effort by the DELPHI collaboration in designing, building, and operating the detector, nor without the foresight of its data preservation team.
The authors are profoundly grateful for the DELPHI collaboration's monumental effort in making these pristine datasets and software infrastructure publicly available, and would especially like to thank Dietrich Liko and Ulrich Schwickerath for their guidance on the DELPHI Open Data.
We thank Anthony Badea, whose comments made a significant contribution to this documentation. J.Z. is deeply grateful to Sang Hyun Ko and Sitian Qian for their substantial work on the software stack for DELPHI Open Data analysis, on which this measurement partly relies. We thank KITP Santa Barbara for its hospitality while this project was initiated.

\clearpage
\bibliographystyle{JHEP}
\bibliography{main} 
\clearpage

\begin{appendices}

\clearpage
\section{Data-driven correction from hadronic events}
\label{sec:hadronic_correction}

The main result of this note relies on the $Z \to \tau^+\tau^-$ tag-and-probe measure of the charge misreconstruction rate. It constrains the component of the charge bias that is antisymmetric under $\theta \to 180^\circ\!-\theta$, and therefore aligns with the parity-odd signal, with a pure external charge tag. This appendix explores complementary strategies in which the hadronic sample corrects itself, with no input from a control region. By symmetry, no such hadronic self-correction can constrain this antisymmetric bias component, so the methods below serve as cross-checks of the main result rather than replacements for it. The two methods capture different components of the bias. The global-imbalance method of Appendix~\ref{app:global_imbalance} captures only the $\theta$-averaged, parity-even part of the charge bias, with no $\theta$ dependence. The parity projection of Appendix~\ref{app:symmetry_projection} captures the parity-even part exactly, including its full $\theta$ dependence, but neither method captures the parity-odd part. A parity-odd bias consists of opposite net charge migrations at mirror angles. A net positive-to-negative migration at $\theta = 30^\circ$, accompanied by a net negative-to-positive migration at $\theta = 150^\circ$, lowers the measured correlator in one hemisphere and raises it in the other. A bias of this form has the same angular structure as the parity-odd signal, and no self-correction alone can separate the two.

The previous measurements controlled this parity-odd detector bias with dedicated detector-level calibration. The early DELPHI, OPAL, and L3 measurements~\cite{DELPHI:1991mqi, OPAL:1992jsm, L3:1998jet} relied on extensive verification with the detector simulation. ALEPH supplemented the detector simulation with external calibrations, deriving a sagitta calibration of the high-momentum tracking from the momentum balance of $Z\to\mu^+\mu^-$ events and a calibration of the detector material from photon conversions~\cite{ALEPH:1991fba, ALEPH:1996qlh}. A later OPAL measurement on $b$-enriched samples likewise measured the forward--backward asymmetry of the detector material with photon conversions, finding a significant asymmetry only near the edge of the acceptance~\cite{OPAL:2002ddm}. At SLD, the charge-dependent forward--backward sagitta bias was studied with dimuon and Bhabha events and constituted the largest systematic uncertainty of the measurement~\cite{SLD:1996gjt}. All of these approaches rest on extensive detector-level calibration and verification, and their counterparts for the DELPHI Open Data are natural next steps toward a precision measurement. In this analysis, the residual charge bias is instead bounded with the single tag-and-probe measurement of Section~\ref{sec:data_driven}, which retains both parity components and therefore addresses the parity-even and parity-odd parts of the detector bias at once. The self-corrections below are nevertheless statistically powerful, and we expect them to become useful ingredients of a future precision $A_{\rm FB}$ measurement. This holds in particular for the second method, which transfers the hemisphere charge-flow moments of the LEP inclusive jet-charge $A_{\rm FB}$ measurements~\cite{DELPHI:1991mqi, OPAL:1992jsm, ALEPH:1996qlh, L3:1998jet} to a differential, bin-by-bin setting.

\subsection{Self-correction exploiting the global charge imbalance}
\label{app:global_imbalance}

The first method directly exploits the global charge imbalance that remains after the MC-based correction of Section~\ref{sec:mc_correction}. Rather than correcting track-by-track, it uses the aggregate imbalance to correct the charge correlator directly.
Two limiting assumptions bracket the possible origin of the charge imbalance, leading to two distinct correction methods:

\noindent\textbf{Shift (additive) correction}
All of the observed charge imbalance is assumed to originate from fake tracks. Under this assumption, the imbalance represents a spurious additive contribution to the charge sums, and the correction is applied as a direct subtraction (shift) from the raw charge correlator in each $\theta$ bin.

\noindent\textbf{Multiplicative correction}
All of the observed charge imbalance is assumed to arise from a mismodelling of the charge-dependent tracking efficiency. In this scenario, the excess (or deficit) of charges is redistributed equally, half assigned to the positive-charge count and half to the negative-charge count, yielding a multiplicative rescaling of the correlator. Conceptually, this is analogous to the zeroth-order polynomial fit variant used in the single-track charge misreconstruction correction of Section~\ref{sec:data_driven}, where a flat residual effectively acts as a global rescaling of the charge balance.

Both corrections are applied independently in the track-$p_T$
bins, $2<p_T<4$, $4<p_T<6$, $6<p_T<8$, and $>8$~GeV. For each bin, the
weights or shifts are derived from that bin's own integrated charge
imbalance, and the corrected $\theta$ distributions in each bin are summed back to get the final result. This captures the $p_T$-dependent
structure in the charge imbalance, but the $\theta$ dependence is not included.

\begin{figure}[htbp]
    \centering
    \includegraphics[width=0.55\textwidth]{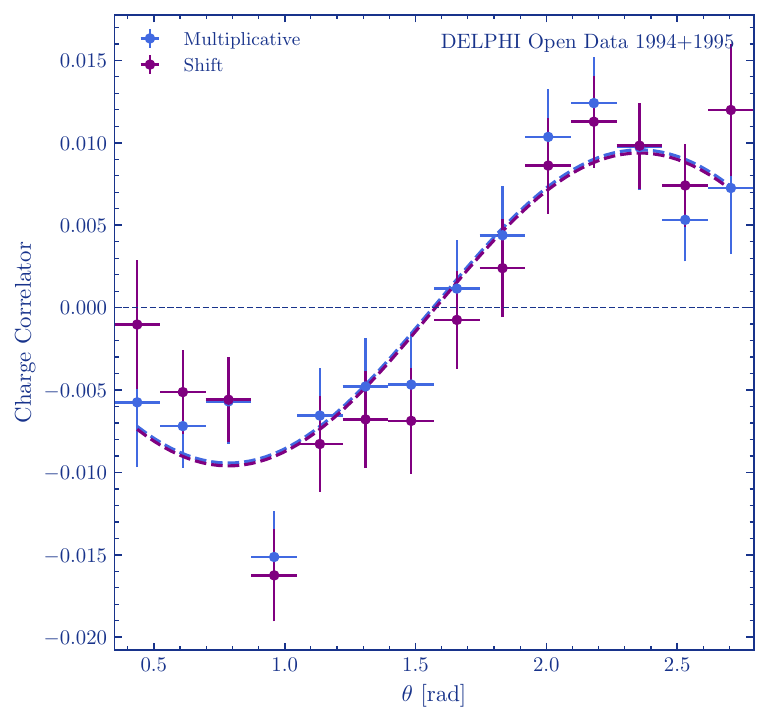}
    \caption{Corrected charge correlator as a function of $\theta$ in the analysis fiducial phase space of $p_{\rm T} > 2$~GeV, using the shift (additive) and multiplicative correction methods applied to DELPHI Open Data (1994). Dashed curves show $A\sin(2\theta) + C$ fits.}
    \label{fig:bruteforce}
\end{figure}

The corrected charge correlator under each method is shown in Figure~\ref{fig:bruteforce}. The dashed curves correspond to $A\sin(2\theta) + C$ fits, confirming the expected $\sin(2\theta)$ modulation. Both correction methods produce consistent results within the statistical uncertainties, confirming that the two limiting assumptions yield compatible $\sin(2\theta)$ amplitudes. The residual difference between the shift and multiplicative corrections is taken as a systematic uncertainty for this approach.

The generator-model dependence is estimated using \textsc{KK}2f as an alternative MC
sample, run through the same correction procedure as \textsc{PYTHIA}~8. The
per-$\theta$ difference between the two MC-corrected results is fit with
a first-order polynomial in $\theta$ to suppress bin-by-bin statistical
fluctuations and extract the coherent generator-level component. The smoothed
envelope is taken as a systematic uncertainty on this approach.

\begin{figure}[htbp]
    \centering
    \includegraphics[width=0.55\textwidth]{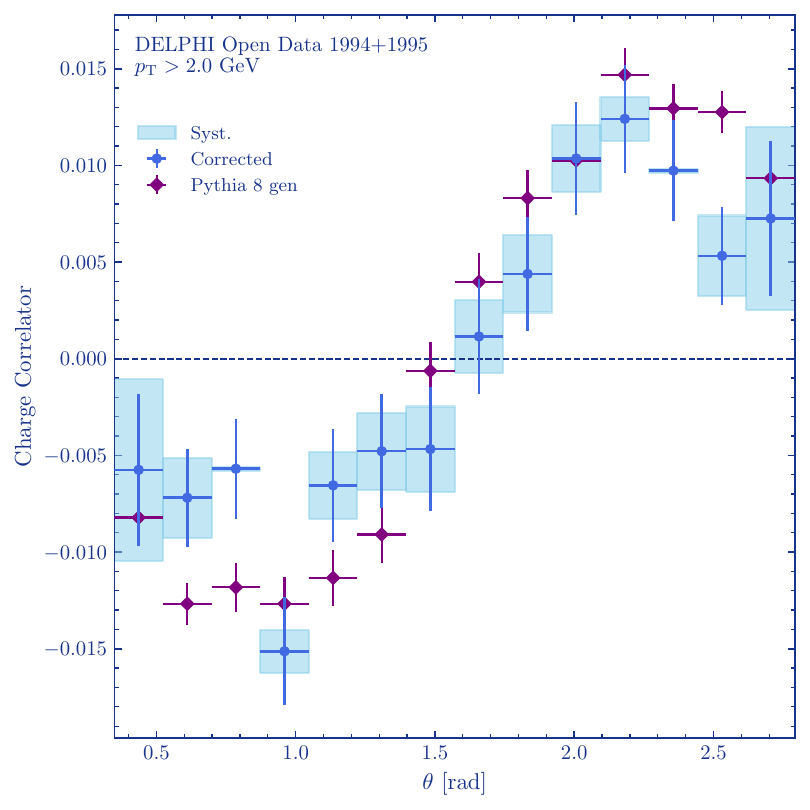}%
    \caption{Final corrected charge correlator (1994$+$1995
    combined), compared to \textsc{PYTHIA}~8 generator level. Error bars are statistical uncertainties. 
    Outer boxes include the method-choice (shift vs multiplicative) and
    \textsc{PYTHIA}~8-vs-\textsc{KK}2f systematic uncertainties combined in quadrature.}
    \label{fig:bruteforce_final}
\end{figure}

The combined 1994$+$1995 result from this approach is shown in
Figure~\ref{fig:bruteforce_final}, compared to the \textsc{PYTHIA}~8
generator-level prediction. A
clear charge asymmetry is observed. The agreement, however, is worse than that obtained from the TNP-based approach in
Section~\ref{sec:data_driven}, consistent with the bias components that this approach does not
capture, namely the $\theta$ dependence of the parity-even part and the
parity-odd part, which is left uncorrected.

\subsection{Self-correction exploiting CP symmetry}
\label{app:symmetry_projection}

The second method exploits how the bias transforms under the reflection $\theta \to 180^\circ\!-\theta$. CP invariance of inclusive hadronic $Z$ decays requires the density of positive hadrons at $\theta$ to equal the density of negative hadrons at $180^\circ\!-\theta$, so on average $\langle N^+(\theta)\rangle = \langle N^-(180^\circ\!-\theta)\rangle$ and the true correlator is exactly odd, $\mathcal{Q}_{\rm true}(\theta) = -\mathcal{Q}_{\rm true}(180^\circ\!-\theta)$. This is a stronger physics assumption than the global charge conservation used in Appendix~\ref{app:global_imbalance} and in the main analysis. Charge conservation fixes only the angular integral of the true correlator, whereas CP invariance fixes its reflection property point by point in $\theta$. The assumption is well justified, since CP violation in inclusive hadronic $Z$ decays is negligible at the present precision. The approach follows the strategy of the LEP and SLD measurements of the inclusive hadronic charge asymmetry~\cite{ALEPH:1991fba, ALEPH:1996qlh, DELPHI:1991mqi, OPAL:1992jsm, L3:1991gfs, L3:1998jet, SLD:1996gjt}, in which the observable is the difference of the forward and backward hemisphere charges $\langle Q_{\rm F} - Q_{\rm B}\rangle$, where $Q_{\rm F}$ and $Q_{\rm B}$ are the momentum-weighted sums of the particle charges in the forward and backward halves of the event, so that the forward--backward-symmetric part of the detector response cancels in the difference while the sum $\langle Q_{\rm F} + Q_{\rm B}\rangle$ monitors its size. Unlike the hemisphere charges, whose hemispheres are defined by the event thrust or sphericity axis, the folding employed here acts on the polar angle of each track with respect to the $e^-$ beam direction, and no event axis enters. The main analysis nevertheless measures the full $\theta$ spectrum of the correlator, rather than only its parity-odd projection, and therefore does not employ the folding.

In place of the data-driven charge-misreconstruction correction of Section~\ref{sec:data_driven}, the MC-corrected distribution $\mathcal{Q}_{\rm eff}(\theta)$, the output of Section~\ref{sec:mc_correction} identical to the main analysis, is folded about $\theta = 90^\circ$:
\begin{equation}\label{eq:folding_correction}
    \mathcal{Q}_{\rm corr}(\theta) = \tfrac{1}{2}\left[\mathcal{Q}_{\rm eff}(\theta) - \mathcal{Q}_{\rm eff}(180^\circ\!-\theta)\right],
    \qquad 90^\circ < \theta \le 160^\circ.
\end{equation}
The folding is the correction step. Writing $\mathcal{Q}_{\rm eff}(\theta) = \mathcal{Q}_{\rm true}(\theta) + b(\theta)$, where $b$ collects the residual data--MC charge bias from any mechanism (charge misreconstruction, charge-dependent tracking efficiency, charge-asymmetric fake rates), the exact oddness of $\mathcal{Q}_{\rm true}$ implies
\begin{equation}\label{eq:folding_result}
    \mathcal{Q}_{\rm corr}(\theta) = \mathcal{Q}_{\rm true}(\theta) + b_{\rm O}(\theta),
    \qquad b_{\rm O}(\theta) \equiv \tfrac{1}{2}\left[b(\theta) - b(180^\circ\!-\theta)\right].
\end{equation}
The parity-even component of the bias, which includes everything targeted by the shift and multiplicative corrections of Appendix~\ref{app:global_imbalance}, cancels identically in Equation~\ref{eq:folding_correction}, independently of its mechanism and differentially in $\theta$, with no input from a control region. The parity-odd component $b_{\rm O}$, which transforms identically to the signal, is not constrained by this method. Constraining it requires an external charge reference, such as the tag-and-probe measurement of the main analysis or the external calibrations used by the previous measurements.

\begin{figure}[htbp]
    \centering
    \includegraphics[width=0.55\textwidth]{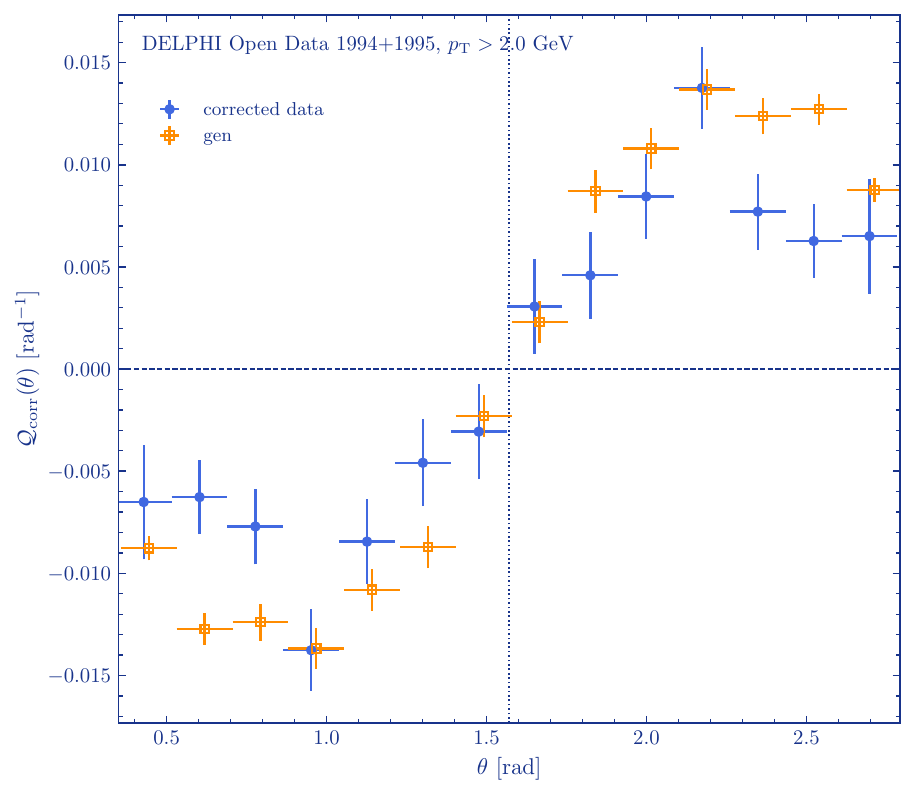}
    \caption{One-point charge correlator obtained with the self-correction exploiting CP symmetry, Equation~\ref{eq:folding_correction}, for the $p_{\rm T} > 2$~GeV selection, combining the 1994 and 1995 DELPHI Open Data samples. The MC-corrected distribution of Section~\ref{sec:mc_correction} is folded about $\theta = 90^\circ$, which removes the parity-even component of the charge bias; the parity-odd component is left uncorrected, and no data-driven charge-misreconstruction correction is applied. The folded \textsc{PYTHIA}~8.3 generator-level prediction is overlaid, with the shape change induced by the hadronic event selection included as a correction to the data (Figure~\ref{fig:evtsel_gating}), as in Section~\ref{sec:results}. The distribution is displayed over the full angular range by antisymmetric mirroring; bins mirrored about $90^\circ$ are fully anticorrelated by construction and carry no independent information. Error bars are statistical.}
    \label{fig:folded_crosscheck}
\end{figure}

The combined 1994$+$1995 result from this approach is shown in Figure~\ref{fig:folded_crosscheck}, compared to the folded \textsc{PYTHIA}~8 generator-level prediction. As in the comparisons of Section~\ref{sec:results}, the shape change induced by the hadronic event selection is included as a correction derived from the generator-level comparison of Figure~\ref{fig:evtsel_gating}, so the folded data are compared to the inclusive generator-level prediction. A clear charge asymmetry is observed. The folding yields the distribution on the half-range $90^\circ < \theta \le 160^\circ$. Since the folded correlator carries the full physics content of the observable, it is displayed over the full angular range by antisymmetric mirroring, with mirrored bins fully anticorrelated by construction. The folding itself is an exact, parameter-free projection. It introduces no additional systematic uncertainty and no control-sample statistical uncertainty, in contrast to the scale-factor uncertainties of the main analysis, which are dominated by the statistical precision of the $\tau$ control sample.

Later LEP measurements introduced one further moment, the hemisphere product $\langle Q_{\rm F}\,Q_{\rm B}\rangle$~\cite{ALEPH:1996qlh}, to determine from the data themselves the charge separation $\delta_q = \langle Q_q\rangle - \langle Q_{\bar{q}}\rangle$, the average jet-charge difference between the quark and antiquark hemispheres, which converts the measured charge flow $\langle Q_{\rm F} - Q_{\rm B}\rangle$ into $A_{\rm FB}$. Its analog here, the per-event product of net charges in mirrored bins, is a slice of the two-point charge correlator. Charge--charge correlations are, however, a sensitive probe of QCD fragmentation dynamics in their own right. It would be interesting both to measure the two-point charge correlator itself with the present dataset and to employ it, as at LEP, as an in-situ calibration of the fragmentation charge dynamics in a future $A_{\rm FB}$ extraction.

\clearpage
\section{Additional data vs MC comparisons}
\label{app:additional_datamc}

\begin{figure}[ht!]
    \centering
    \includegraphics[width=0.49\textwidth]{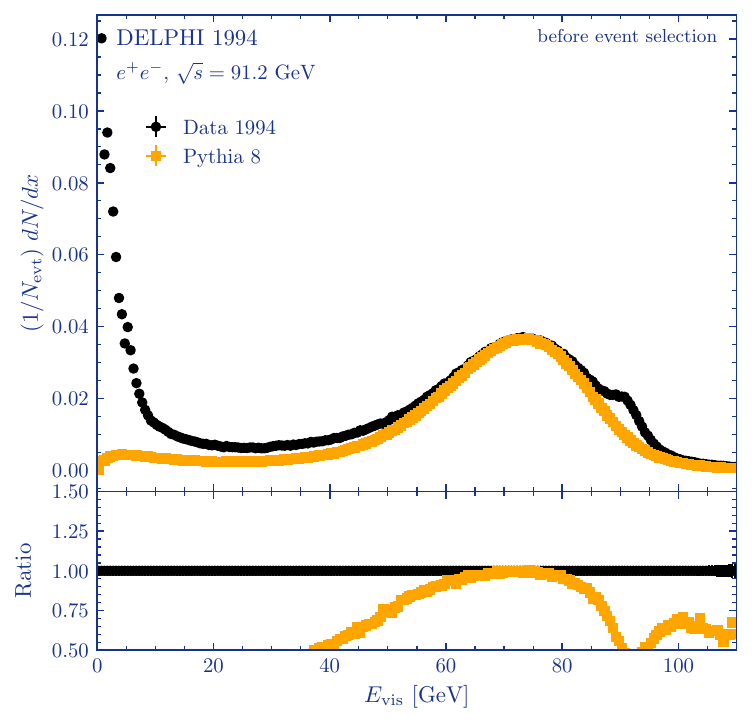}
    \includegraphics[width=0.49\textwidth]{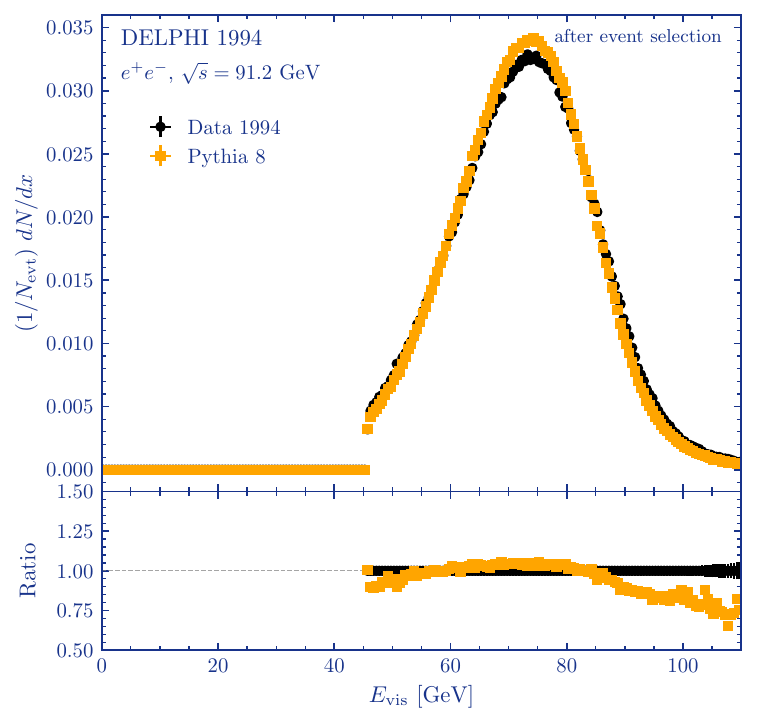}
    \caption{Total visible energy $E_{\rm vis}$ in the 1994 DELPHI Open Data (black) compared to the reconstructed hadronic \textsc{PYTHIA}~8.3 simulation (orange), normalized per event, before (left) and after (right) the hadronic event selection of Table~\ref{tab:SelectionSummary}. The lower panels show the ratio to the data. Before the selection, the data show a large low-$E_{\rm vis}$ population from two-photon collisions and leptonic $Z$ decays, and a full-energy shoulder from $e^+e^-$ and $\mu^+\mu^-$ final states, none of which are modeled by the hadronic simulation. The $E_{\rm vis} \ge 0.5\,E_{\rm cm}$ requirement, visible as the sharp boundary at $45.6$~GeV, together with the track-multiplicity and thrust-axis requirements, removes these populations; after the selection the data agree well with the hadronic simulation.}
    \label{fig:app_evis}
\end{figure}

Figure~\ref{fig:app_evis} shows the total visible energy $E_{\rm vis}$ in the 1994 data compared to the reconstructed hadronic \textsc{PYTHIA}~8.3 simulation, before and after the hadronic event selection of Table~\ref{tab:SelectionSummary}. Before the selection, the data exhibit two populations that are absent from the hadronic simulation: a large excess at low $E_{\rm vis}$, dominated by two-photon collisions and leptonic $Z$ decays with undetected neutrinos (in particular $\tau^+\tau^-$), and a full-energy shoulder near $E_{\rm vis} \approx \sqrt{s}$ from $e^+e^-$ and $\mu^+\mu^-$ final states. The requirement $E_{\rm vis} \ge 0.5\,E_{\rm cm}$, visible as the sharp boundary at $45.6$~GeV in the right panel, removes these backgrounds together with the tail of radiative-return events in which a hard initial-state photon dramatically lowers the effective annihilation energy. 
This protects the physics interpretation of the measurement: at $\sqrt{s'}$ away from the $Z$ pole, the charge asymmetry acquires contributions unrelated to the pole parity-violating couplings, most notably from the $\gamma$--$Z$ interference term, which vanishes at $\sqrt{s'} = M_Z$, and from QED radiative effects that are unsuppressed off resonance. Hard initial-state radiation is in any case strongly suppressed on the resonance, since radiating a hard photon moves the annihilation off the peak of the steeply falling cross section; the visible-energy requirement removes the residual tail. 
After the full event selection, the data agree well with the hadronic simulation across the accepted $E_{\rm vis}$ range, confirming the high purity of the selected hadronic sample.

\begin{figure}[ht!]
    \centering
    \includegraphics[width=0.49\textwidth]{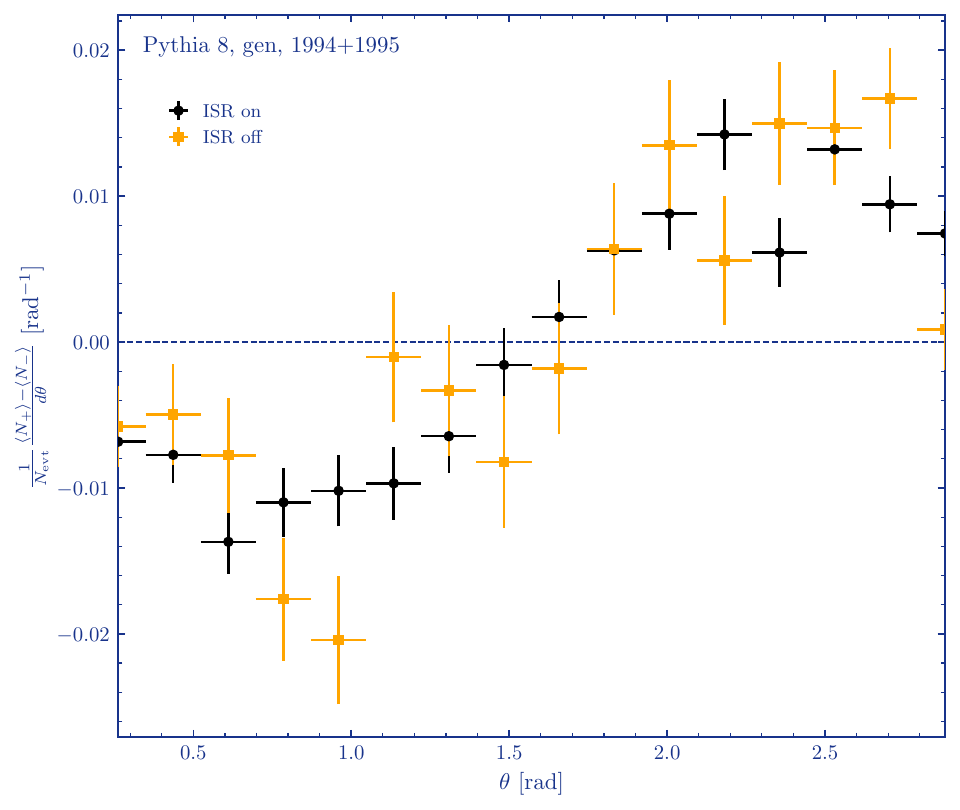}
    \includegraphics[width=0.49\textwidth]{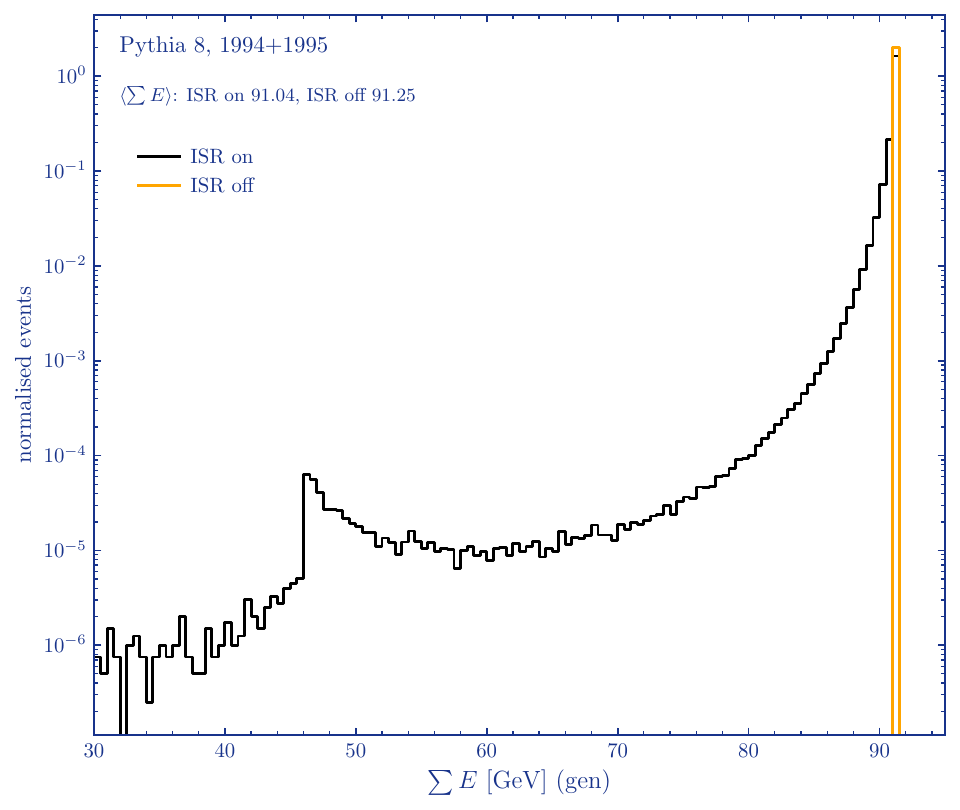}
    \caption{Left: generator-level one-point charge correlator from \textsc{PYTHIA}~8.3 with QED initial-state radiation enabled (black) and disabled (orange), combining the 1994 and 1995 configurations, with no requirement imposed on the radiated ISR energy. The two are consistent within the statistical precision, indicating a negligible ISR effect at the current precision of the measurement. Right: sum of the generator-level final-state particle energies, $\sum E$, excluding the ISR photons. Enabling ISR produces a radiative tail toward lower $\sum E$, but shifts the mean by only $\approx 0.2$~GeV (91.04~GeV versus 91.25~GeV).}
    \label{fig:app_isr}
\end{figure}

The residual effect of QED initial-state radiation on the observable itself is quantified directly at the generator level. Figure~\ref{fig:app_isr} compares the \textsc{PYTHIA}~8.3 generator-level one-point charge correlator with ISR enabled and disabled, for the combined 1994+1995 configuration. The two distributions are consistent within the statistical precision of the samples, implying that ISR effects on the charge correlator are negligible at the current level of precision.

\begin{figure}[ht!]
    \centering
    \includegraphics[width=0.75\textwidth]{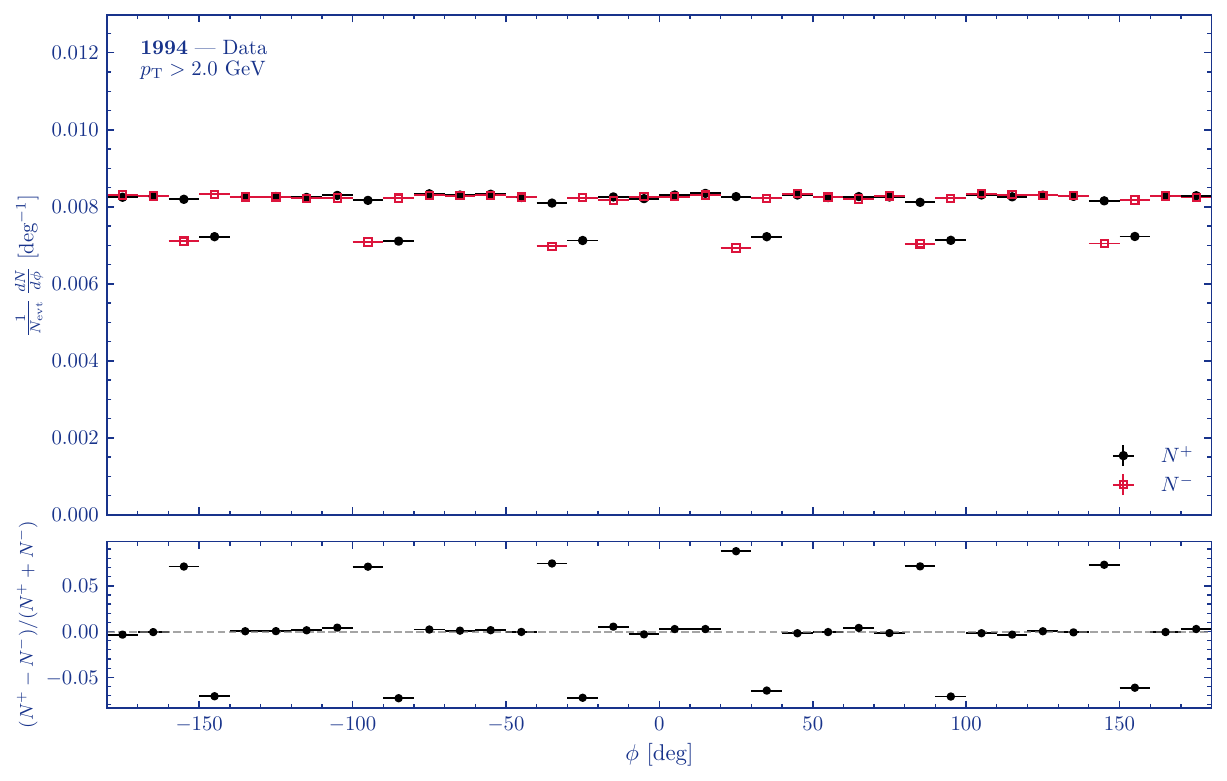}\\[0.5em]
    \includegraphics[width=0.75\textwidth]{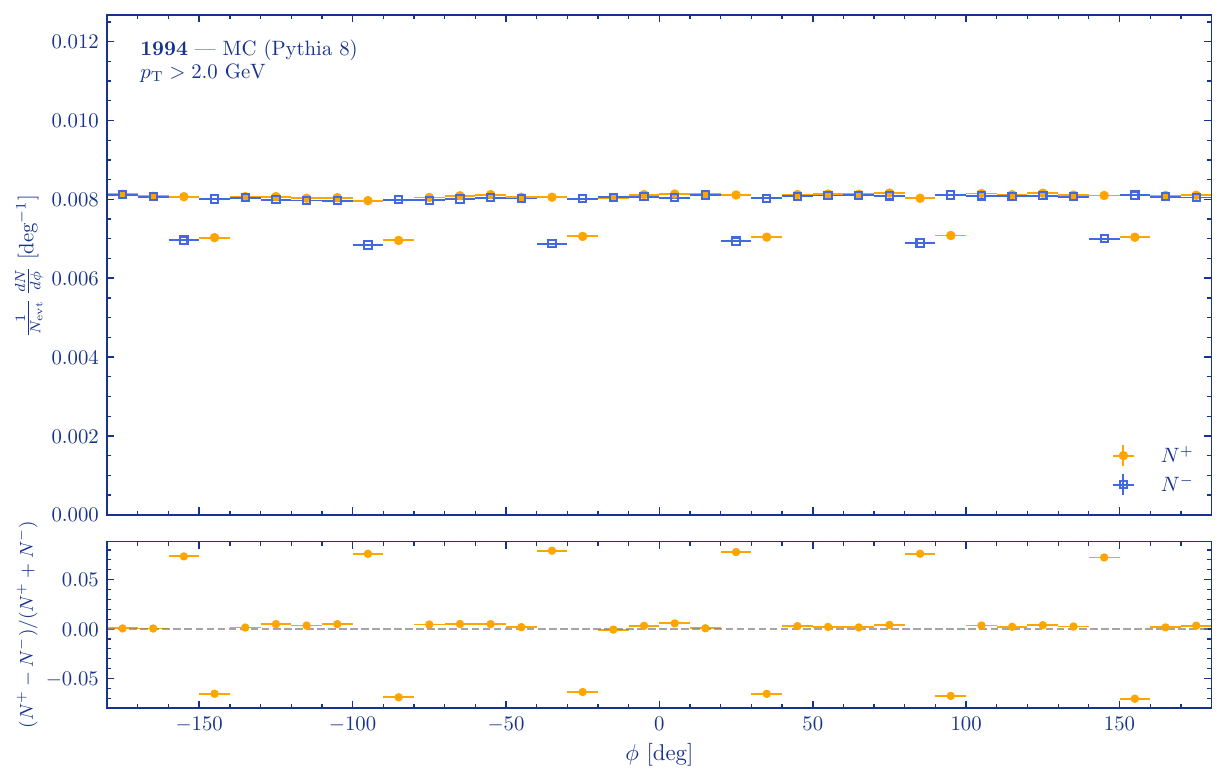}
    \caption{Azimuthal angle ($\phi$) distribution of charged tracks in 1994 data (top) and reconstructed \textsc{PYTHIA}~8.3 MC (bottom), separated by reconstructed charge $q > 0$ (red) and $q < 0$ (blue). The lower panels show the ratio of positive to negative tracks.}
    \label{fig:app_trackphi_pos_vs_neg}
\end{figure}

The azimuthal-angle distribution of Figure~\ref{fig:app_trackphi_pos_vs_neg} shows the same charge-asymmetry trend already seen in the inclusive data--MC comparisons of Section~\ref{sec:data_samples} (Figures~\ref{fig:data_mc_fbcc} and~\ref{fig:data_mc_theta}): the positive-to-negative ratio in data is consistent with unity across $\phi$ within statistical fluctuations, while the reconstructed \textsc{PYTHIA}~8.3 MC carries a systematic excess of positive over negative tracks.

\begin{figure}[ht!]
    \centering
    \includegraphics[width=0.45\textwidth]{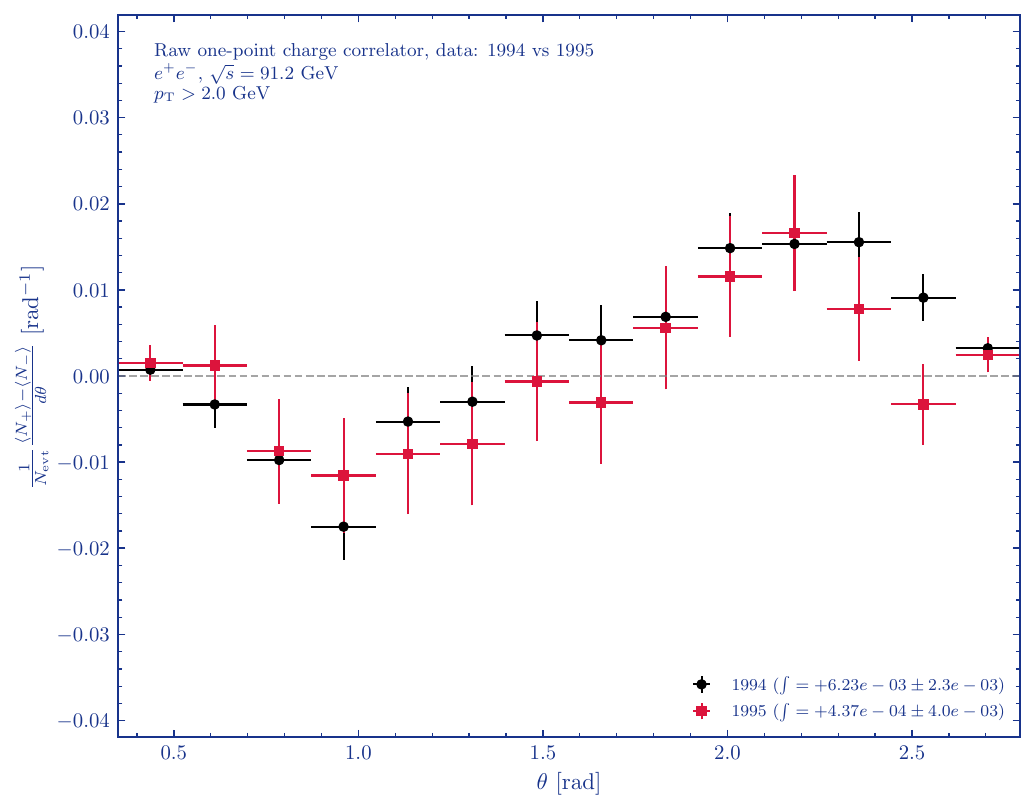}
    \includegraphics[width=0.45\textwidth]{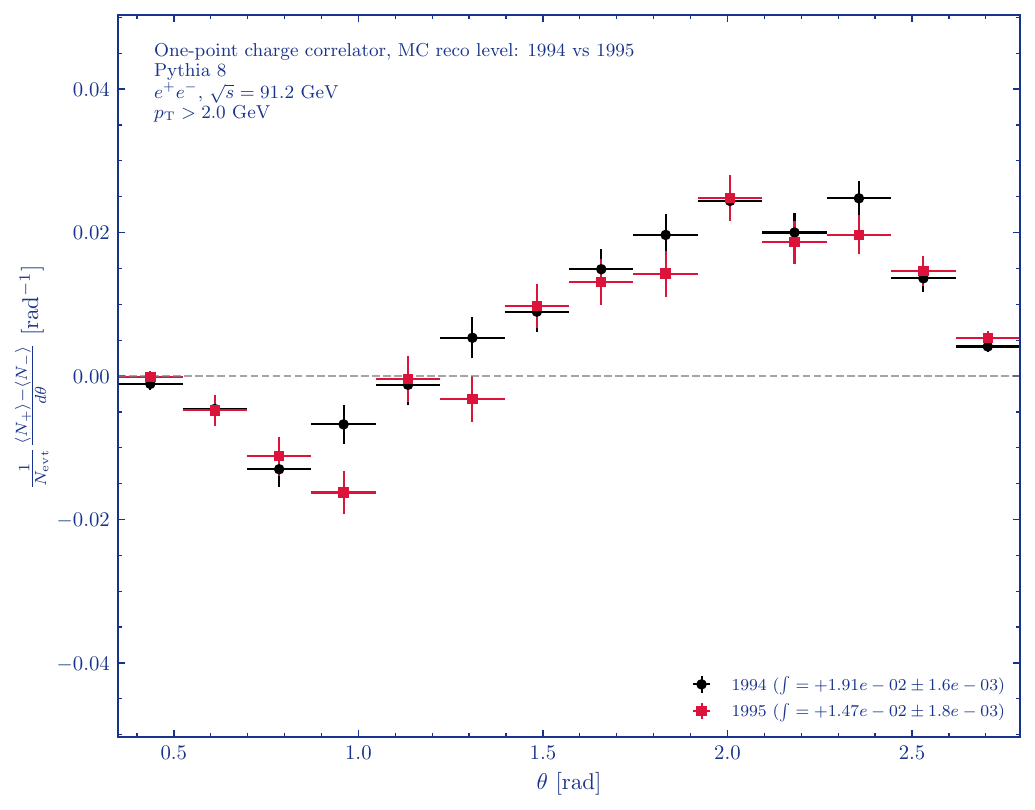}
    \caption{Comparison of the raw one-point charge correlator across different data-taking periods, 1994 and 1995, for the DELPHI Open Data sample (left) and the corresponding reconstructed \textsc{PYTHIA}~8.3 MC (right).}
    \label{fig:app_data_period_compare}
\end{figure}

The data-taking-period comparison of Figure~\ref{fig:app_data_period_compare} shows that the raw charge correlator measured in the 1994 and 1995 sub-periods of the DELPHI Open Data sample is mutually consistent within statistical uncertainties, both in data (left) and in the corresponding reconstructed \textsc{PYTHIA}~8.3 MC (right). This stability across running periods supports the combined treatment of the two years used to obtain the final result.

\begin{figure}[ht!]
    \centering
    \includegraphics[width=0.45\textwidth]{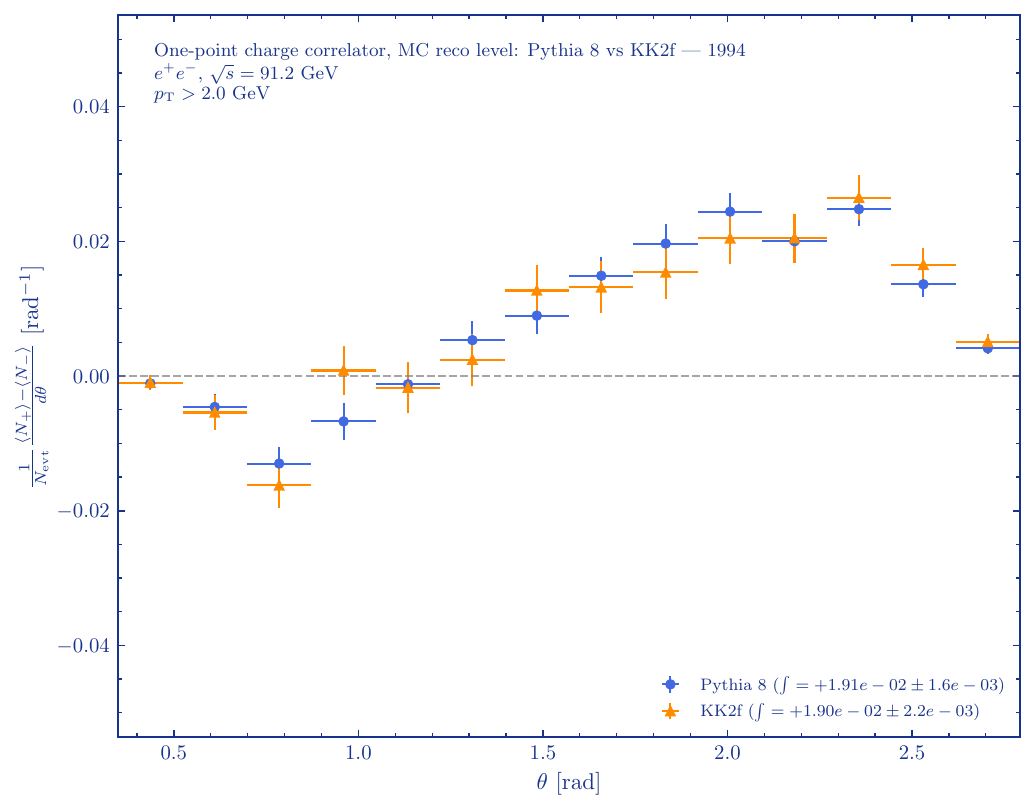}
    \includegraphics[width=0.45\textwidth]{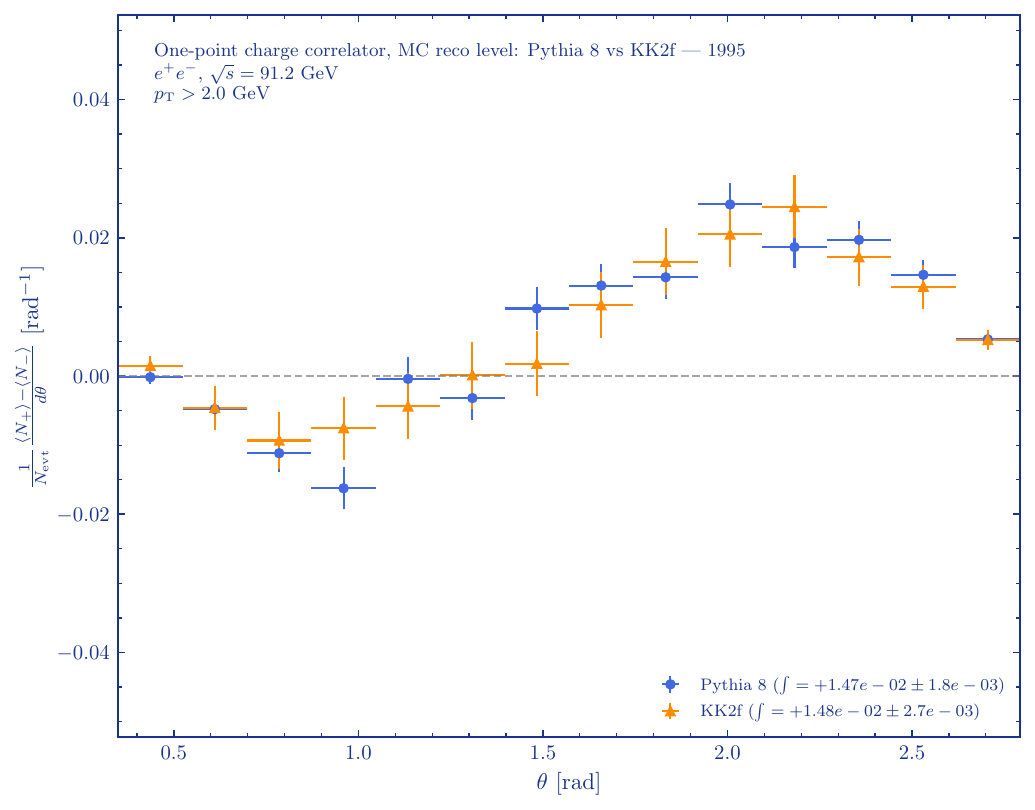}
    \caption{Comparison of the reconstructed-level one-point charge correlator between the self-generated \textsc{PYTHIA}~8.3 sample and the legacy \textsc{KK}2f+\textsc{PYTHIA}~6 sample, for the 1994 (left) and 1995 (right) datasets.}
    \label{fig:app_reco_pythia_vs_kk2f}
\end{figure}

Figure~\ref{fig:app_reco_pythia_vs_kk2f} shows the comparison of the reconstructed-level one-point charge correlator between the self-generated \textsc{PYTHIA}~8.3 sample and the legacy \textsc{KK}2f+\textsc{PYTHIA}~6 sample. The agreement validates that the self-generated sample reproduces the legacy \textsc{KK}2f+\textsc{PYTHIA}~6 sample.

\begin{figure}[ht!]
    \centering
    \includegraphics[width=0.49\textwidth]{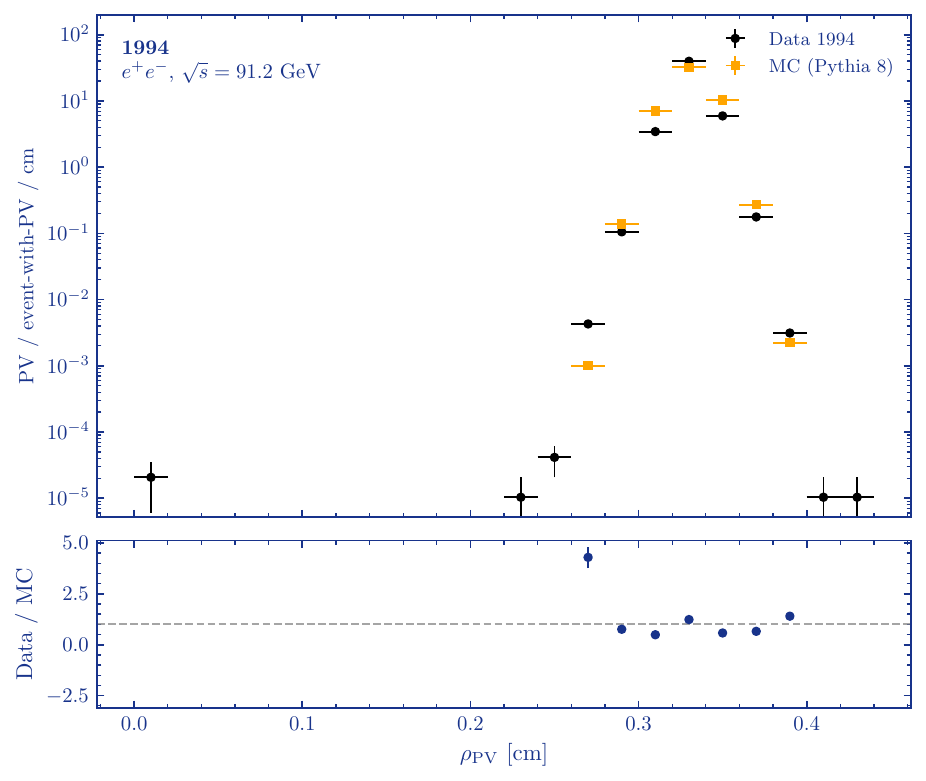}
    \includegraphics[width=0.49\textwidth]{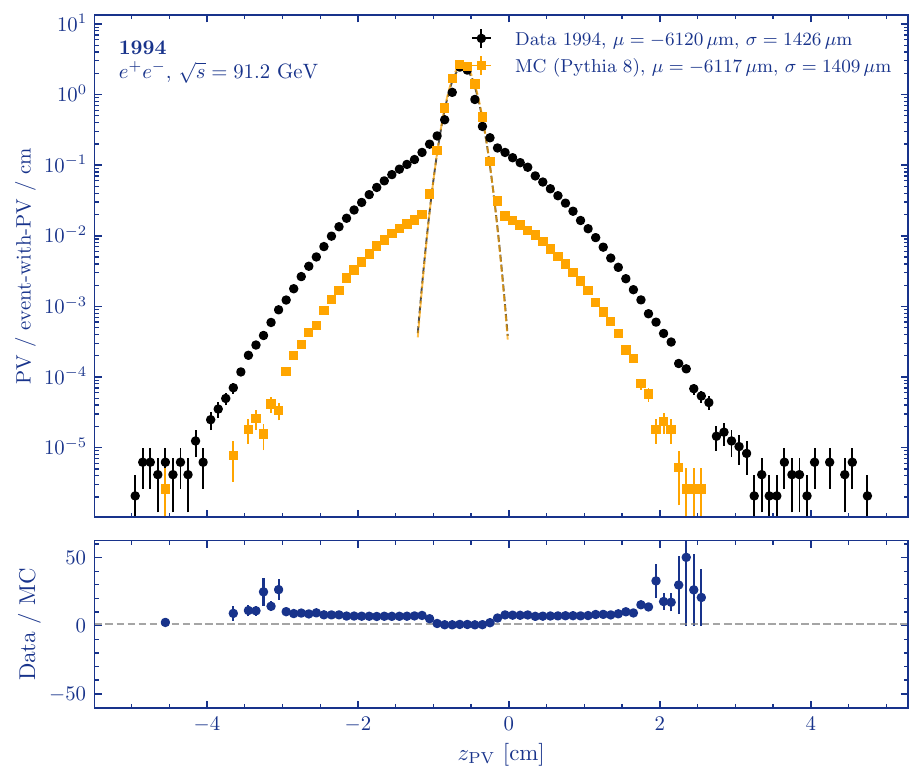}
    \caption{Reconstructed primary-vertex position in the 1994 DELPHI Open Data sample compared to reconstructed \textsc{PYTHIA}~8.3 MC: the radial coordinate $\rho_{\rm PV}$ (left) and the longitudinal coordinate $z_{\rm PV}$ (right), each normalized per event with a reconstructed primary vertex. The lower panels show the data/MC ratio.}
    \label{fig:app_pv}
\end{figure}

The reconstructed primary-vertex position is shown in Figure~\ref{fig:app_pv} for the radial ($\rho_{\rm PV}$) and longitudinal ($z_{\rm PV}$) coordinates. The bulk of the distributions is well described by the simulation. The fitted longitudinal width of the luminous region agrees with data at the percent level ($\sigma = 1426~\mu$m in data versus $1409~\mu$m in MC).

\begin{figure}[ht!]
    \centering
    \includegraphics[width=0.95\textwidth]{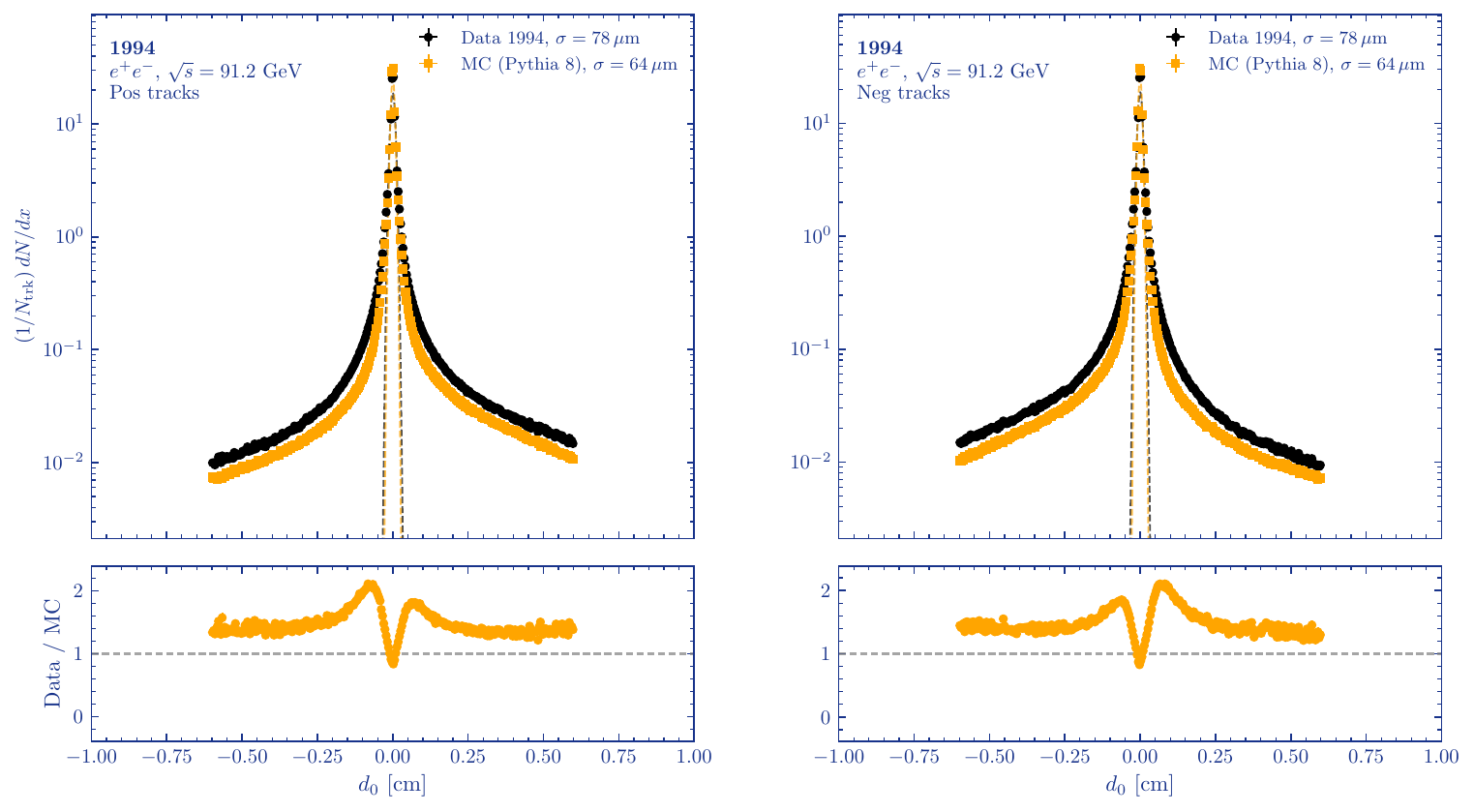}
    \caption{Transverse impact parameter $d_0$ of charged tracks with respect to the reconstructed primary vertex in the 1994 DELPHI Open Data sample compared to reconstructed \textsc{PYTHIA}~8.3 MC, for positively (left) and negatively (right) charged tracks. The lower panels show the data/MC ratio.}
    \label{fig:app_d0}
\end{figure}

\begin{figure}[ht!]
    \centering
    \includegraphics[width=0.95\textwidth]{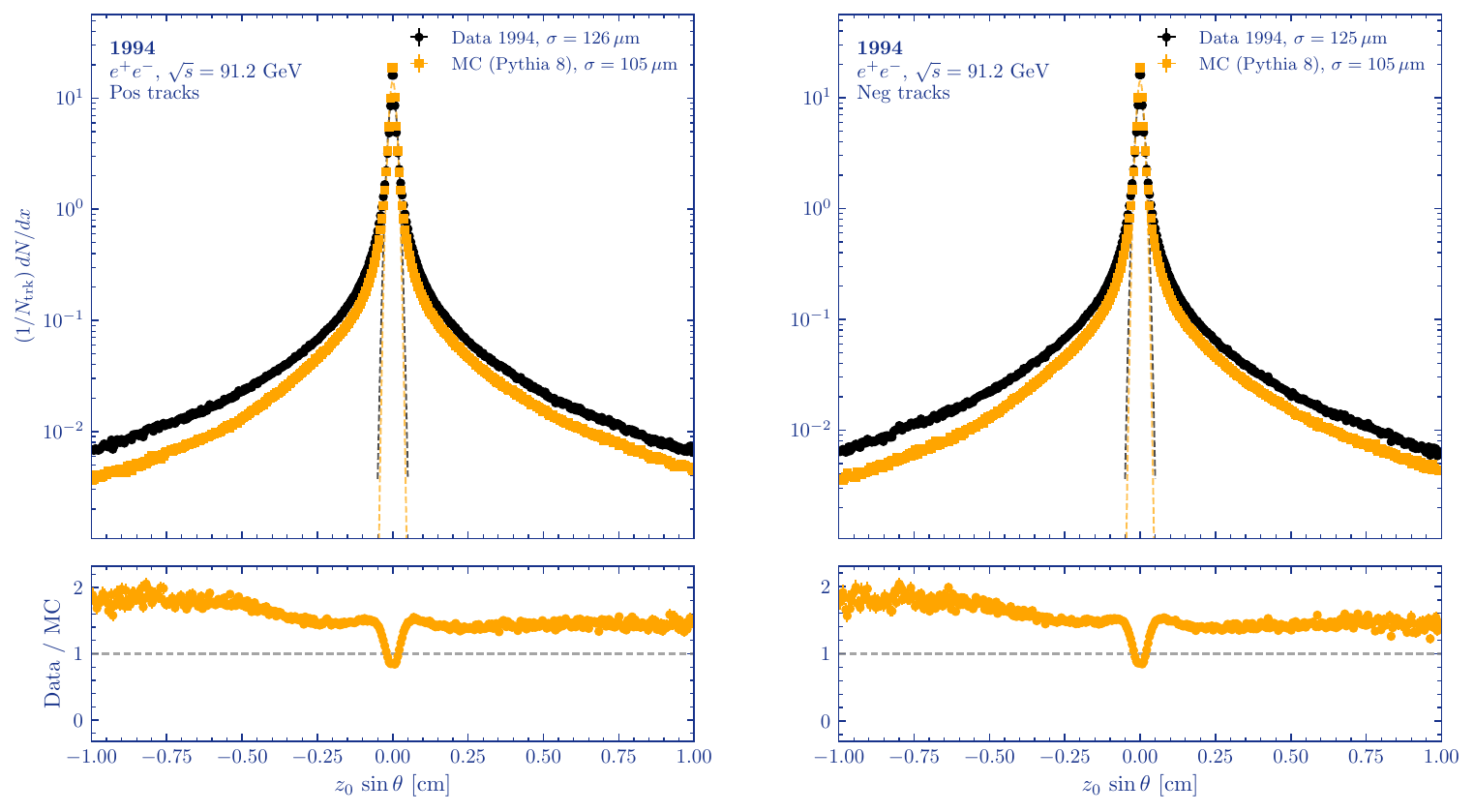}
    \caption{Longitudinal impact parameter projected onto the polar direction, $z_0\sin\theta$, of charged tracks with respect to the reconstructed primary vertex in the 1994 DELPHI Open Data sample compared to reconstructed \textsc{PYTHIA}~8.3 MC, for positively (left) and negatively (right) charged tracks. The lower panels show the data/MC ratio.}
    \label{fig:app_z0sin}
\end{figure}

\begin{figure}[ht!]
    \centering
    \includegraphics[width=0.95\textwidth]{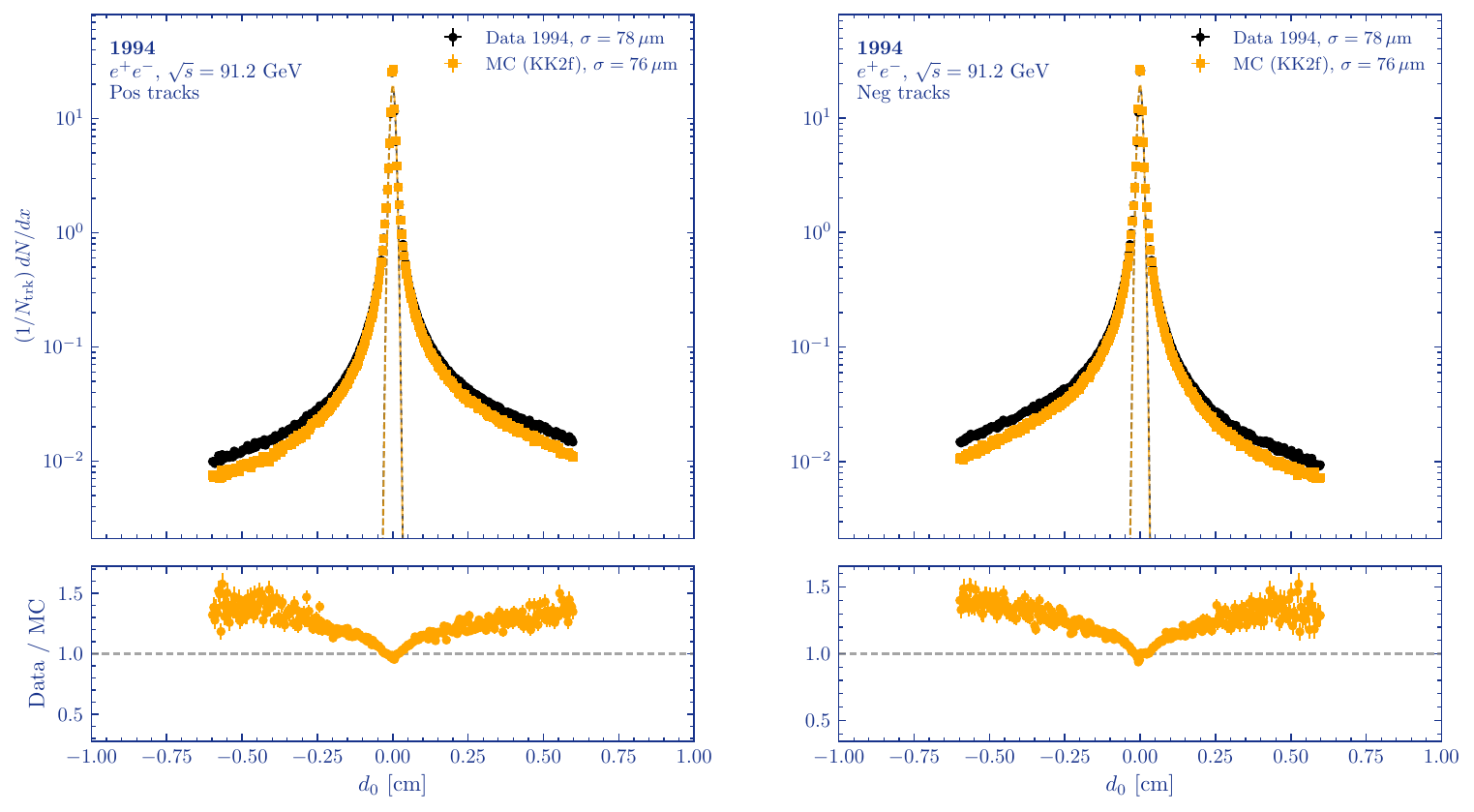}
    \caption{Same transverse impact parameter $d_0$ comparison as Figure~\ref{fig:app_d0}, but with the reconstructed \textsc{KK}2f+\textsc{PYTHIA}~6 sample in place of the self-generated \textsc{PYTHIA}~8.3 MC. The \textsc{KK}2f sample reproduces the measured core width ($\sigma = 76~\mu$m, versus $78~\mu$m in data) substantially better than \textsc{PYTHIA}~8.3.}
    \label{fig:app_d0_kk2f}
\end{figure}

\begin{figure}[ht!]
    \centering
    \includegraphics[width=0.95\textwidth]{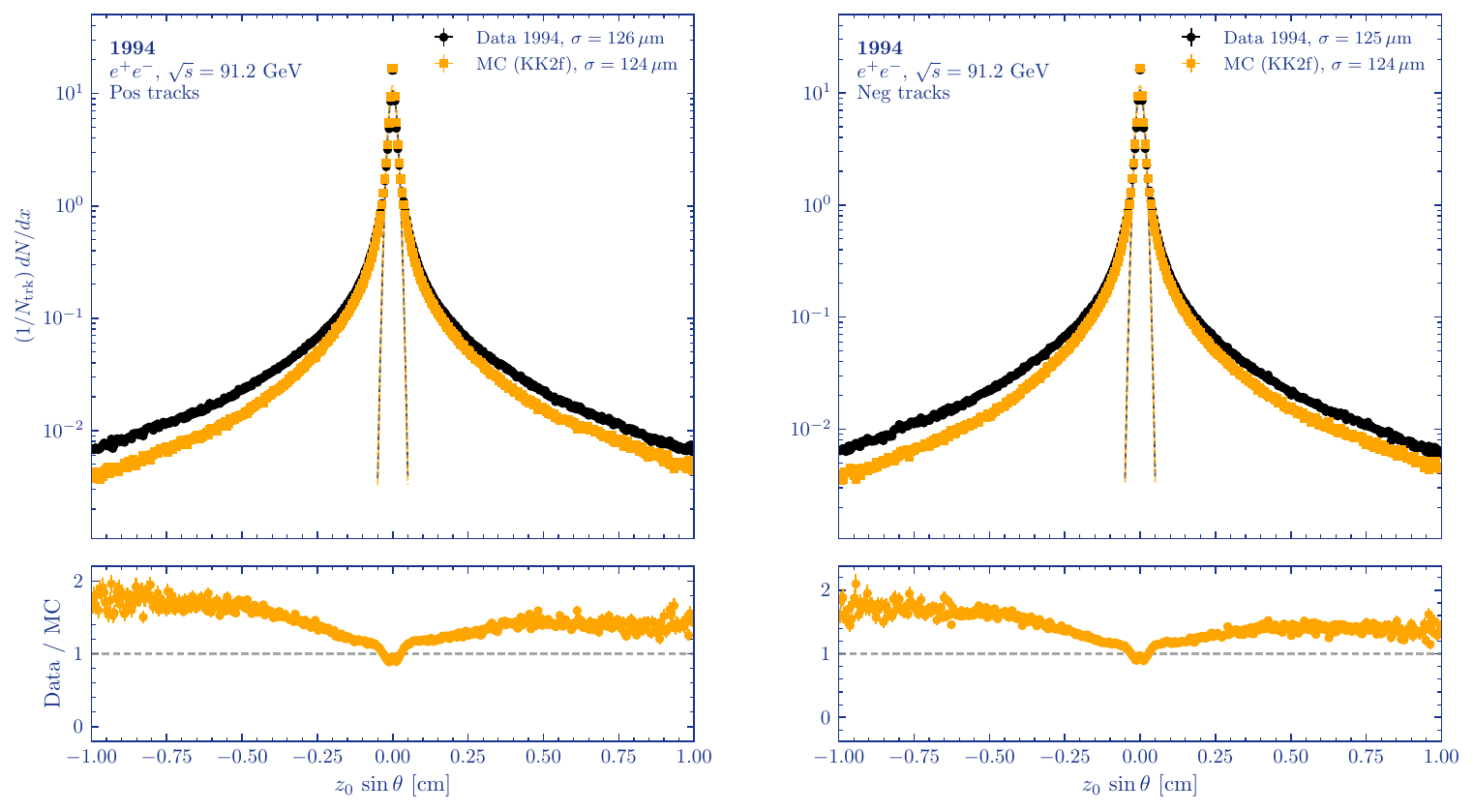}
    \caption{Same longitudinal impact parameter $z_0\sin\theta$ comparison as Figure~\ref{fig:app_z0sin}, but with the reconstructed \textsc{KK}2f+\textsc{PYTHIA}~6 sample in place of the self-generated \textsc{PYTHIA}~8.3 MC.}
    \label{fig:app_z0sin_kk2f}
\end{figure}

Figures~\ref{fig:app_d0} and~\ref{fig:app_z0sin} show the transverse ($d_0$) and longitudinal ($z_0\sin\theta$) impact parameters of charged tracks with respect to the reconstructed primary vertex, separated by reconstructed charge, for both 1994 data and reconstructed \textsc{PYTHIA}~8.3 MC. The self-generated \textsc{PYTHIA}~8.3 sample produces a visibly narrower impact-parameter core than the data ($\sigma_{d_0} = 64~\mu$m versus $78~\mu$m), so the data/MC ratio dips at the core and rises in the tails. The same comparison against the legacy \textsc{KK}2f+\textsc{PYTHIA}~6 sample, shown in Figures~\ref{fig:app_d0_kk2f} and~\ref{fig:app_z0sin_kk2f}, reproduces the measured core width much more closely ($\sigma_{d_0} = 76~\mu$m). The origin of this difference has not been isolated; candidate explanations include differences between the two production chains (geometry tag, alignment, or luminous-region size) as well as generator-level physics modeling. Its residual impact on the corrected charge correlator is covered by the \textsc{PYTHIA}-vs-\textsc{KK}2f generator-model systematic uncertainty (Section~\ref{sec:systematicsSource}).

\begin{figure}[ht!]
    \centering
    \includegraphics[width=0.95\textwidth]{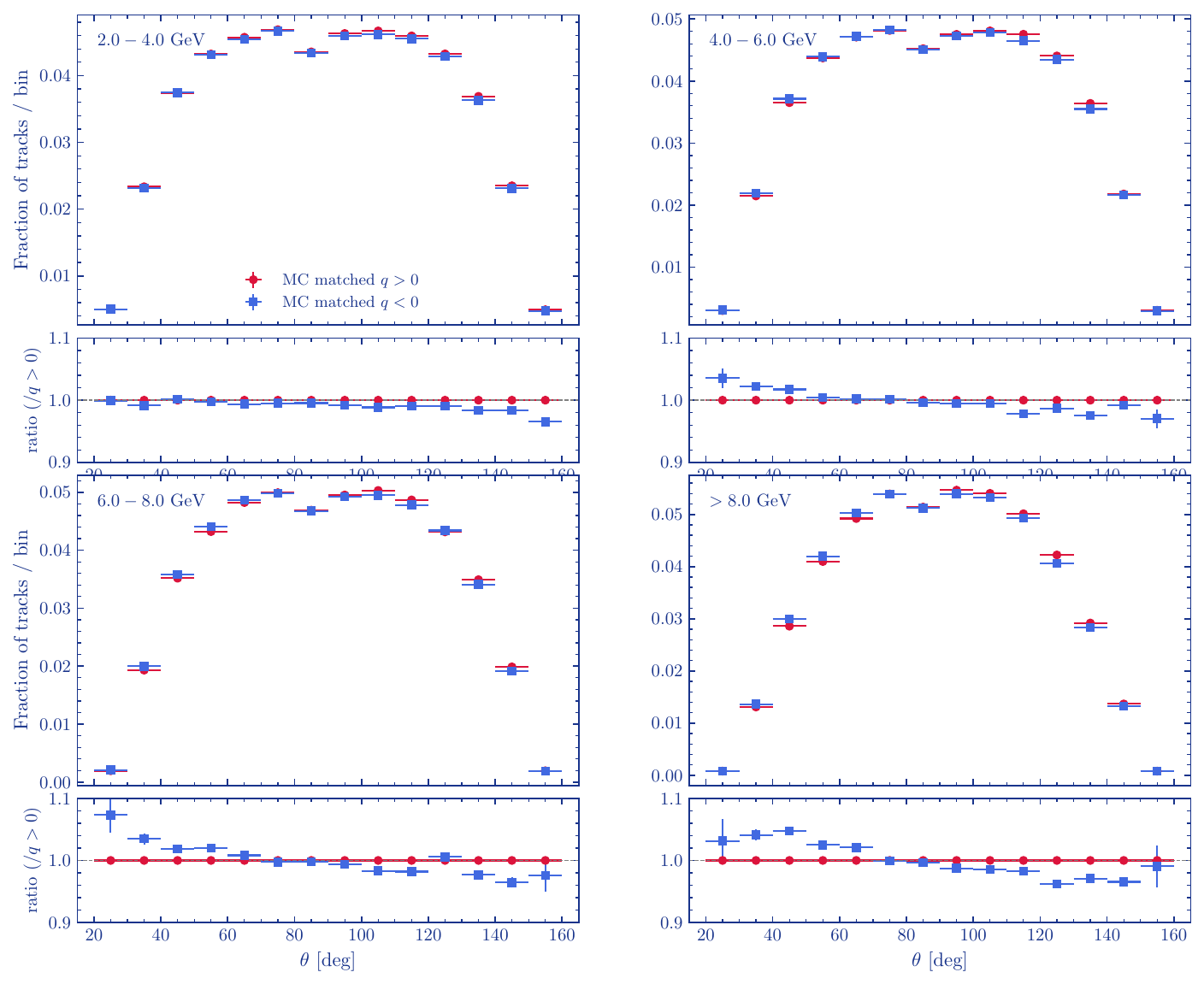}
    \caption{Polar-angle distributions of positively ($q>0$, red) and negatively ($q<0$, blue) truth-matched reconstructed tracks in \textsc{PYTHIA}~8.3 simulation, broken down into transverse-momentum bins. This is the $p_{\rm T}$ breakdown of the truth-matched component shown in the top-left panel of Figure~\ref{fig:mc_fake_match}.}
    \label{fig:app_matched_per_pt}
\end{figure}

Figure~\ref{fig:app_matched_per_pt} shows the polar-angle distributions of positively and negatively charged truth-matched tracks in \textsc{PYTHIA}~8.3 simulation, broken down into transverse-momentum bins. The bin-by-bin view shows that the reconstructed-level charge imbalance progressively shrinks as $p_{\rm T}$ increases, with the highest-$p_{\rm T}$ bins exhibiting near-symmetric $q>0$ and $q<0$ distributions.

\begin{figure}[ht!]
    \centering
    \includegraphics[width=0.45\textwidth]{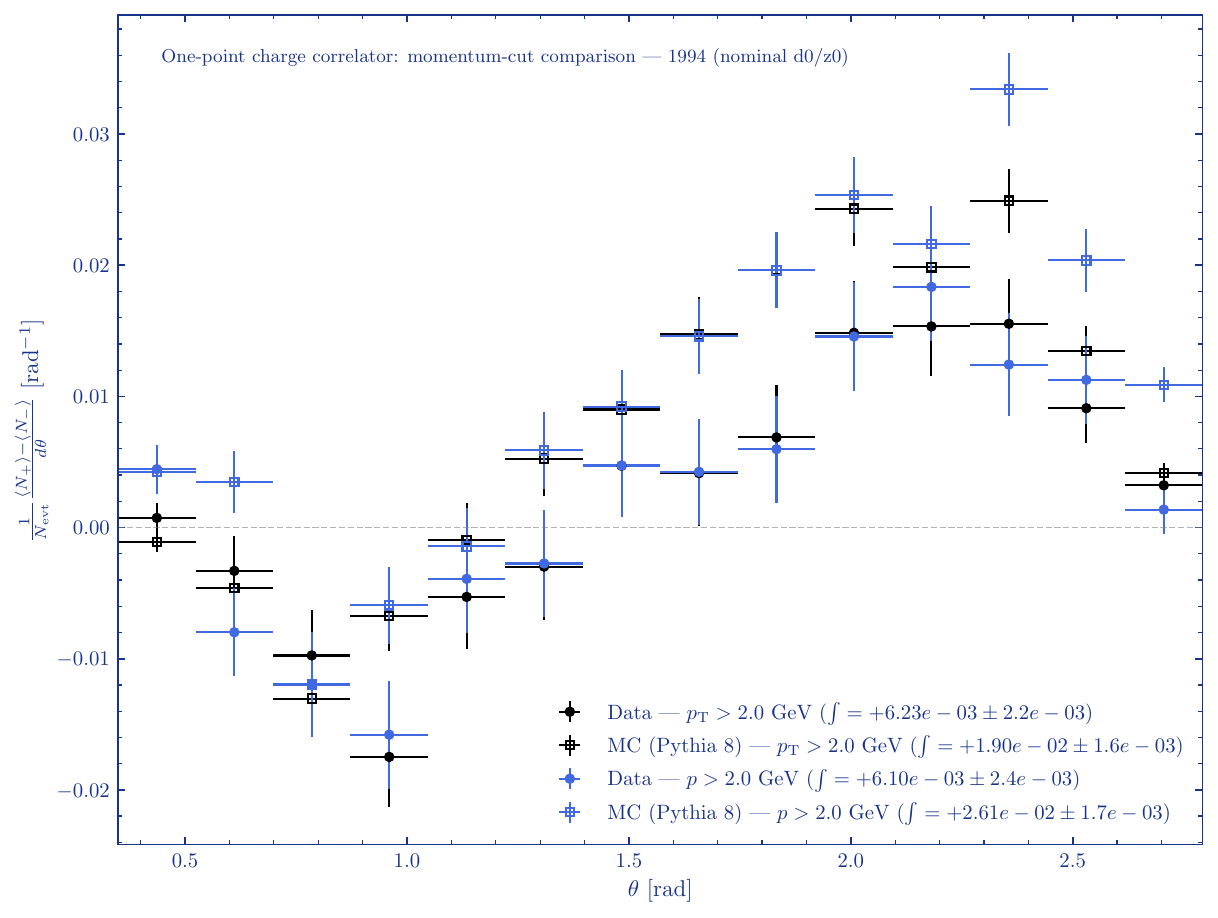}
    \includegraphics[width=0.45\textwidth]{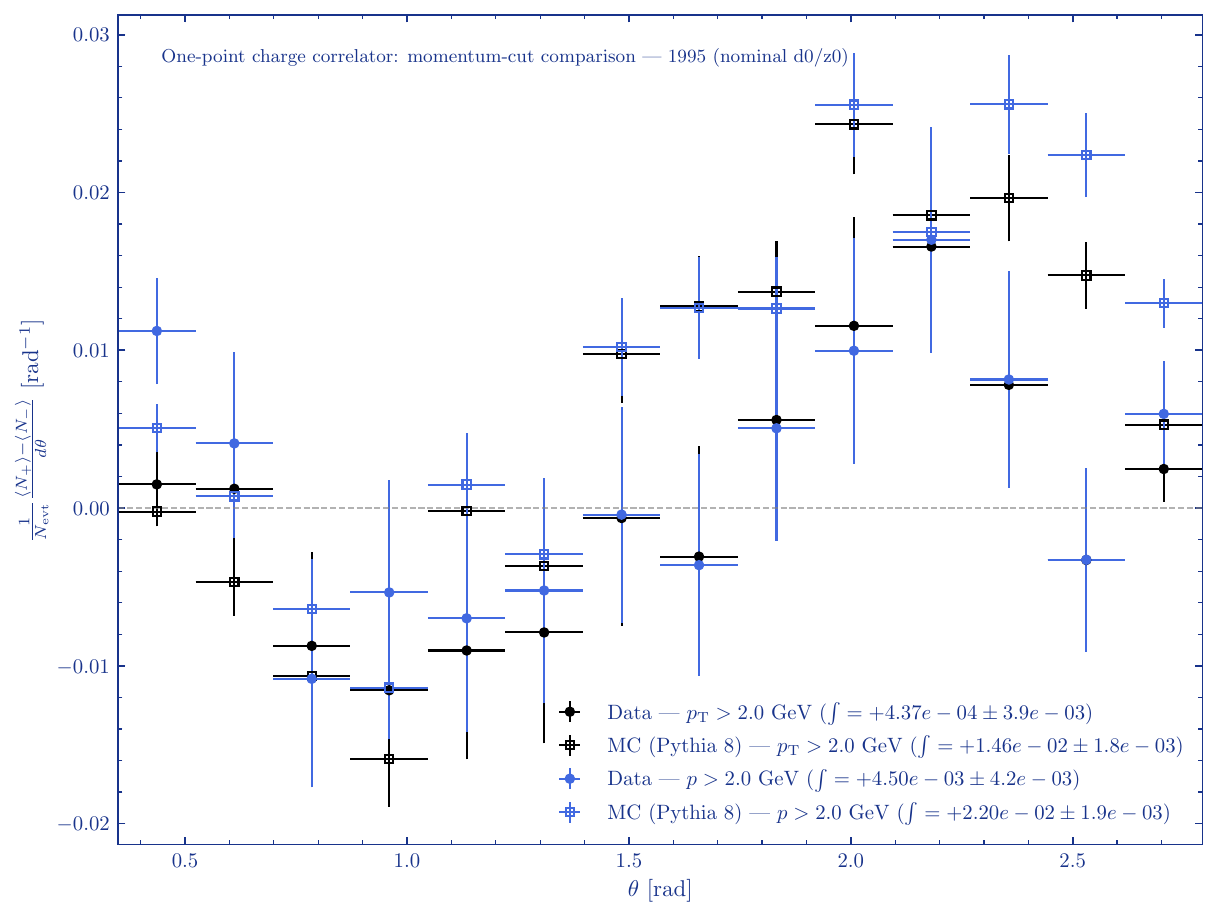}
    \caption{Comparison of the raw one-point charge correlator $\mathcal{Q}(\theta) = N^+(\theta) - N^-(\theta)$ at reconstructed level between the nominal $p_{\rm T} > 2$~GeV and the alternative $|p| > 2$~GeV selections, for data and reconstructed \textsc{PYTHIA}~8.3 MC, in the 1994 (left) and 1995 (right) datasets. The nominal $p_{\rm T}$ selection produces a better charge balance, particularly in data.}
    \label{fig:app_pcut_compare}
\end{figure}

\begin{figure}[ht!]
    \centering
    \includegraphics[width=0.45\textwidth]{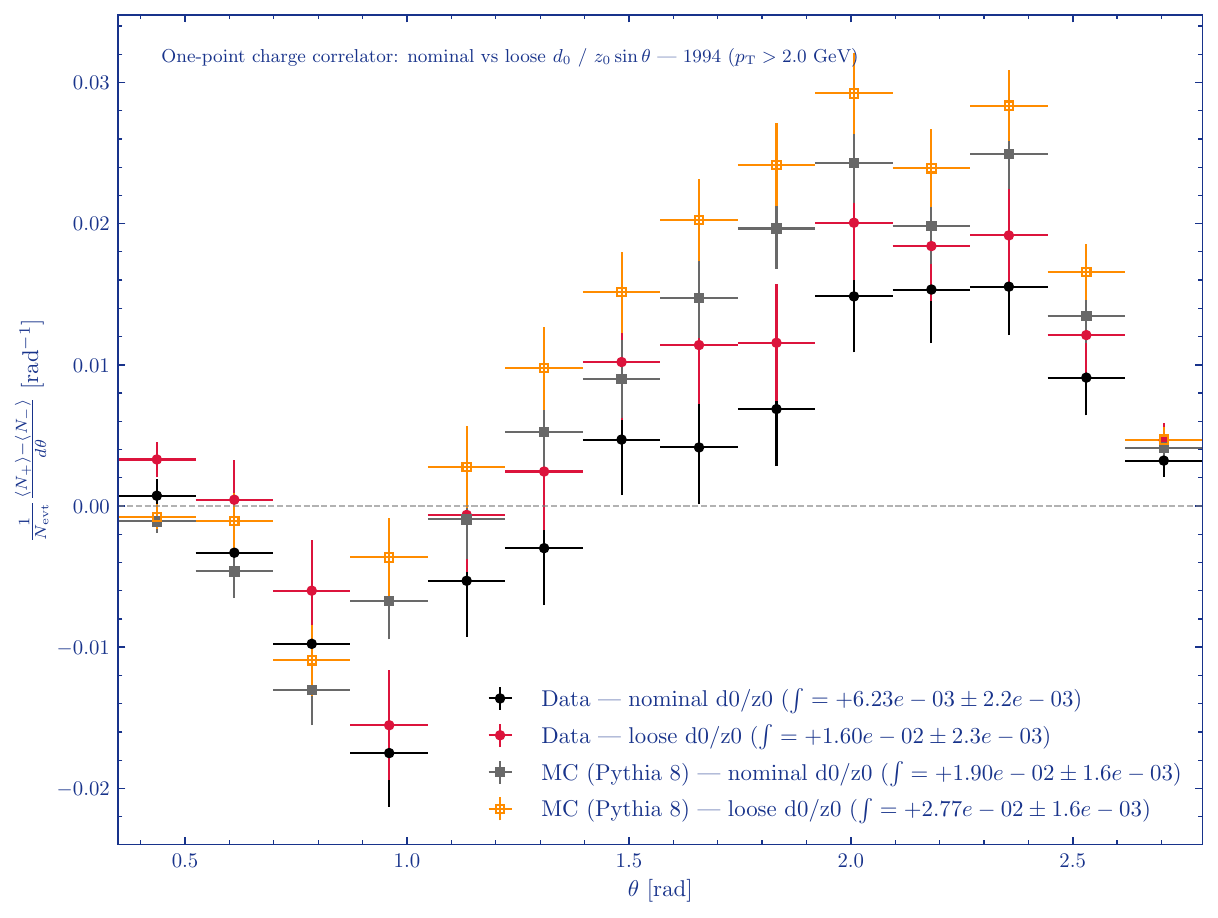}
    \includegraphics[width=0.45\textwidth]{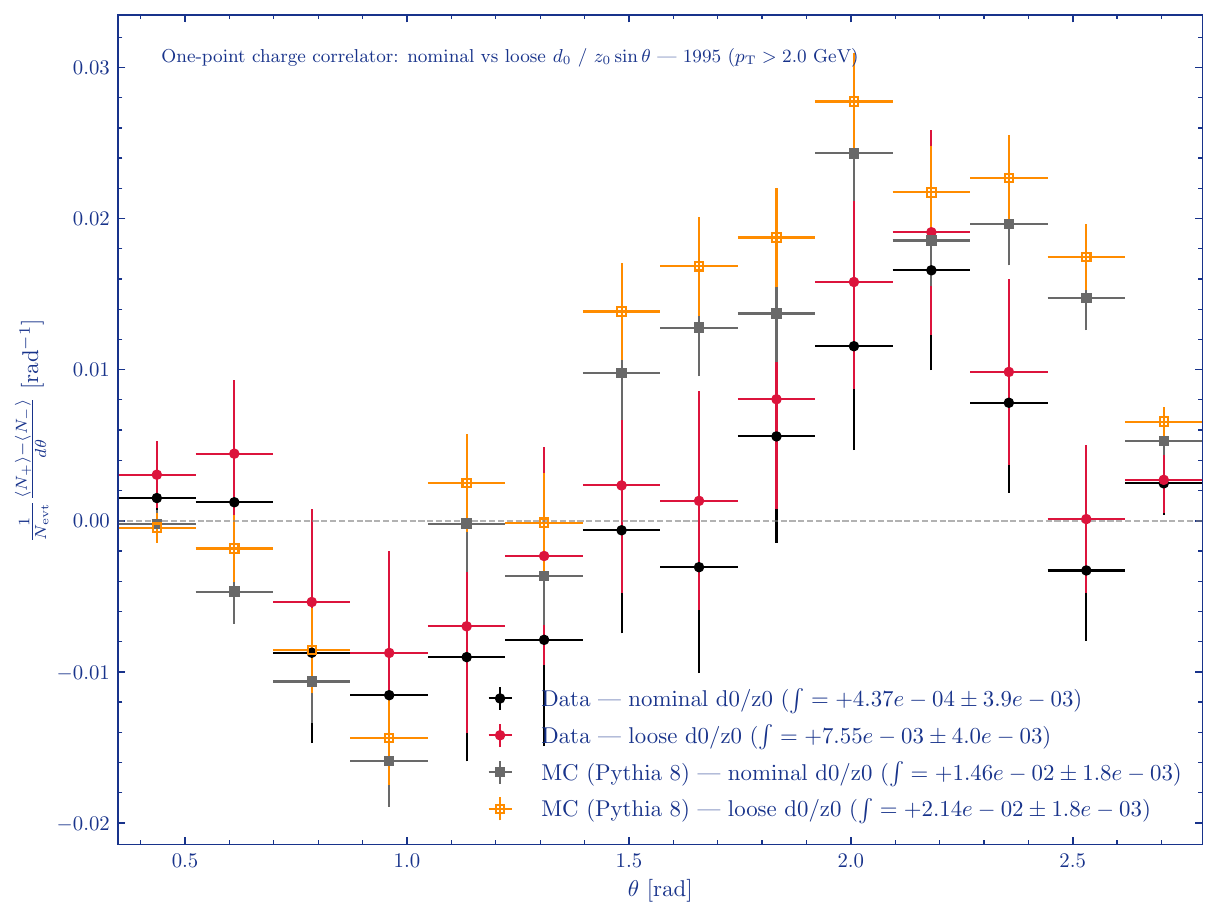}
    \caption{Comparison of the raw one-point charge correlator $\mathcal{Q}(\theta) = N^+(\theta) - N^-(\theta)$ at reconstructed level between the nominal tight impact-parameter requirement ($d_0 \le 0.6$~cm, $z_0 \le 1.0$~cm) and a loosened one ($d_0 \le 4$~cm, $z_0 \le 10$~cm), for data and reconstructed \textsc{PYTHIA}~8.3 MC, in the 1994 (left) and 1995 (right) datasets. The tight impact-parameter selection produces a better charge balance, particularly in data.}
    \label{fig:app_d0z0_compare}
\end{figure}

Finally, Figures~\ref{fig:app_pcut_compare} and~\ref{fig:app_d0z0_compare} illustrate how the selection choices affect the reconstructed charge balance. Figure~\ref{fig:app_pcut_compare} compares the raw one-point charge correlator under the nominal $p_{\rm T} > 2$~GeV selection with the alternative $|p| > 2$~GeV selection, and Figure~\ref{fig:app_d0z0_compare} compares the nominal tight impact-parameter requirement ($d_0 \le 0.6$~cm, $z_0 \le 1.0$~cm) with a loosened one ($d_0 \le 4$~cm, $z_0 \le 10$~cm), in both cases for data and reconstructed \textsc{PYTHIA}~8.3 MC. The nominal $p_{\rm T}$ and tight impact-parameter selections produce a visibly better charge balance, an improvement most pronounced in data. The tighter cuts remove soft and displaced tracks, whose charge is the most prone to mismodeling: soft tracks suffer from multiple Coulomb scattering, and displaced tracks complicate the track fit. They also reject more secondary tracks from photon conversions and hadronic interactions. Together, these effects sharpen the overall charge balance of the input sample and support the nominal fiducial and track-quality requirements adopted in Section~\ref{sec:data_samples}.

\clearpage
\section{Tag-and-probe selection bias and sample purity studies}
\label{app:tnp_check}

The cut-flow table for the $Z \to \tau^+\tau^-$ selection is shown below in Table~\ref{tab:tau_cutflow}. 

\begin{table}[h]
\centering
\caption{Tau-pair selection cut-flow for the 1994 sample.
         Counts are unweighted event yields. The simulated columns correspond to the samples processed for this study; in particular, only part of the hadronic $q\bar{q}$ sample is used. The ``Topology'' requirement selects events classified as a $1{+}1$ or $1{+}3$ prong configuration; ``Exact $1p{+}1p$'' subsequently retains only the $1{+}1$ configuration.}
\label{tab:tau_cutflow}
\begin{tabular}{rlrrrrr}
\toprule
\# & Bin label & data (1994) & $\tau\tau$ & $ee$ & $\mu\mu$ & $q\bar q$ \\
\midrule
1 & Total                  & 5{,}753{,}343 & 1{,}347{,}300 & 1{,}260{,}900 & 1{,}239{,}300 & 2{,}872{,}846 \\
2 & Good $\geq 2$          & 4{,}453{,}213 & 1{,}334{,}413 & 1{,}257{,}719 & 1{,}167{,}625 & 2{,}872{,}846 \\
3 & $N_{\mathrm{ch}} \leq 4$ &   606{,}250 & 1{,}185{,}349 & 1{,}033{,}407 & 1{,}002{,}991 &    65{,}684 \\
4 & $E_{\mathrm{vis}}$     &   260{,}178 & 1{,}059{,}640 &   286{,}971 &    53{,}562 &    51{,}919 \\
5 & Thrust                 &   129{,}080 & 1{,}025{,}132 &   273{,}955 &    44{,}325 &     8{,}066 \\
6 & Acollinearity          &    78{,}306 &   908{,}690 &   216{,}646 &    23{,}466 &     6{,}821 \\
7 & Topology                 &    59{,}148 &   789{,}710 &   199{,}710 &    23{,}402 &     2{,}301 \\
8 & Lepton veto                   &    19{,}924 &   325{,}712 &    20{,}308 &     2{,}855 &       623 \\
9 & Exact $1p{+}1p$              &    16{,}339 &   256{,}151 &    19{,}753 &     2{,}855 &       370 \\
\bottomrule
\end{tabular}
\end{table}

The generator-level topology composition of the reconstructed $1p{+}1p$ selection for the TNP track selection is shown below in Table~\ref{tab:reco1p1p_genfrac}. The TNP angular and quality cuts reduce the gen-$1p{+}3p$ contamination from $\sim 3\%$ to $\sim 1.7\%$.

\begin{table}[h]
\centering
\caption{Generator-level topology composition of the reconstructed $1p{+}1p$
         selection in the $\tau^+\tau^-$ Monte Carlo (1994).}
\label{tab:reco1p1p_genfrac}
\begin{tabular}{lcc}
\toprule
Selection & gen $1p{+}1p$ & gen $1p{+}3p$ \\
\midrule
raw reco $1p{+}1p$          & 96.87\% & 3.01\% \\
reco $1p{+}1p$ after tag-and-probe selection   & 98.24\% & 1.73\% \\
\bottomrule
\end{tabular}
\end{table}

The data-driven correction extracted from the $Z \to \tau^+\tau^-$ control 
sample relies on the TNP charge misreconstruction rate measured in $1{+}1$ 
prong $\tau$-pair events, with the track of smaller $\Delta p/p$ in each event designated as the tag.

\begin{figure}[h]
  \centering
  \includegraphics[width=0.75\textwidth]{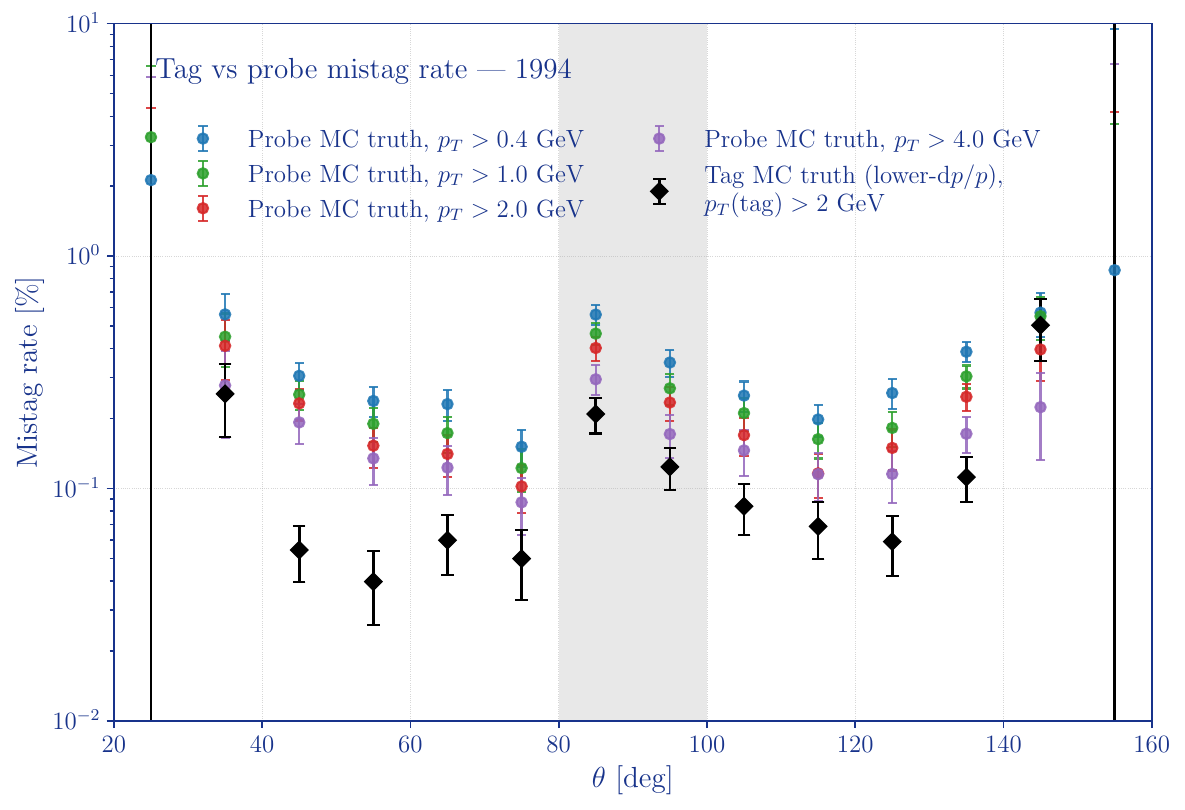}
  \caption{Per-track charge-misreconstruction rate from MC truth as a function of 
  $\theta$. Open circles: probe rate at four $p_{\rm T}$ thresholds. 
  Black diamonds: tag rate (smaller-$\Delta p/p$ track in the $1{+}1$ 
  selection) in the analysis $p_{\rm T}$ range. The shaded band between $80^\circ$ and $100^\circ$ marks the bins containing the cathode plane of the DELPHI TPC. The
  tag rate is a factor of $\sim$2--4 smaller than the probe
  rate over the bulk of the $\theta$ range, with the separation shrinking in the
  outermost bins at either end.}
  \label{fig:tag_mistag_from_mc}
\end{figure}

Figure~\ref{fig:tag_mistag_from_mc} shows the per-track charge-misreconstruction rate 
estimated directly from MC truth as a function of $\theta$. The probe rate 
is shown for several $p_{\rm T}$ thresholds (open circles), and the tag 
rate, restricted to the same $1{+}1$ prong selection but always taken 
as the smaller-$\Delta p/p$ track, is overlaid as black diamonds. The
$\Delta p/p$ ordering suppresses the misreconstruction rate by a factor of two to
four across the bulk of the $\theta$ range. The separation is smaller in the
outermost bins at either end of the acceptance, and in the most backward bin the
tag rate is comparable to the probe rate, so the tag advantage cannot be assumed to
hold uniformly across $\theta$; the residual effect of the finite tag purity is
covered by the tag-and-probe selection-bias systematic uncertainty discussed below.
Over the bulk of the acceptance this validates the tag definition used in the data-driven correction: the 
tag is genuinely higher-quality than the probe, so the TNP same-sign rate is dominated by the probe and provides a relatively clean measurement of its misreconstruction rate.

\begin{figure}[h]
  \centering
  \includegraphics[width=0.85\textwidth]{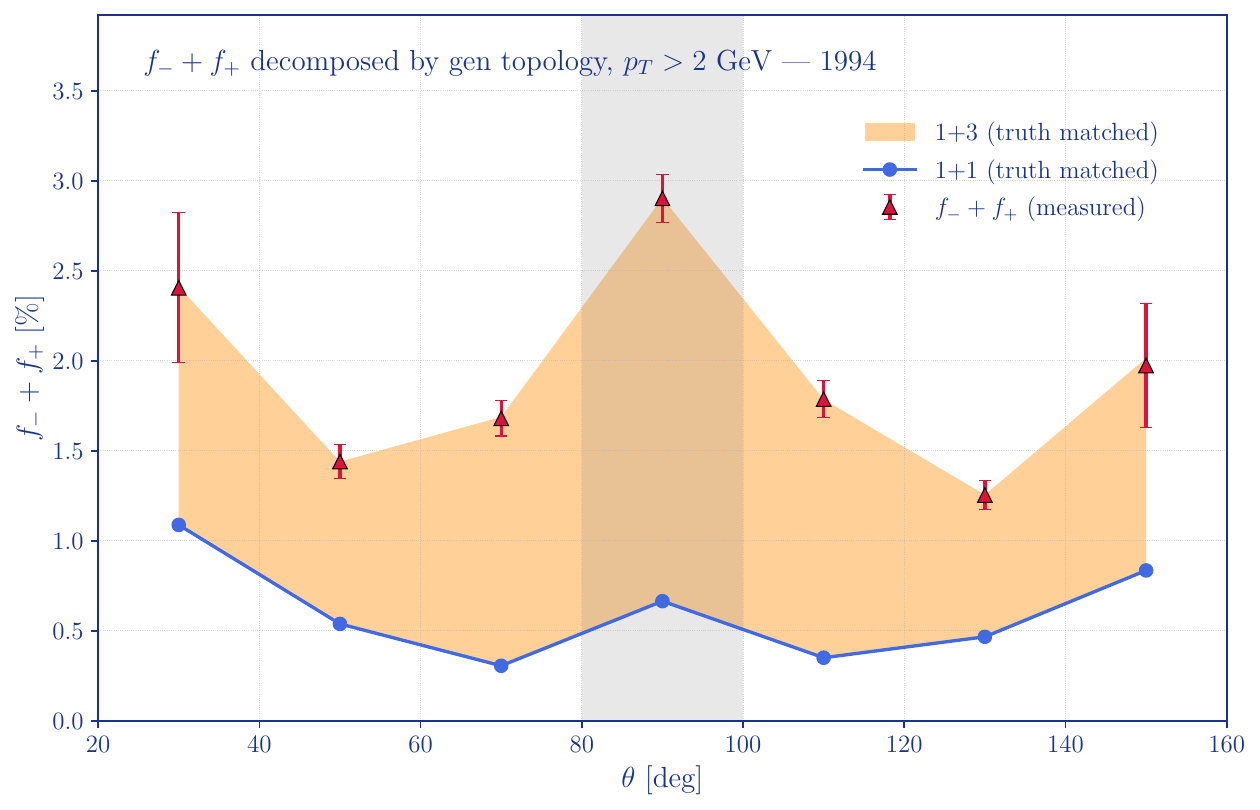}
  \caption{Comparison of the TNP-measured charge misreconstruction at 
  $p_{\rm T}>2$~GeV with the generator-level truth-matched rate, decomposed by generator-level event topology. Blue line: $1{+}1$ prong 
  contribution ($w_{1{+}1}\,\delta_{1{+}1}$). Orange band: multi-prong 
  leakage contribution ($w_{\rm leak}\,\delta_{\rm leak}$, dominantly 
  $1{+}3$). Red triangles: total $\delta_{\rm sum}^{\rm TNP}$ measured on 
  the full $1{+}1$ reco sample. }
  \label{fig:delta_sum_stacked_pt2}
\end{figure}

\begin{figure}[h]
  \centering
  \includegraphics[width=0.85\textwidth]{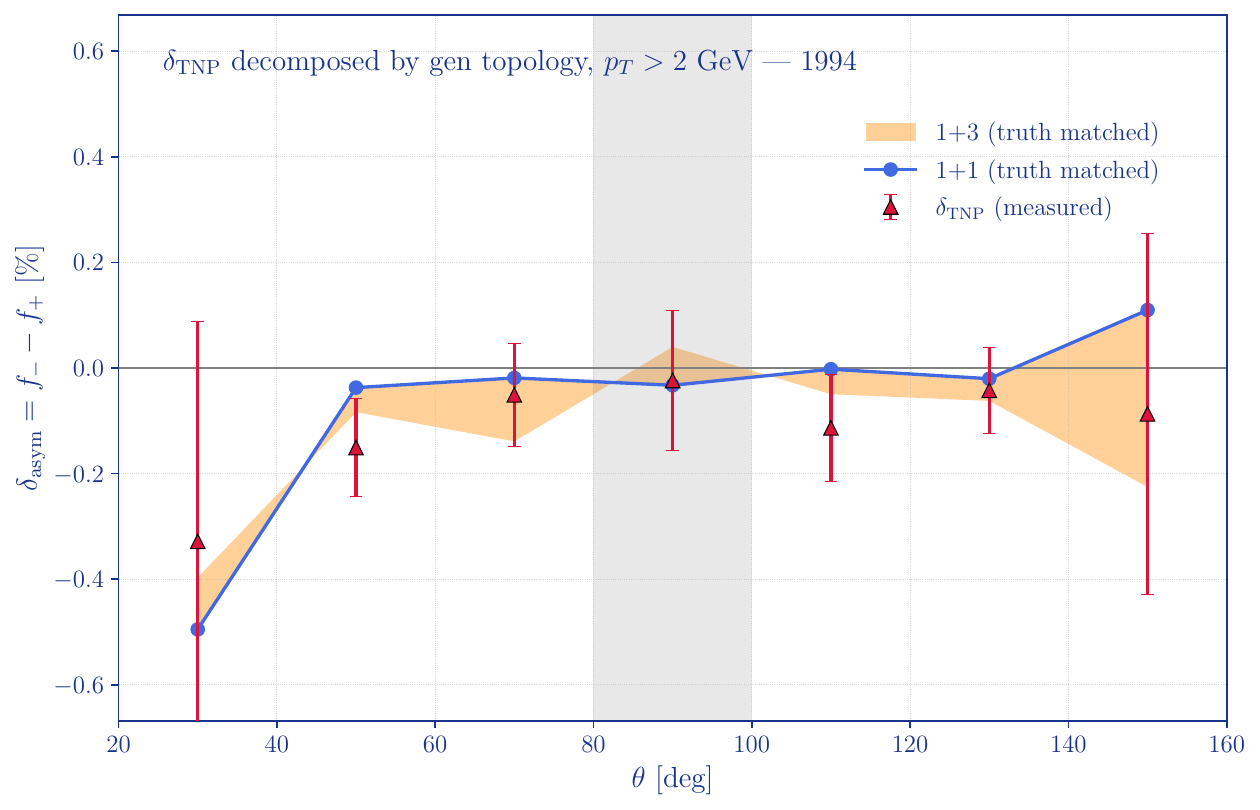}
  \caption{Comparison of $\delta_{\rm asym}^{\rm TNP}$ at 
  $p_{\rm T}>2$~GeV with the generator-level truth-matched rate, decomposed by generator-level event topology. Blue line: $1{+}1$ prong 
  contribution ($w_{1{+}1}\,\delta_{1{+}1}$). Orange band: multi-prong 
  leakage contribution ($w_{\rm leak}\,\delta_{\rm leak}$, dominantly 
  $1{+}3$). Red triangles: total $\delta_{\rm asym}^{\rm TNP}$ measured on 
  the full $1{+}1$ reco sample.}
  \label{fig:delta_asym_stacked_pt2}
\end{figure}

Using MC truth information, we compare the TNP-measured sum $\delta_{\rm sum}^{\rm TNP}=f^-+f^+$ and asymmetric 
residual $\delta_{\rm asym}^{\rm TNP}=f^--f^+$ to the corresponding generator-level 
truth matched value, as shown in Figure~\ref{fig:delta_sum_stacked_pt2} and Figure~\ref{fig:delta_asym_stacked_pt2}, respectively. 

We further decompose $\delta_{\rm asym}^{\rm TNP}$ in MC by generator-level event topology. We find that the dominant contribution to $\delta_{\rm asym}^{\rm TNP}$ comes 
from $1{+}3$ prong leakage into the $1{+}1$ reco sample. Although the 
multi-prong leakage fraction after the tag-and-probe selection is only $1.7\%$, the 
leaked events carry an intrinsically charge-asymmetric structure that 
survives in the TNP selection. The data-driven correction, therefore, depends on the accuracy of the modeling of the $1{+}3$ topology: if the modeling is perfect, the leakage contribution 
cancels in the data--MC residual. 
To account for imperfect modeling, we assign a $10\%$ uncertainty on the $1{+}3$ topology contribution as an additional systematic.

A discrepancy at the $0.1\%$ level is observed between $\delta_{\rm asym}^{\rm TNP}$ and the corresponding generator-level truth. This 
reflects residual selection biases, including the limited tag purity mentioned above. The difference between $\delta_{\rm asym}^{\rm TNP}$ and the generator-level truth is taken as an additional systematic uncertainty.

\begin{figure}[h]
  \centering
  \includegraphics[width=0.6\textwidth]{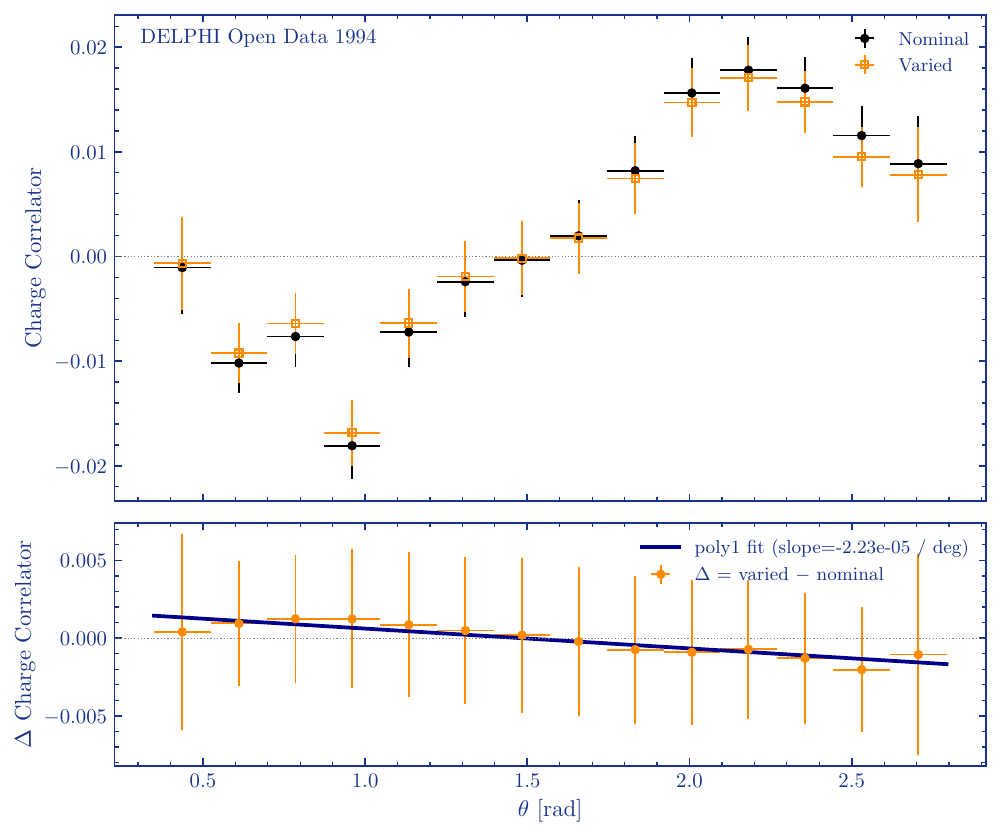}
  \caption{Propagated impact on the corrected one-point charge correlator $\mathcal{Q}(\theta)$ of the residual difference between the TNP-extracted $\delta_{\rm asym}^{\rm TNP}$ and the corresponding generator-level truth. The bin-by-bin difference between the TNP-extracted and the generator-level $\delta_{\rm asym}$ is fitted by a first-order polynomial fit in $\theta$.}
  \label{fig:tnp_nonclosure}
\end{figure}

The 0.1\% absolute discrepancy between $\delta_{\rm asym}^{\rm TNP}$ and the generator-level truth is propagated through the full data-driven correction chain to obtain its actual impact on the corrected one-point charge correlator $\mathcal{Q}(\theta)$. The resulting $\theta$-dependent shift, shown in Figure~\ref{fig:tnp_nonclosure}, is the quantity propagated as the tag-and-probe selection-bias systematic in the corresponding paragraph of Section~\ref{sec:systematics}. The shift is amplified relative to the underlying 0.1\% mistag-rate discrepancy by the total track yield $N_{\rm tot}(\theta) = N^+ + N^-$.

\clearpage
\section{Key plots from the 1995 analysis}
\label{app:1995_check}

\begin{figure}[ht!]
    \centering
    \includegraphics[width=0.55\textwidth]{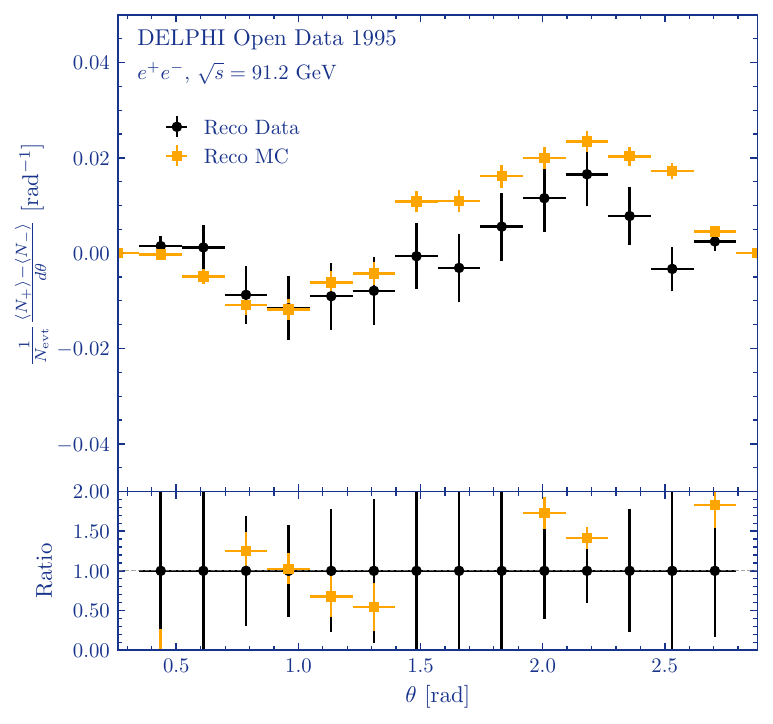}
    \caption{Comparison of the raw one-point charge correlator $\mathcal{Q}(\theta) = N^+(\theta) - N^-(\theta)$ between 1995 data (black) and reconstructed \textsc{PYTHIA}~8.3 MC (yellow), for the $p_{\rm T} > 2$~GeV selection. The lower panel shows the data/MC ratio.}
    \label{fig:app_data_mc_pt_95}
\end{figure}

\begin{figure}[ht!]
    \centering
    \includegraphics[width=0.45\textwidth]{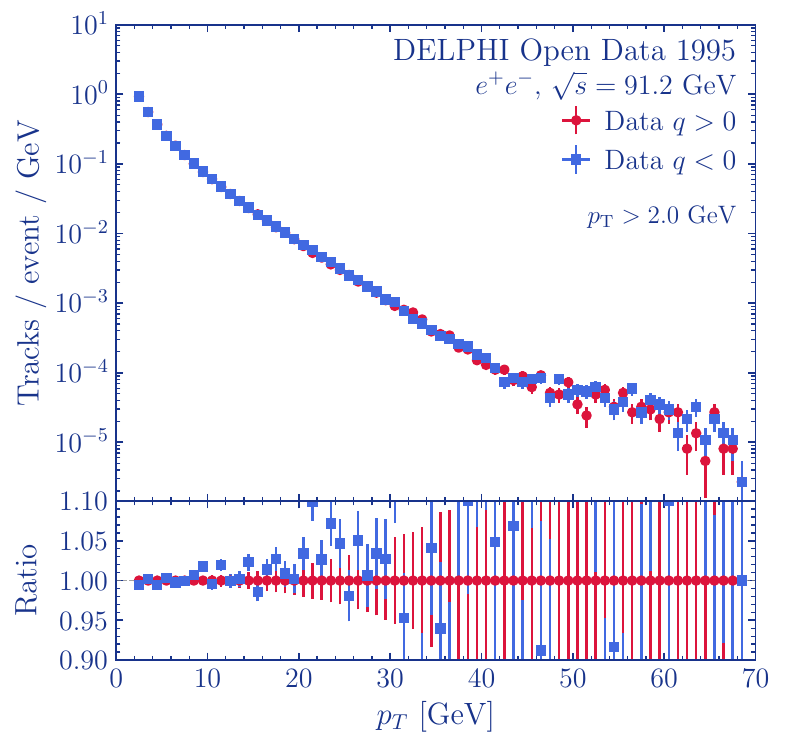}
    \includegraphics[width=0.45\textwidth]{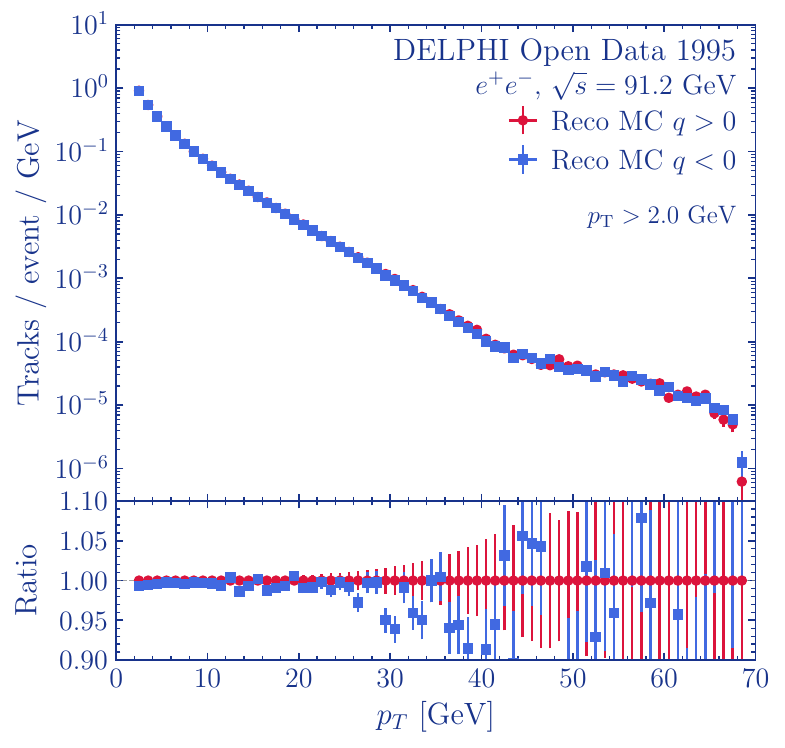}
    \caption{Cross-check counterpart of Figure~\ref{fig:data_mc_theta} for the 1995 sample: charged-track momentum spectra in 1995 data (left) and reconstructed \textsc{PYTHIA}~8.3 MC (right), separated by reconstructed charge, as a function of $p_{\rm T}$.}
    \label{fig:app_data_mc_theta_95}
\end{figure}

\begin{figure}[ht!]
    \centering
    \includegraphics[width=0.48\textwidth]{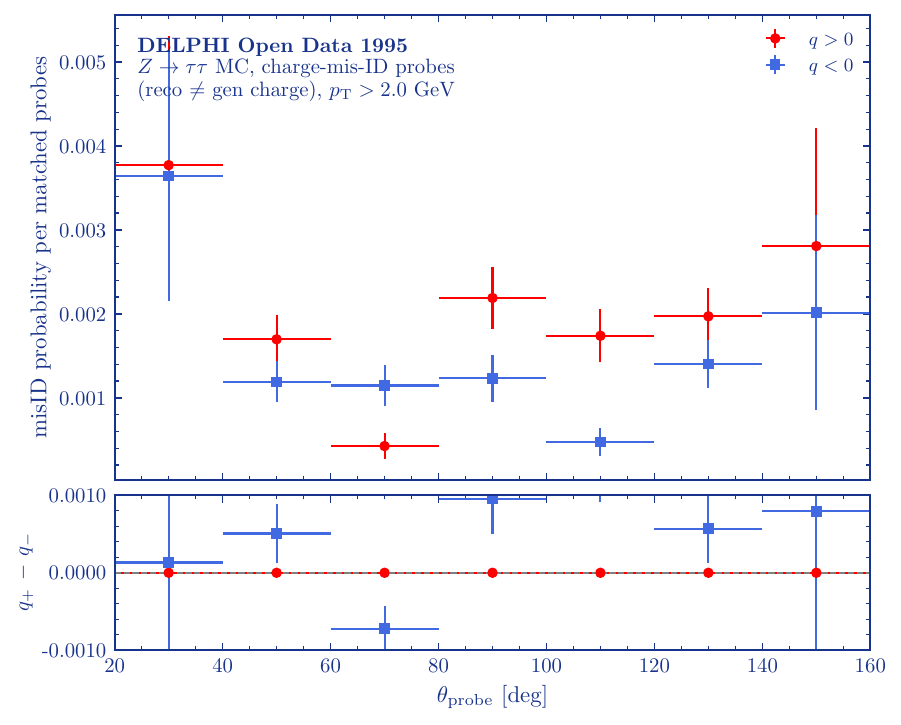}
    \includegraphics[width=0.48\textwidth]{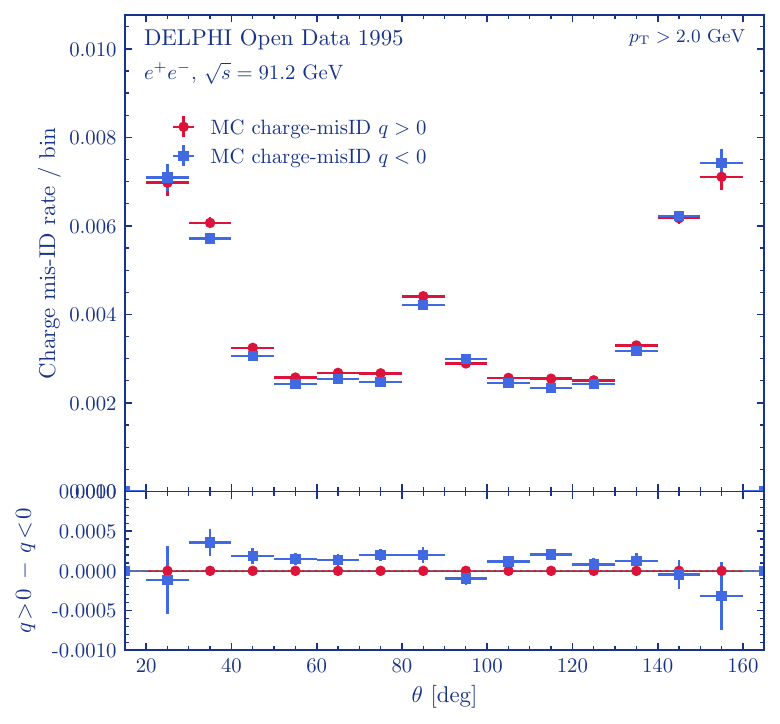}
    \caption{Cross-check counterpart of Figure~\ref{fig:tau_misid_rate_mc} for the 1995 sample: truth-matched per-track charge-misreconstruction rate, shown separately for the two reconstructed charges ($q>0$ and $q<0$), in the $\tau$ control-region MC (left) and the inclusive hadronic MC (right).}
    \label{fig:app_misid_rate_mc_95}
\end{figure}

\begin{figure}[ht!]
    \centering
    \includegraphics[width=\textwidth]{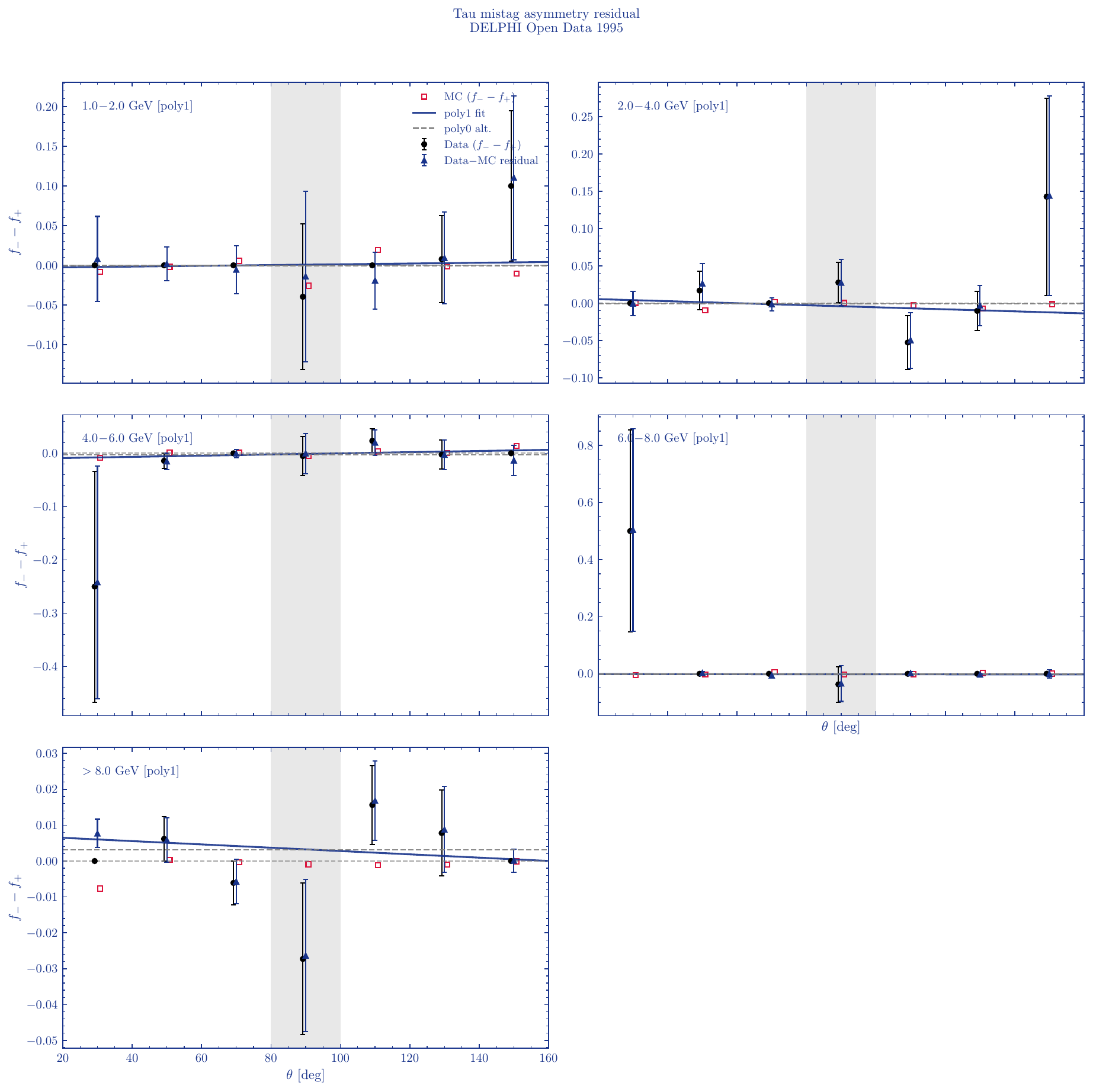}
    \caption{Data--MC residual of the charge misreconstruction asymmetry $\delta^i_{\rm asym}(\theta)$ measured on the 1995 DELPHI Open Data sample, in the same five exclusive probe $p_{\rm T}$ bins as Figure~\ref{fig:misid_asymmetry}. The residuals carry larger statistical uncertainties than for 1994 due to the smaller integrated luminosity ($\approx 15$~pb$^{-1}$).}
    \label{fig:app_misid_asym_95}
\end{figure}

\begin{figure}[ht!]
    \centering
    \includegraphics[width=\textwidth]{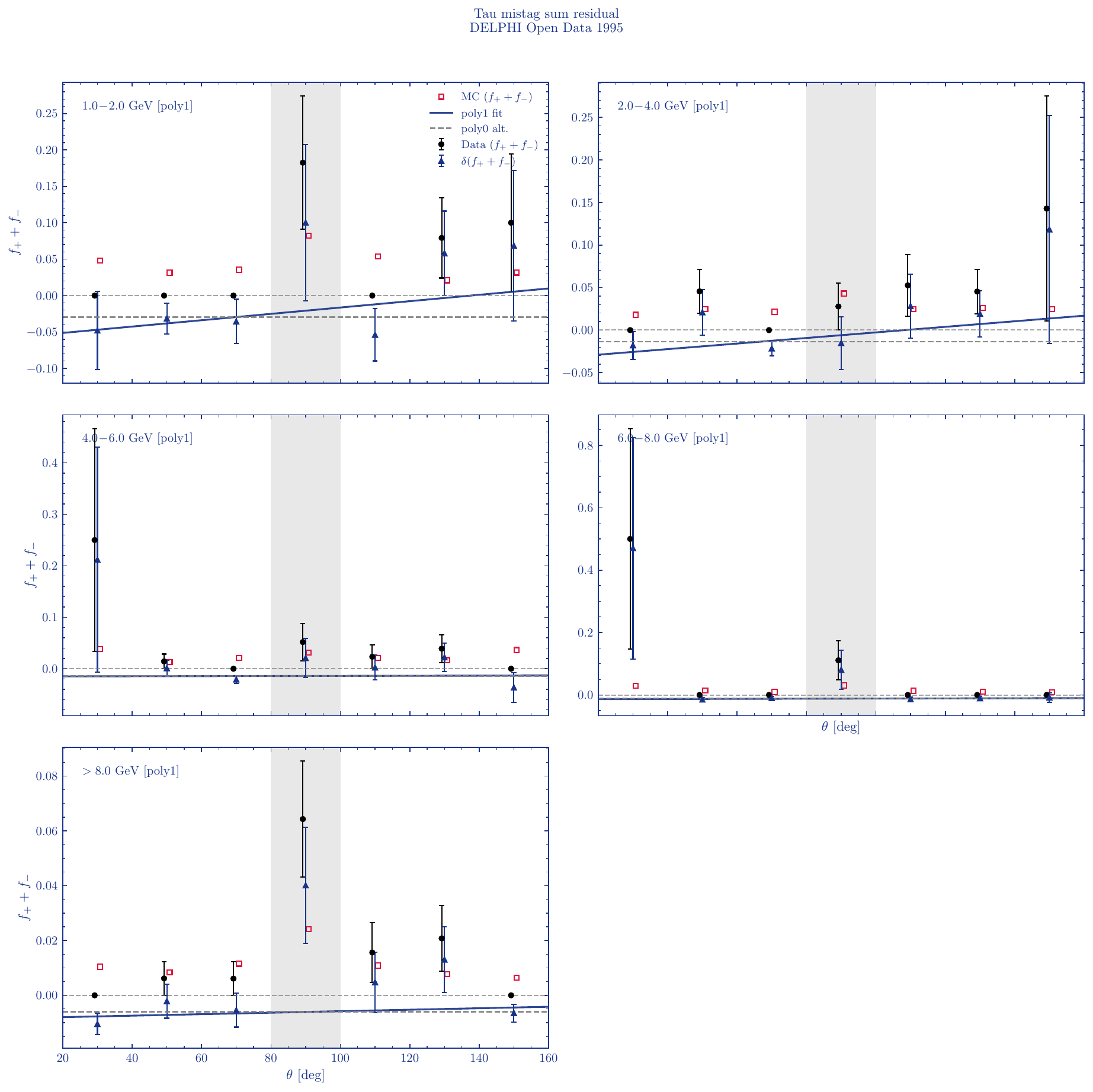}
    \caption{Data--MC residual of the charge misreconstruction sum $\delta^i_{\rm sum}(\theta)$ measured on the 1995 DELPHI Open Data sample, in the same five exclusive probe $p_{\rm T}$ bins as Figure~\ref{fig:misid_sum}.}
    \label{fig:app_misid_sum_95}
\end{figure}

\begin{figure}[htbp]
    \centering
    \includegraphics[width=0.85\textwidth]{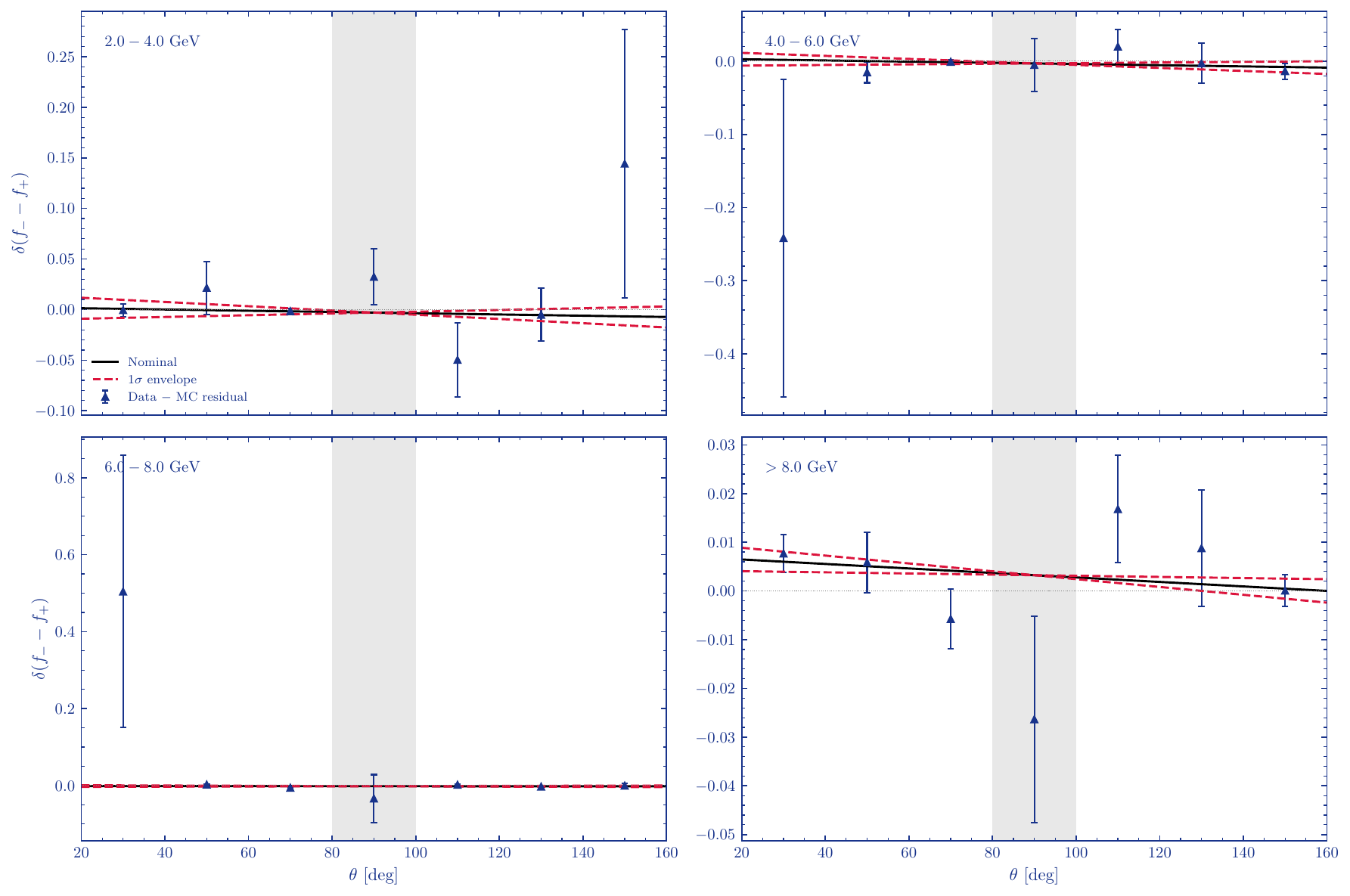}\\[0.5em]
    \includegraphics[width=0.85\textwidth]{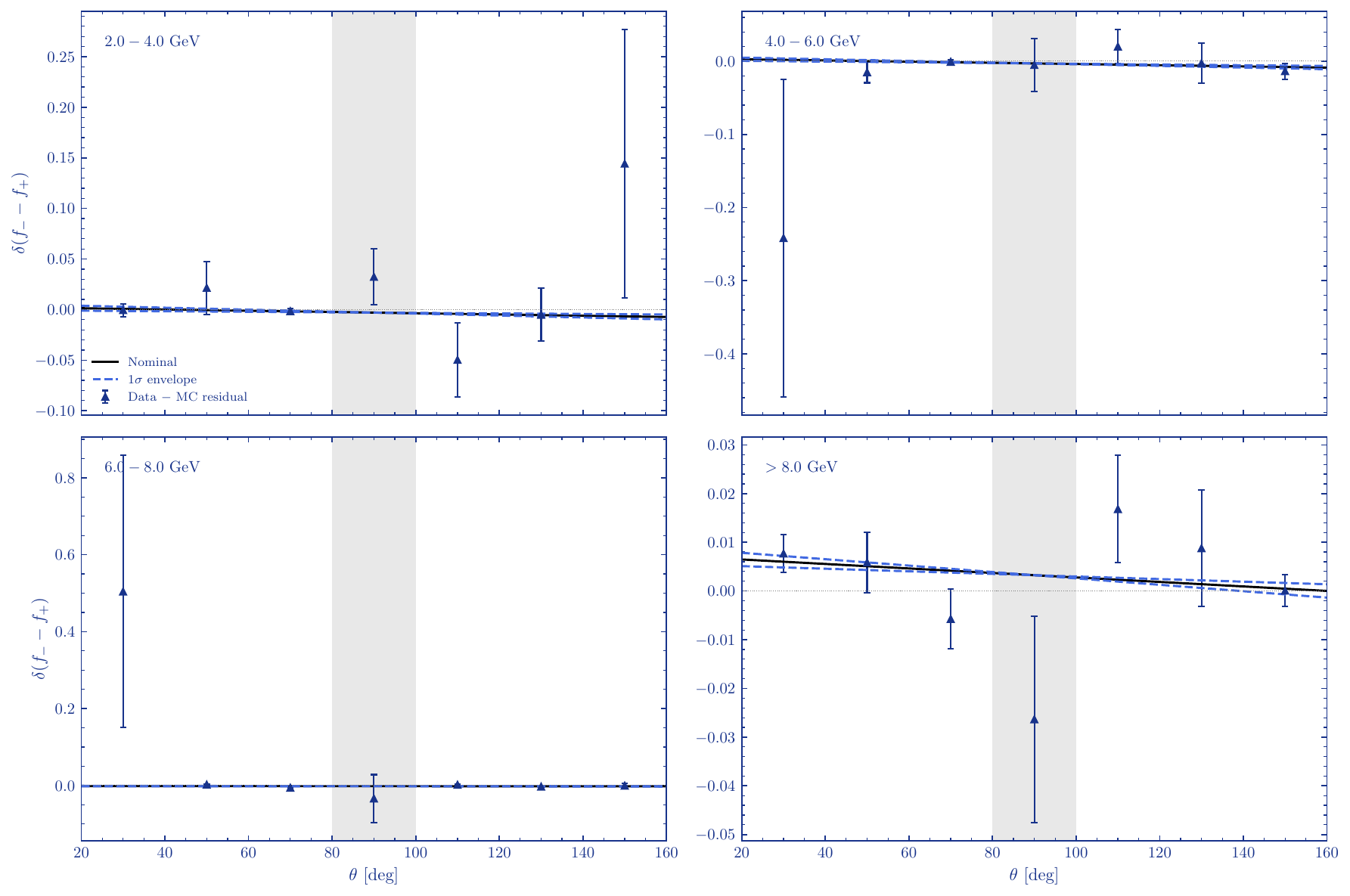}
    \caption{Cross-check counterpart of Figure~\ref{fig:misid_eigenmodes} for the 1995 sample: $\pm 1\sigma$ envelopes of the two eigenmodes ($\lambda_+$, top; $\lambda_-$, bottom) of the covariance matrix of the linear parametrization of $\delta_{\rm asym}(\theta)$, propagated as independent shift vectors representing the full statistical uncertainty of the data-driven charge-misreconstruction bias scale factor. The envelopes are visibly wider than in 1994, reflecting the smaller integrated luminosity of the 1995 $\tau$ control sample. The shaded band between $80^\circ$ and $100^\circ$ marks the bins containing the cathode plane of the DELPHI TPC.}
    \label{fig:app_misid_eigenmodes_95}
\end{figure}

\begin{figure}[ht!]
    \centering
    \includegraphics[width=0.6\textwidth]{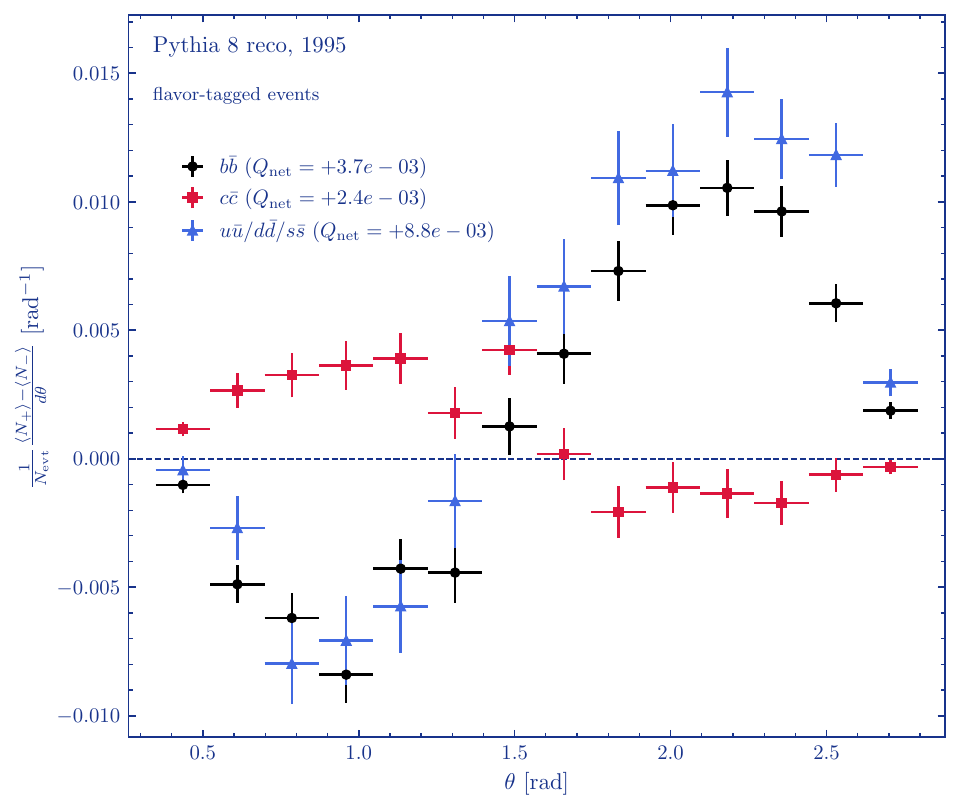}
    \caption{Same as Figure~\ref{fig:flavor_universality}, for the 1995 detector configuration.}
    \label{fig:app_flavor_95}
\end{figure}

\begin{figure}[ht!]
    \centering
    \includegraphics[width=0.45\textwidth]{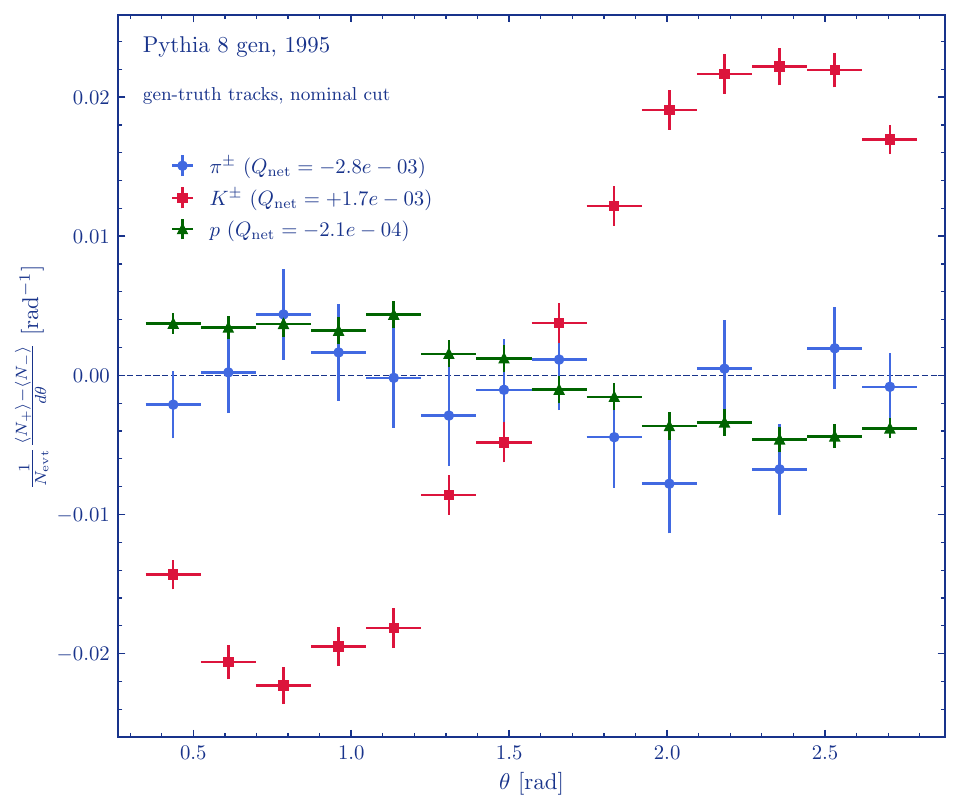}
    \includegraphics[width=0.45\textwidth]{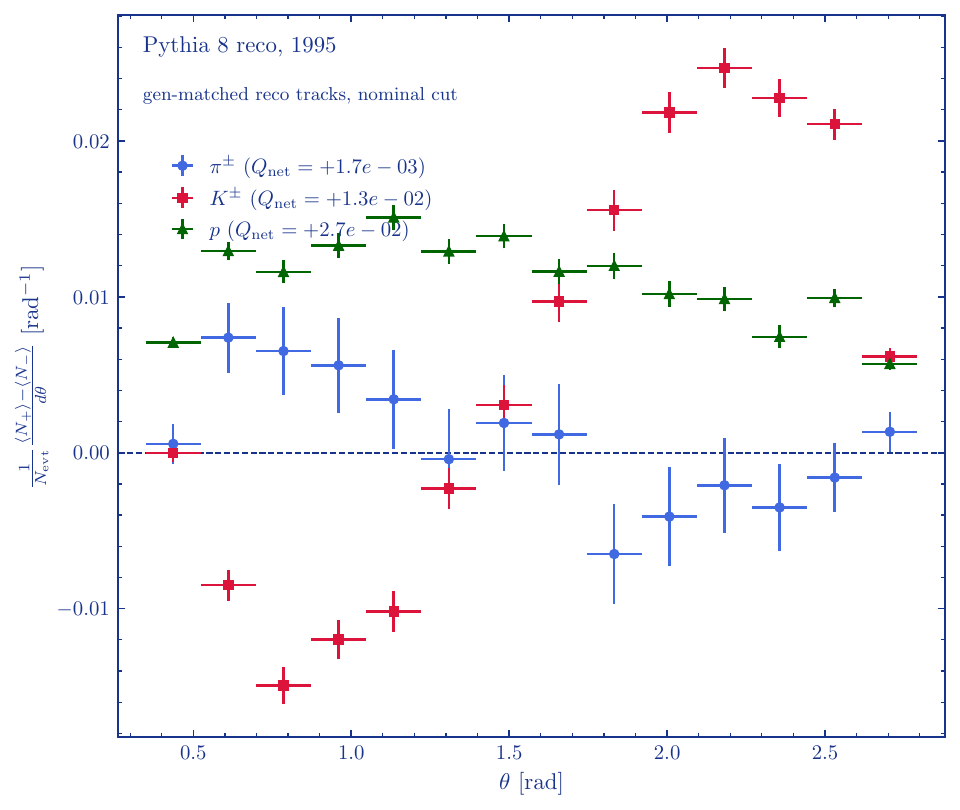}
    \caption{Same as Figure~\ref{fig:species_universality}, for the 1995 detector configuration.}
    \label{fig:app_species_95}
\end{figure}

\begin{figure}[ht!]
    \centering
    \includegraphics[width=0.48\textwidth]{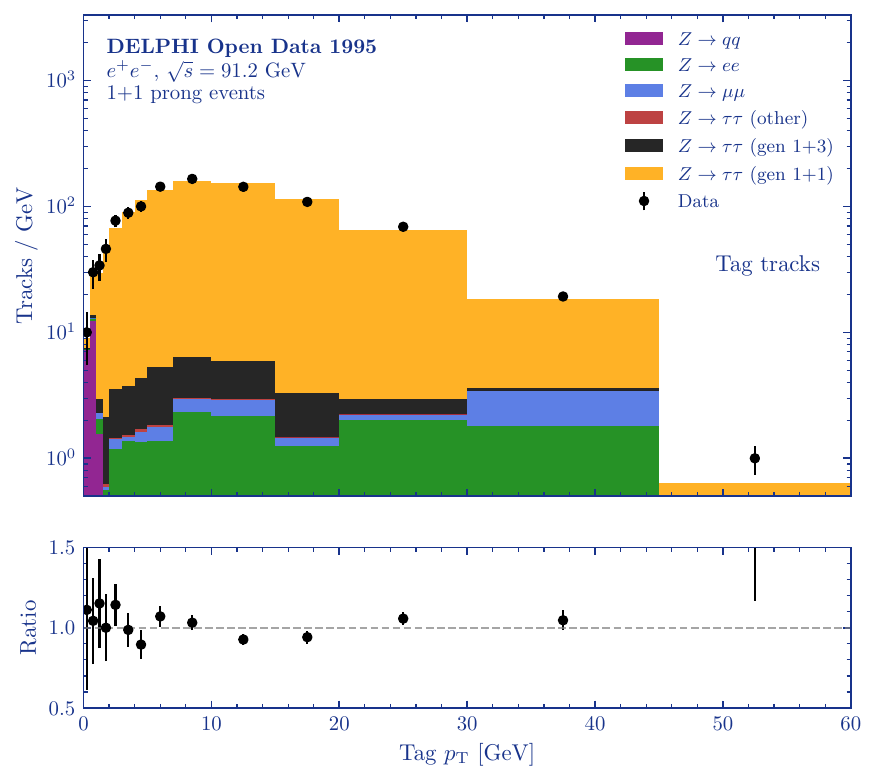}
    \includegraphics[width=0.48\textwidth]{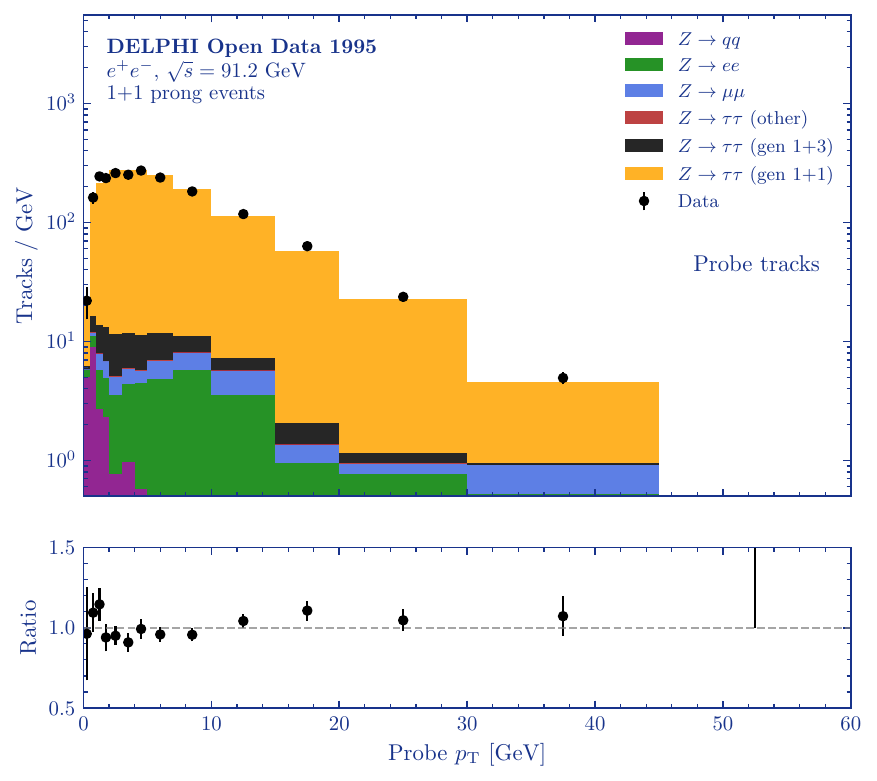}\\
    \includegraphics[width=0.48\textwidth]{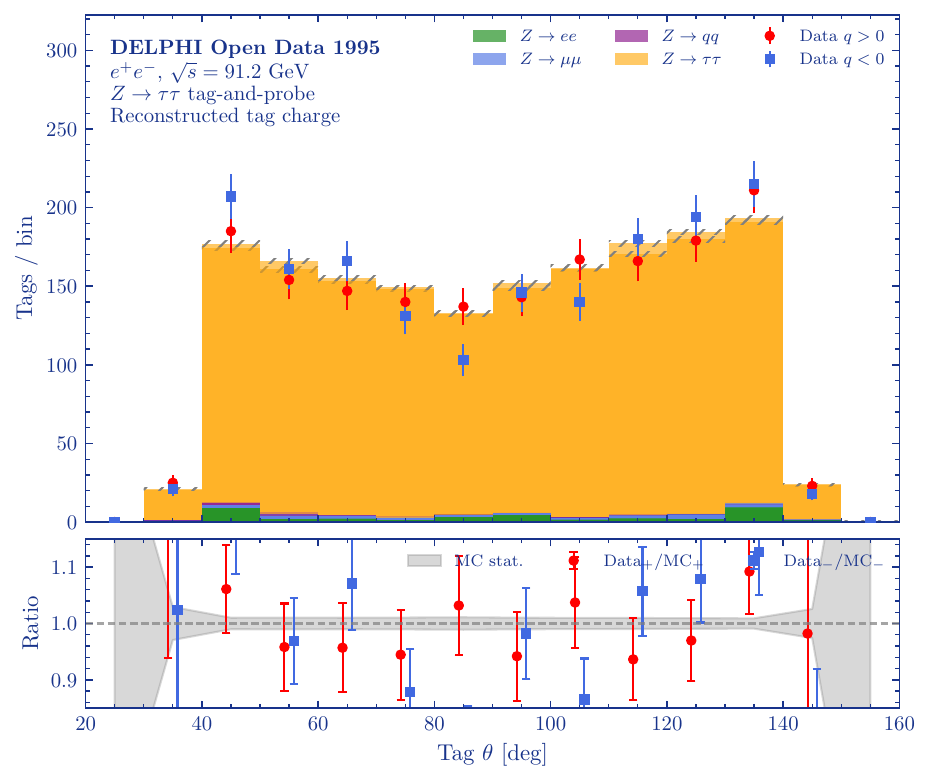}
    \includegraphics[width=0.48\textwidth]{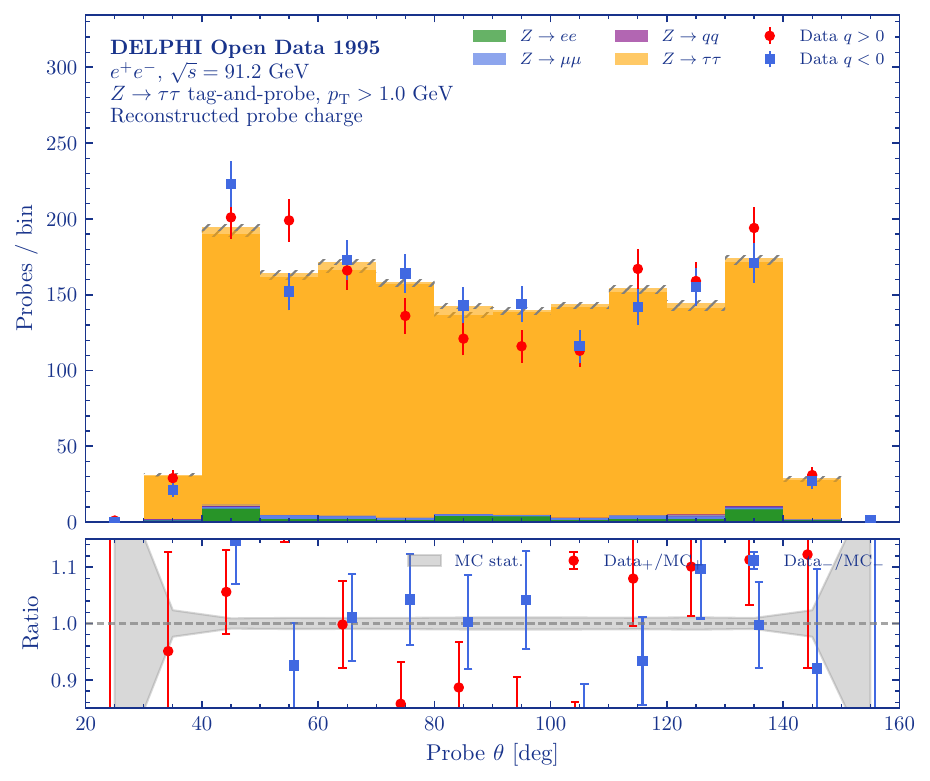}
    \caption{Cross-check counterpart of Figure~\ref{fig:tau_tnp_kin} for the 1995 sample: kinematic distributions of the tag and probe tracks in the $\tau$ control sample. Top row: track $p_{\rm T}$ distribution of the tag (left) and probe (right), with stacked MC contributions from $Z \to \tau^+\tau^-$, $Z \to \mu^+\mu^-$, $Z \to e^+e^-$, and $Z \to q\bar{q}$. Bottom row: polar-angle distribution of the tag (left) and probe (right) tracks separated by reconstructed charge, $q > 0$ (red circles) and $q < 0$ (blue squares); lower panels show the data/MC ratio. It is interesting to note that in the 1994 sample most of the high-quality tag tracks populate the forward hemisphere, whereas in the 1995 sample shown here the opposite is observed, with the tag population concentrated in the backward hemisphere. This year-to-year asymmetry is likely to reflect differences in the operating conditions of individual DELPHI tracking subdetectors between the two running periods.}
    \label{fig:app_tau_theta_95}
\end{figure}

The 1995 DELPHI Open Data sample corresponds to roughly 15~pb$^{-1}$ at the $Z$ resonance, a factor of three smaller than the 1994 sample.
This limited integrated luminosity translates directly into a much weaker statistical constraint on the data-driven charge misreconstruction correction extracted from the $\tau$ control region: the per-bin residuals shown in Figures~\ref{fig:app_misid_asym_95} and \ref{fig:app_misid_sum_95} are statistically much noisier than their 1994 counterparts (Figures~\ref{fig:misid_asymmetry} and \ref{fig:misid_sum}). 
The fully corrected 1995 result is consistent with the \textsc{PYTHIA}~8.3 prediction within its larger statistical uncertainties; a direct comparison between the fully corrected 1994 and 1995 samples is shown in Figure~\ref{fig:year_comparison}.
For completeness, the cross-check counterparts of the 1994 data--MC and TNP kinematic comparisons are shown in Figures~\ref{fig:app_data_mc_pt_95}, \ref{fig:app_data_mc_theta_95}, and \ref{fig:app_tau_theta_95}. The generator-level matching comparison of the per-track charge-misreconstruction rates between the $\tau$ and hadronic environments (Figure~\ref{fig:tau_misid_rate_mc}) is repeated for 1995 in Figure~\ref{fig:app_misid_rate_mc_95}. The 1995 counterpart of the eigenmode envelopes of the data-driven correction (Figure~\ref{fig:misid_eigenmodes}) is shown in Figure~\ref{fig:app_misid_eigenmodes_95}, and the corresponding systematic uncertainty summary is given in Table~\ref{tab:systematic_summary_95}. The 1995 flavor- and species-resolved MC decompositions, mirroring Figures~\ref{fig:flavor_universality} and~\ref{fig:species_universality}, are shown in Figures~\ref{fig:app_flavor_95} and~\ref{fig:app_species_95}. 

\begin{table}[h]
\centering
\caption{Relative systematic uncertainties (\%) on the corrected one-point charge correlator at two representative $\theta$ bins ($p_{\mathrm{T}} > 2.0$ GeV, 1995). Each per-source entry is obtained by imposing the charge-conservation constraint on that source alone, whereas the total is obtained by imposing it on all sources simultaneously; the entries are therefore not additive.}
\label{tab:systematic_summary_95}
\begin{tabular}{lcc}
\hline\hline
Source & $\theta = 55^\circ$ & $\theta = 125^\circ$ \\
\hline
Charged-track finding efficiency & 16.74 & 13.13 \\
Charge misreconstruction asymmetry residual & 47.12 & 40.75 \\
Charge misreconstruction dilution residual & 0.23 & 0.16 \\
Tag-and-probe selection bias & 3.34 & 3.08 \\
Tag-and-probe sample contamination modeling & 1.39 & 1.17 \\
MC generator model (\textsc{PYTHIA}~8 vs \textsc{KK}2f) & 1.17 & 4.17 \\
Tag-and-probe sample contamination bias & 6.79 & 3.55 \\
\hline
Total systematic (post-constraint) & 50.27 & 39.50 \\
Statistical & 34.87 & 27.18 \\
\hline\hline
\end{tabular}
\end{table}

\clearpage
\section{General cross-checks}
\label{app:fiducial_check}

This appendix collects a set of cross-checks that verify the robustness of the nominal result against alternative analysis choices, including alternative fiducial regions (the total-momentum cut $|p| > 2$~GeV and a tighter momentum-resolution requirement $\Delta p/p < 0.5$), a loosened impact-parameter selection ($d_0 \le 4$~cm, $z_0 \le 10$~cm), an alternative quadratic-in-$\theta$ parametrization of the data-driven charge-misreconstruction residual, and alternative angular and probe-$p_{\rm T}$ binnings of the control-region extraction. In every case the corrected charge correlator remains consistent with the \textsc{PYTHIA}~8.3 prediction within the combined statistical and systematic uncertainties of the cross-check configuration.

\begin{figure}[ht!]
    \centering
    \includegraphics[width=0.45\textwidth]{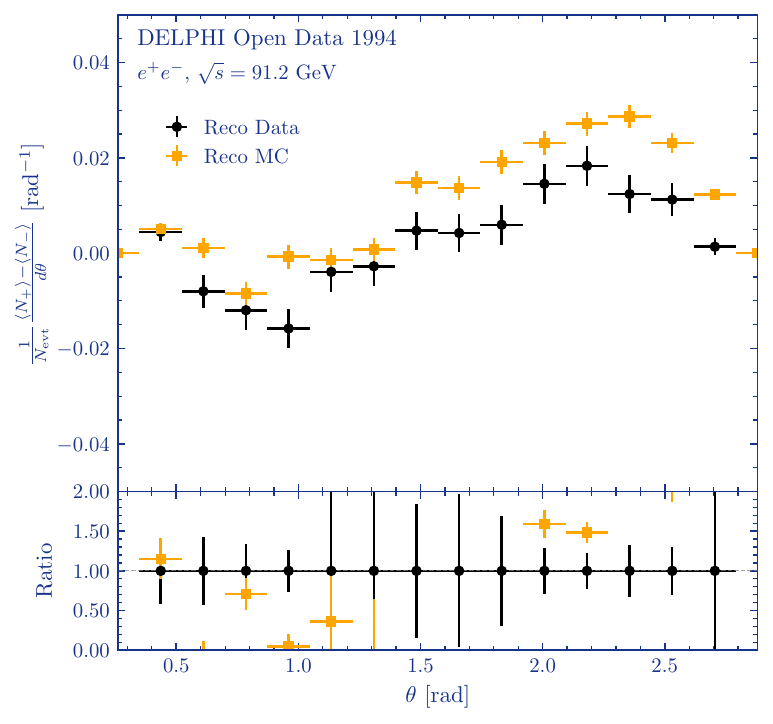}
    \includegraphics[width=0.45\textwidth]{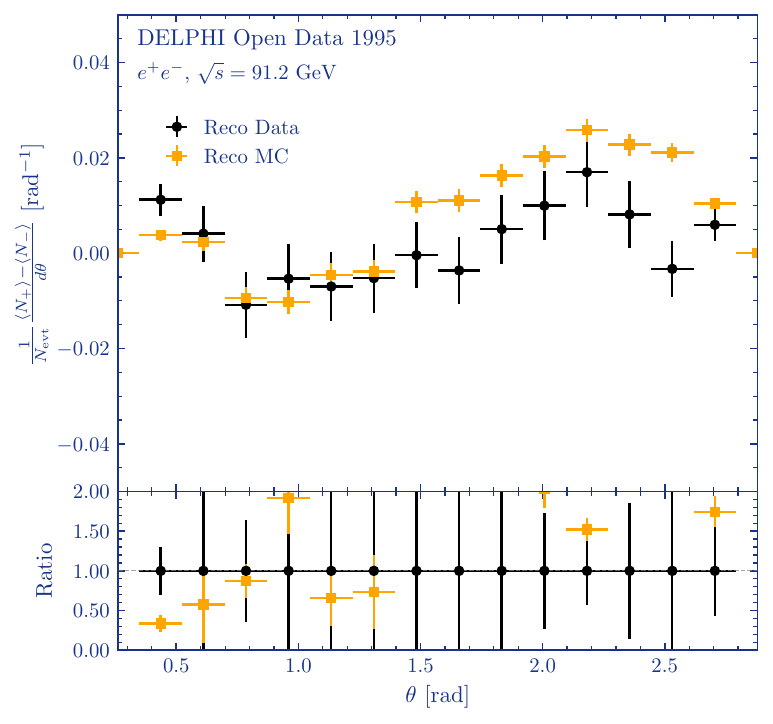}
    \caption{Cross-check counterpart of Figure~\ref{fig:data_mc_fbcc} for the $|p| > 2$~GeV selection: raw one-point charge correlator $\mathcal{Q}(\theta) = N^+(\theta) - N^-(\theta)$ between data (black) and reconstructed \textsc{PYTHIA}~8.3 MC (yellow), for the 1994 (left) and 1995 (right) datasets, with the data/MC ratio shown in the lower panels.}
    \label{fig:app_pmag2_inputs}
\end{figure}
\begin{figure}[ht!]
    \centering
    \includegraphics[width=0.45\textwidth]{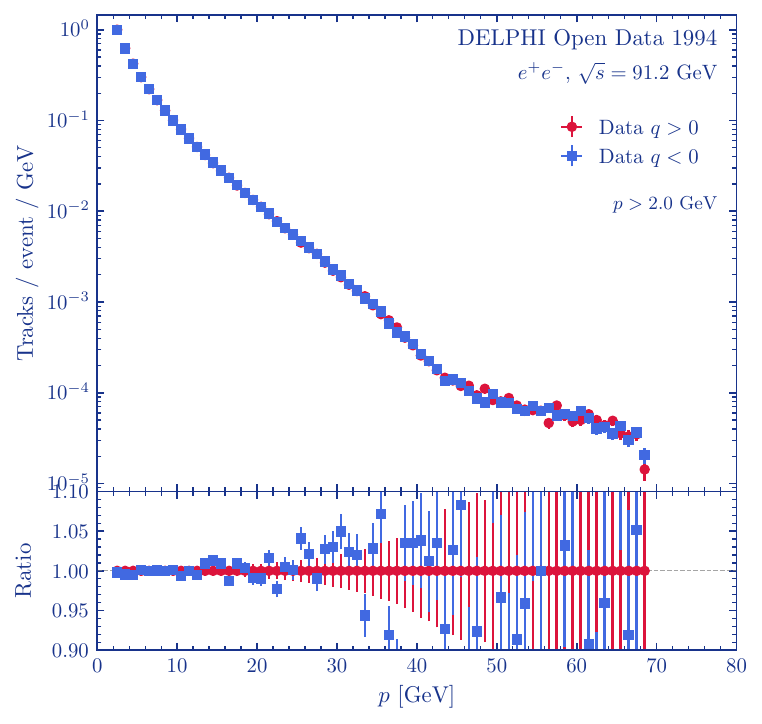}
    \includegraphics[width=0.45\textwidth]{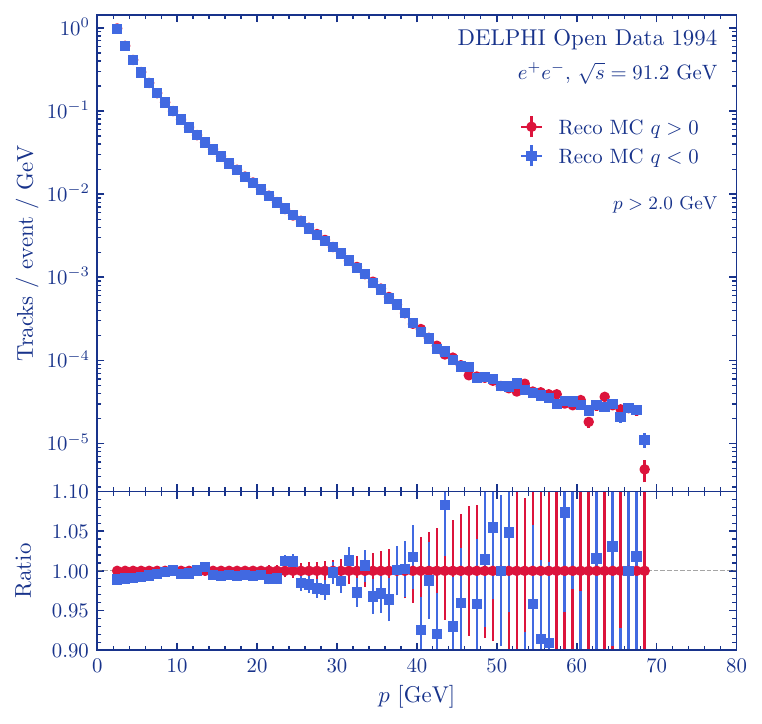}
    \caption{Charged-track momentum spectra in 1994 data (left) and reconstructed \textsc{PYTHIA}~8.3 MC (right), separated by reconstructed charge, for the $|p| > 2$~GeV selection. Compared to the nominal $p_{\rm T} > 2$~GeV selection (Figure~\ref{fig:data_mc_theta}), the reconstructed MC exhibits a larger charge imbalance under the $|p|$ cut, while the data remain charge-symmetric.}
    \label{fig:app_pmag2_trackp}
\end{figure}
\begin{figure}[ht!]
    \centering
    \includegraphics[width=0.55\textwidth]{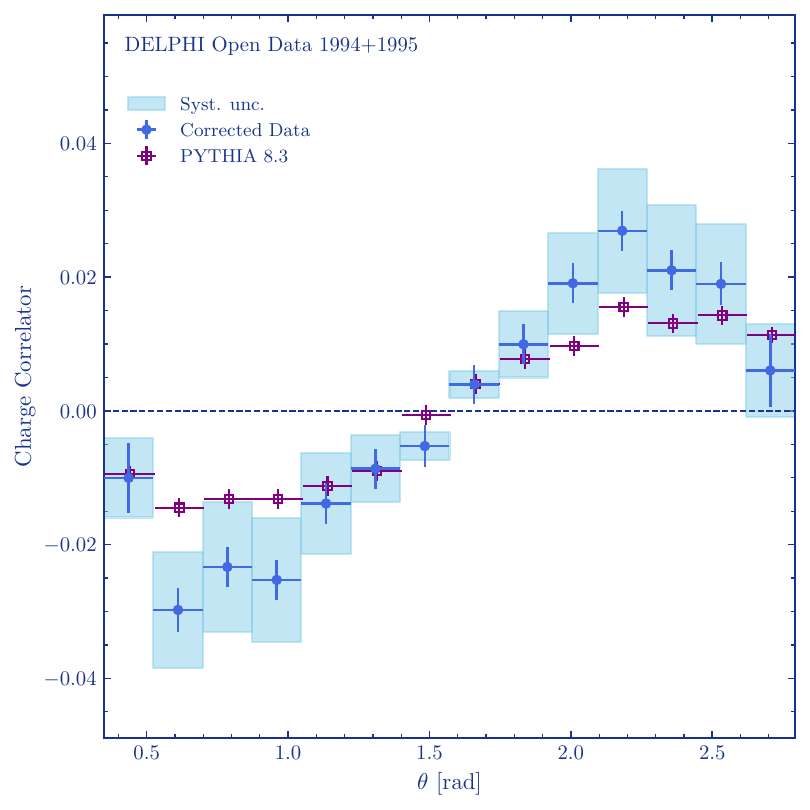}
    \caption{Cross-check counterpart of Figure~\ref{fig:final_result_pythia} for the $|p| > 2$~GeV selection: fully corrected one-point charge correlator (blue) using the DELPHI 1994 and 1995 datasets, compared to the \textsc{PYTHIA}~8.3 generator-level prediction (purple dashed). Data points are shown with statistical uncertainties; the light blue shaded boxes indicate the systematic uncertainties.}
    \label{fig:app_pmag2_corrected}
\end{figure}

As discussed in Section~\ref{sec:data_samples}, the nominal fiducial selection for this analysis is $p_{\rm T} > 2$~GeV, which is chosen because it effectively removes soft tracks that are sensitive to charge mismodeling due to multiple Coulomb scattering, without distorting the $\sin(2\theta)$ modulation. As a cross-check, the analysis is repeated using the less optimal $|p| > 2$~GeV selection, which retains more soft tracks. The same correction chain (Sections~\ref{sec:mc_correction} and~\ref{sec:data_driven}) is applied. Figure~\ref{fig:app_pmag2_inputs} shows the input data--MC comparison, Figure~\ref{fig:app_pmag2_trackp} shows the corresponding track-momentum spectra, and Figure~\ref{fig:app_pmag2_corrected} shows the fully corrected charge correlator compared to the \textsc{PYTHIA}~8.3 generator-level prediction for this selection. The corrected result remains consistent with the prediction, albeit with a larger residual discrepancy than in the nominal selection, as expected from the increased sensitivity of the soft-track population to charge-reconstruction mismodeling.

\begin{figure}[ht!]
    \centering
    \includegraphics[width=0.45\textwidth]{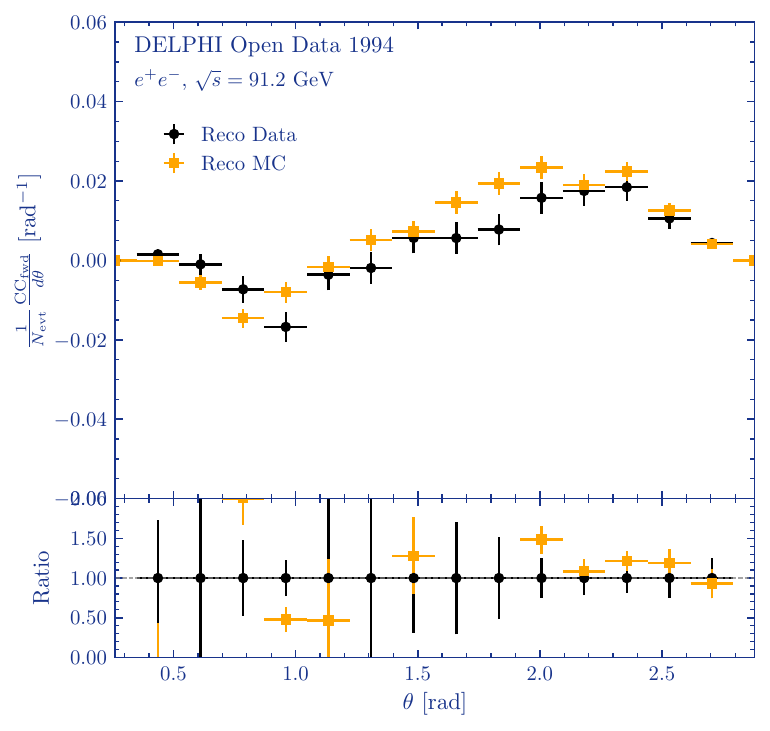}
    \includegraphics[width=0.45\textwidth]{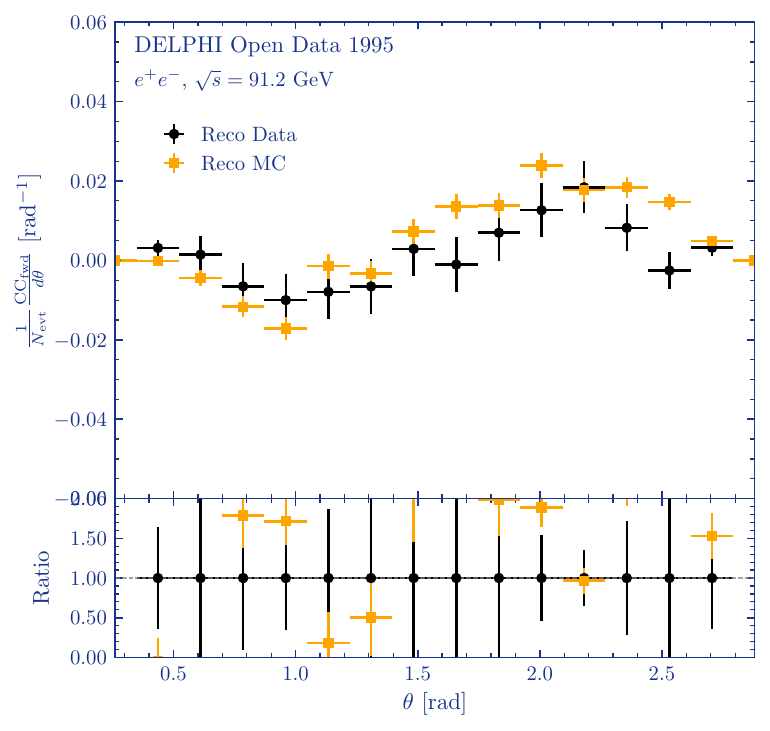}
    \caption{Cross-check counterpart of Figure~\ref{fig:data_mc_fbcc} with the additional $\Delta p/p < 0.5$ momentum-resolution requirement: raw one-point charge correlator $\mathcal{Q}(\theta) = N^+(\theta) - N^-(\theta)$ between data (black) and reconstructed \textsc{PYTHIA}~8.3 MC (yellow), for the 1994 (left) and 1995 (right) datasets, with the data/MC ratio shown in the lower panels.}
    \label{fig:app_dpp_inputs}
\end{figure}

\begin{figure}[ht!]
    \centering
    \includegraphics[width=0.55\textwidth]{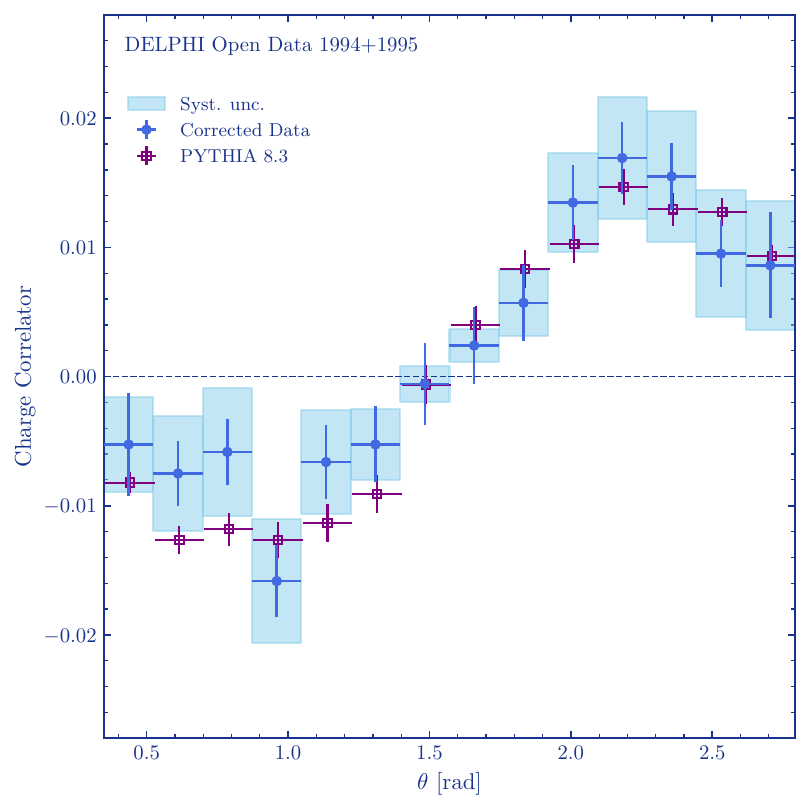}
    \caption{Cross-check counterpart of Figure~\ref{fig:final_result_pythia} with the additional $\Delta p/p < 0.5$ momentum-resolution requirement: fully corrected one-point charge correlator (blue) using the DELPHI 1994 and 1995 datasets, compared to the \textsc{PYTHIA}~8.3 generator-level prediction (purple dashed). Data points are shown with statistical uncertainties; the light blue shaded boxes indicate the systematic uncertainties.}
    \label{fig:app_dpp_corrected}
\end{figure}

As an additional cross-check on the track-quality requirements, the analysis is repeated with a tighter relative momentum-resolution cut of $\Delta p/p < 0.5$ applied on top of the nominal selection, while keeping all other elements of the analysis identical. Figure~\ref{fig:app_dpp_inputs} shows the input data--MC comparison of the raw one-point charge correlator for the 1994 and 1995 datasets, and Figure~\ref{fig:app_dpp_corrected} shows the fully corrected charge correlator combining the two years, compared to the \textsc{PYTHIA}~8.3 generator-level prediction. The corrected result is consistent with the nominal extraction within statistical uncertainties, confirming the robustness of the measurement against the choice of momentum-resolution requirement.

As a complementary check on the track-quality requirements, the analysis is repeated with the impact-parameter cuts loosened from the nominal $d_0 \le 0.6$~cm, $z_0 \le 1.0$~cm to $d_0 \le 4$~cm, $z_0 \le 10$~cm, while keeping all other elements of the analysis identical. The looser selection admits more displaced and secondary tracks, which degrade the input charge balance (Figure~\ref{fig:app_d0z0_compare} of Appendix~\ref{app:additional_datamc}). Figure~\ref{fig:app_d0z0_loose_corrected} shows the fully corrected charge correlator combining the two years, compared to the \textsc{PYTHIA}~8.3 generator-level prediction. The corrected result remains consistent with the prediction, with a residual difference expected from the degraded charge reconstruction for this selection.

\begin{figure}[ht!]
    \centering
    \includegraphics[width=0.55\textwidth]{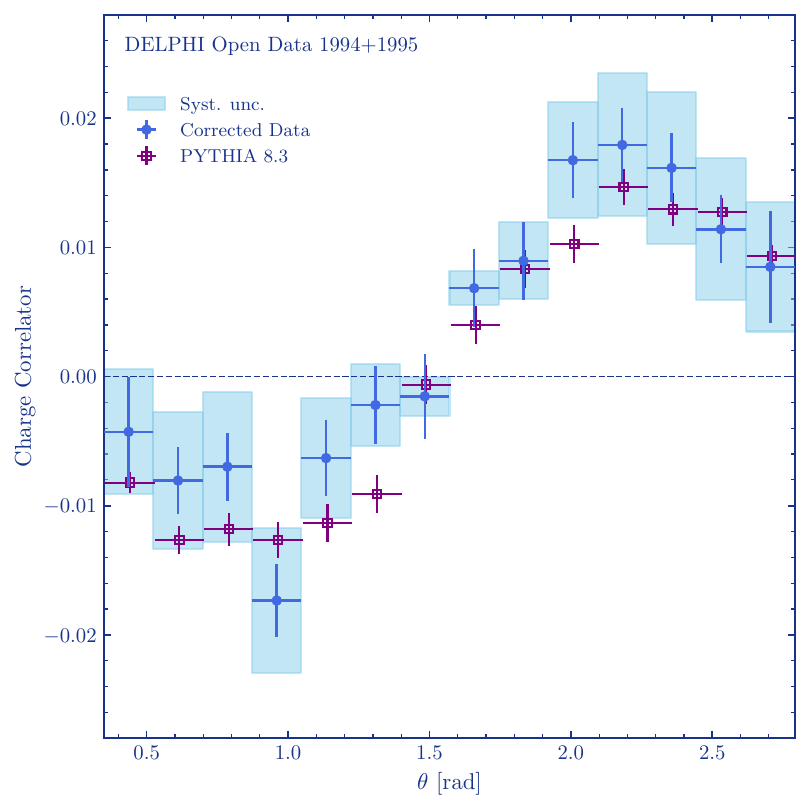}
    \caption{Cross-check counterpart of Figure~\ref{fig:final_result_pythia} with the loosened impact-parameter selection ($d_0 \le 4$~cm, $z_0 \le 10$~cm): fully corrected one-point charge correlator (blue) using the DELPHI 1994 and 1995 datasets, compared to the \textsc{PYTHIA}~8.3 generator-level prediction (purple dashed). Data points are shown with statistical uncertainties; the light blue shaded boxes indicate the systematic uncertainties.}
    \label{fig:app_d0z0_loose_corrected}
\end{figure}

\begin{figure}[ht!]
    \centering
    \includegraphics[width=\textwidth]{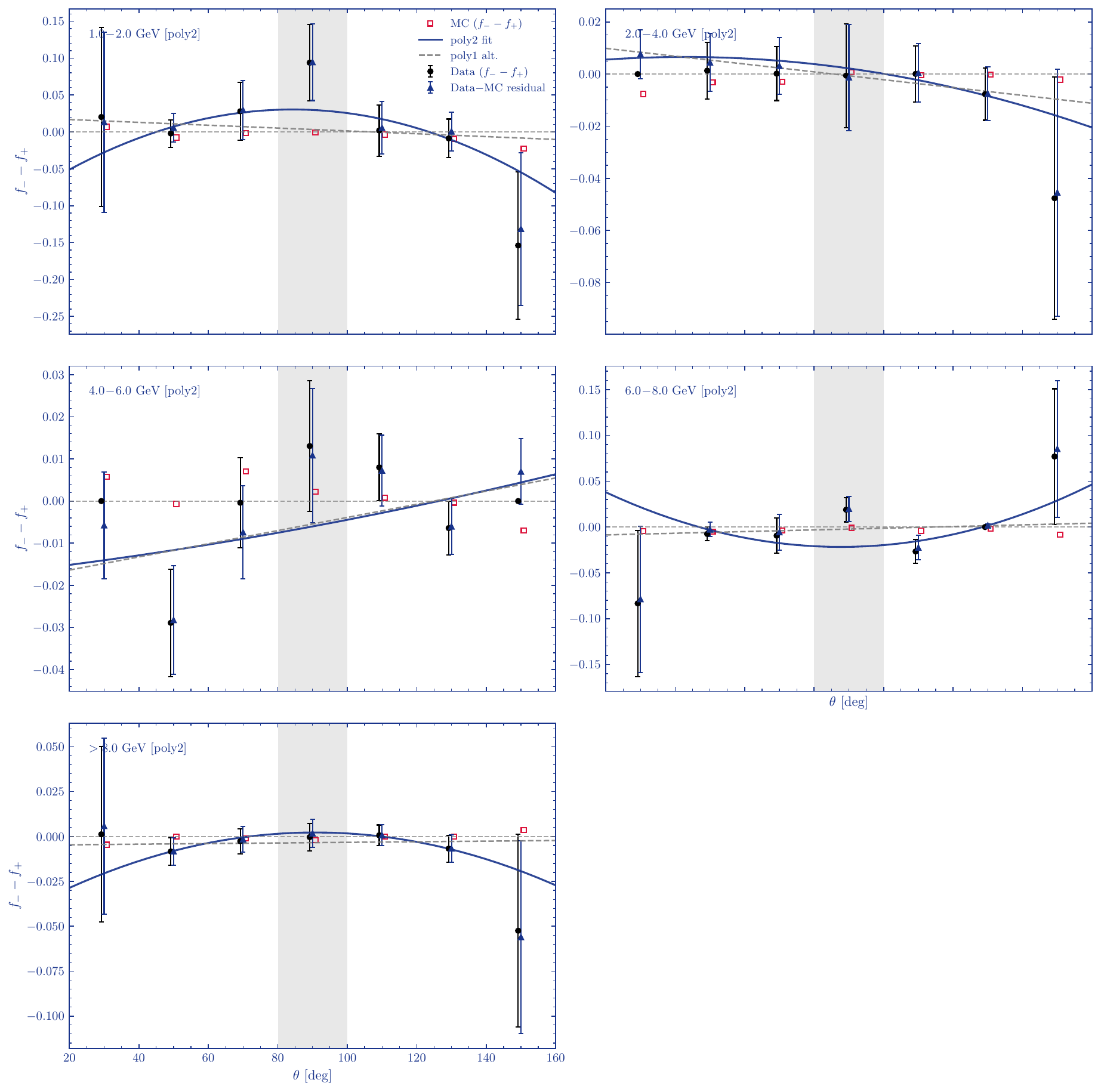}
    \caption{Cross-check counterpart of Figure~\ref{fig:misid_asymmetry}: data--MC residual of the charge misreconstruction asymmetry $\delta_{\rm asym}^i(\theta)$ in the five exclusive probe-$p_{\rm T}$ bins, overlaid with a quadratic-in-$\theta$ fit instead of the nominal linear-in-$\theta$ fit.}
    \label{fig:app_misid_poly2}
\end{figure}
\begin{figure}[ht!]
    \centering
    \includegraphics[width=0.55\textwidth]{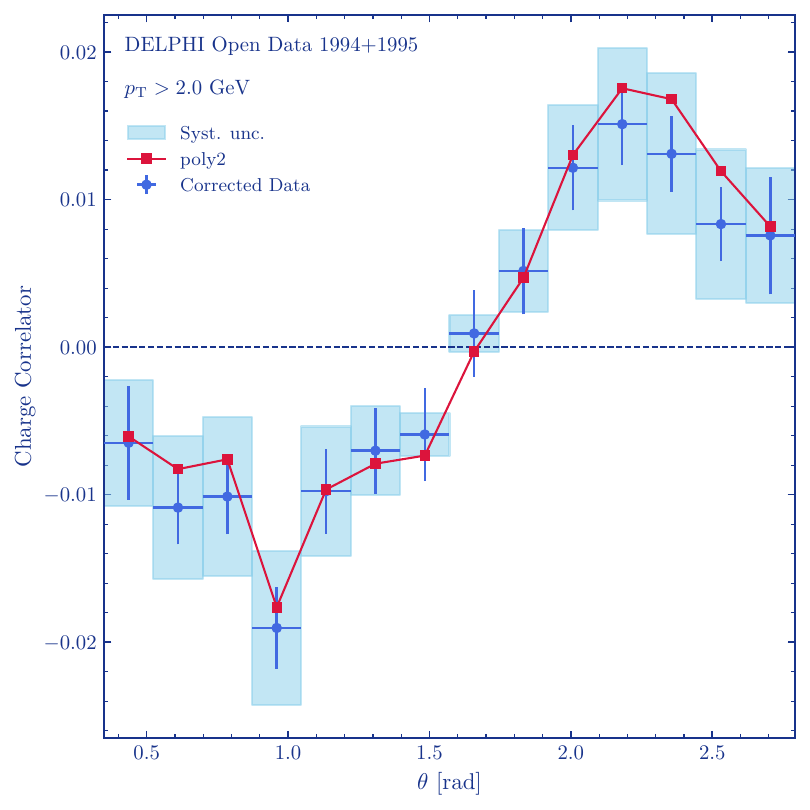}
    \caption{Comparison of the fully corrected one-point charge correlator (1994$+$1995 combined) obtained with the nominal linear-in-$\theta$ charge-misreconstruction residual parametrization and the alternative quadratic-in-$\theta$ parametrization, compared to the \textsc{PYTHIA}~8.3 generator-level prediction (purple dashed). The two parametrizations yield consistent results, with differences covered by the systematic uncertainty.}
    \label{fig:app_misid_model_comparison}
\end{figure}

The nominal data-driven correction parametrizes the data--MC residual $\delta_{\rm asym}^i(\theta)$ as a linear function of $\theta$ within each probe-$p_{\rm T}$ bin (Section~\ref{sec:data_driven}). To test the sensitivity of the corrected result to this functional-form choice, the residual fit is repeated using a second-order polynomial in $\theta$, while keeping all other elements of the analysis identical. Figure~\ref{fig:app_misid_poly2} shows the per-$p_{\rm T}$-bin residuals overlaid with the quadratic fit, and Figure~\ref{fig:app_misid_model_comparison} compares the fully corrected charge correlator between the nominal linear and the alternative quadratic parametrizations. The corrected result is consistent with the nominal extraction within statistical uncertainties, demonstrating that the choice of charge-misreconstruction residual parametrization does not bias the final result.

The data-driven charge-misreconstruction correction is extracted from the $\tau$ control region in bins of the probe polar angle $\theta$ and of the probe transverse momentum $p_{\rm T}$ (Section~\ref{sec:data_driven}). The robustness of the final result against these binning choices is tested by repeating the full extraction and correction with alternative binnings, keeping all other elements of the analysis identical.

\begin{figure}[ht!]
    \centering
    \includegraphics[width=0.55\textwidth]{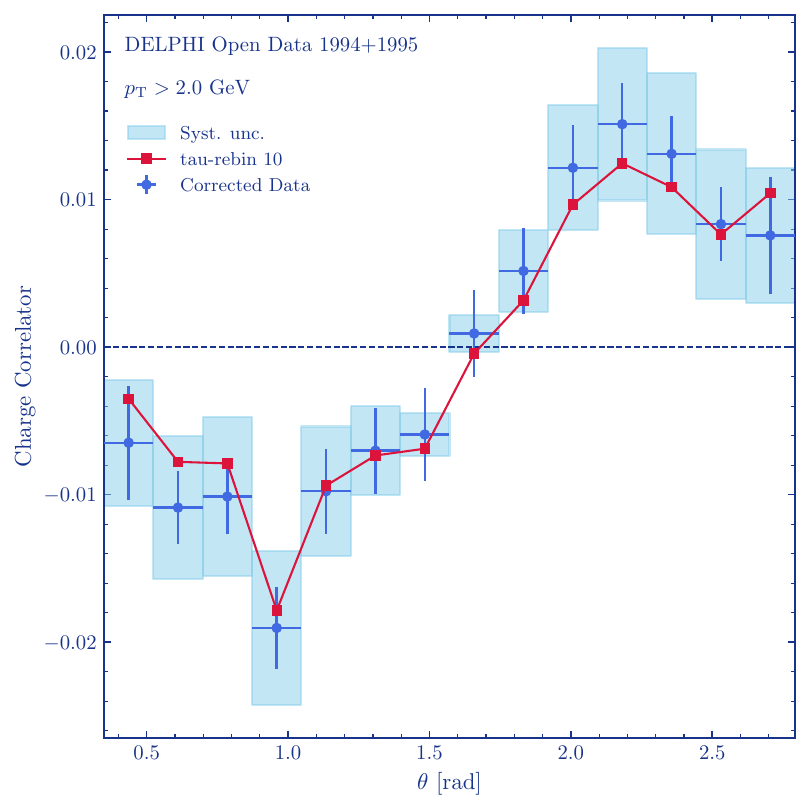}
    \caption{Fully corrected one-point charge correlator (1994$+$1995 combined) obtained with the control-region charge-misreconstruction extraction performed in $10^\circ$-wide angular bins, compared to the nominal $20^\circ$-wide binning and to the \textsc{PYTHIA}~8.3 generator-level prediction. The two binnings yield consistent results, with differences covered by the systematic uncertainty.}
    \label{fig:app_tau_rebin}
\end{figure}

In the first variation, the angular granularity of the residual extraction is increased from the nominal $20^\circ$-wide bins to $10^\circ$-wide bins, doubling the number of $\theta$ bins in which the data--MC residuals $\delta_{\rm asym}^i(\theta)$ and $\delta_{\rm sum}^i(\theta)$ are sampled before the linear fit. Figure~\ref{fig:app_tau_rebin} compares the shape of the fully corrected charge correlator obtained with the refined angular binning to that of the nominal result. The difference in shape between the two is covered by the total systematic uncertainty.

\begin{figure}[ht!]
    \centering
    \includegraphics[width=0.55\textwidth]{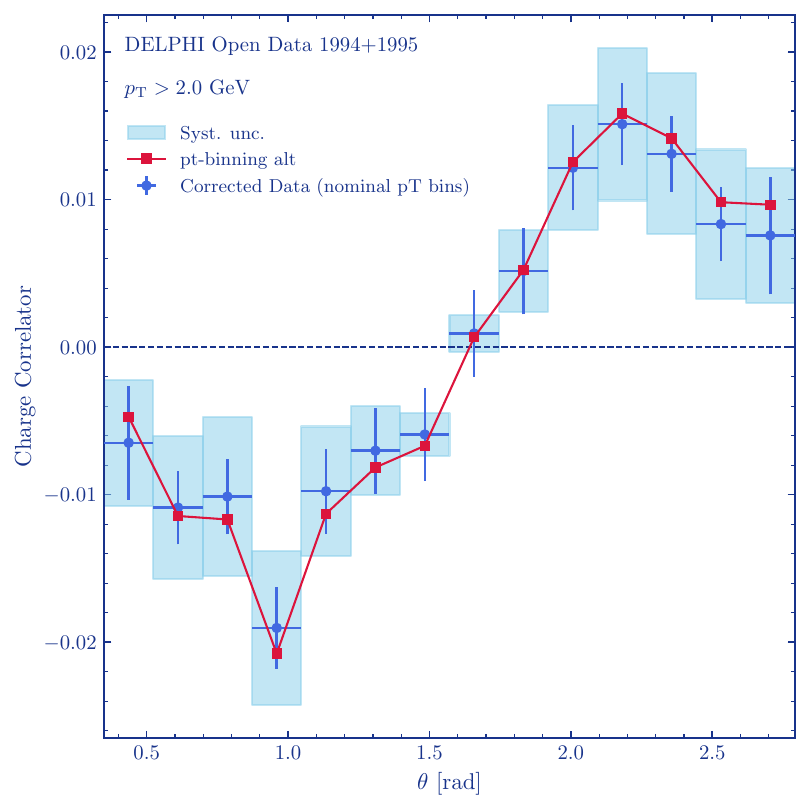}
    \caption{Fully corrected one-point charge correlator (1994$+$1995 combined) obtained with the finer probe-$p_{\rm T}$ binning ($1$--$2$, $2$--$3$, $3$--$4$, $4$--$5$, $5$--$6$, $>6$~GeV) used in the control-region charge-misreconstruction extraction, compared to the nominal binning ($1$--$2$, $2$--$4$, $4$--$6$, $6$--$8$, $>8$~GeV) and to the \textsc{PYTHIA}~8.3 generator-level prediction. The two binnings yield consistent results, with differences covered by the systematic uncertainty.}
    \label{fig:app_pt_binning}
\end{figure}

In the second variation, the probe-$p_{\rm T}$ binning is changed from the nominal scheme ($1$--$2$, $2$--$4$, $4$--$6$, $6$--$8$, $>8$~GeV) to a finer set of bins ($1$--$2$, $2$--$3$, $3$--$4$, $4$--$5$, $5$--$6$, $>6$~GeV), altering how the $\tau$-derived residual is sampled in momentum before being reweighted to the hadronic-event spectrum (Equations~\ref{eq:pt_reweight}--\ref{eq:had_weights}). Figure~\ref{fig:app_pt_binning} compares the shape of the resulting fully corrected charge correlator to that of the nominal result. The difference in shape between the two is covered by the total systematic uncertainty.

\end{appendices}

\end{document}